\documentclass[a4paper]{book}

\usepackage{ucs}
\usepackage[utf8x]{inputenc}
\usepackage[catalan,american]{babel}
\usepackage{syntonly}
\usepackage{fancyhdr}
\usepackage[body={6.0in, 8.2in},left=1.25in,right=1.25in]{geometry}
\makeatletter
    \renewcommand{\chaptermark}[1]{         % Lower Case Chapter marker style
      \markboth{\@chapapp\ %\chaptername\ 
\thechapter.\ #1}{}} %
\makeatother
\makeatletter
    \def\cleardoublepage{\clearpage\if@twoside \ifodd\c@page\else%
        \hbox{}%
        \thispagestyle{empty}%              % Empty header styles
        \newpage%
        \if@twocolumn\hbox{}\newpage\fi\fi\fi}
    \makeatother
\usepackage{graphicx}
\usepackage{amssymb}
\usepackage{pifont}

\def\bfnabla{\mbox{\boldmath $\nabla$}}
\def\bfsigma{\mbox{\boldmath $\sigma$}}
\def\lQ{\Lambda_{\rm QCD}}
\newcommand{\be}{\begin{equation}}{\bf }
\newcommand{\ee}{\end{equation}}
\newcommand{\bea}{\begin{eqnarray}}
\newcommand{\eea}{\end{eqnarray}}
\def\als{\alpha_{\rm s}}
\def\siml{{\ \lower-1.2pt\vbox{\hbox{\rlap{$<$}\lower6pt\vbox{\hbox{$\sim$}}}}\ }}

\begin{document}

%\maketitle

%%%%%%%%%%%%%%%%%%%%%%%%%%%%%%%%%%%%%%%%%%%%%%%%%%%%%%%%%%%%%%%%%%%%%%%%%%%%%%%
          %   Title Page                                                                %
          %%%%%%%%%%%%%%%%%%%%%%%%%%%%%%%%%%%%%%%%%%%%%%%%%%%%%%%%%%%%%%%%%%%%%%%%%%%%%%%

          \begin{titlepage}
          \begin{center}

                                        \vspace*{\stretch{6}}

                                        {\bf {\Huge Applications of effective field
                                        theories to the strong interactions of
                                        heavy quarks}}\\
                                        %\vspace{1mm}{\bf {\Huge of random surface models}}\\
                                        %\vspace{3mm}{\bf {\Huge and the QCD string}}

                                        \vspace*{\stretch{4}}
                                        
                                        {\bf {\Large Xavier Garcia i Tormo}}
                                        
                                        \vspace*{\stretch{1}}

                                        {\bf May 1, 2006}

                                        \vspace*{\stretch{3}}

%                                       \includegraphics[]{Figures/Logo.ps}
%                                       
%                                       \vspace*{\stretch{1}}

                                        { {\Large {  Universitat de Barcelona}}}
                                        
                                        \vspace*{\stretch{1}}
                                        
                                        { {\Large { Departament d'Estructura i Constituents de la Mat\`eria}}}
                                        \vspace{3mm}
                                        \begin{center}
                                               \includegraphics[width=73mm]{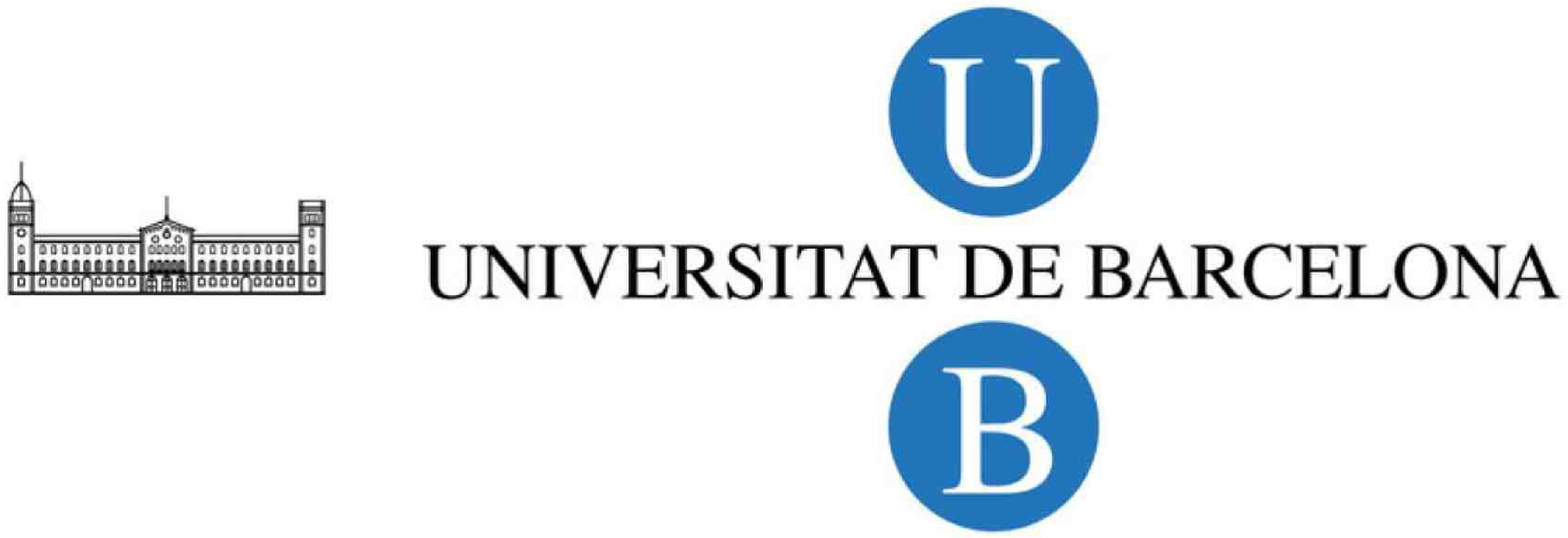}
                                        \end{center}

          \end{center}

          \end{titlepage}
                                        
                                        \pagebreak

                                        \chapter*{   }
                                        \thispagestyle{empty}                                   

                                        \begin{center}
                                        
                                        \vspace*{\stretch{1}}                                   

                                        {\bf {\Huge Applications of effective field
                                        theories to the strong interactions of
                                        heavy quarks}}\\
%                                        \vspace{1mm}{\bf {\Huge of random surface models}}\\
%                                        \vspace{3mm}{\bf {\Huge and the QCD string}}
                                        
                                        \vspace*{\stretch{2}}                                   
                                        
                                        Mem\`oria de la tesi presentada 

                                        per en Xavier Garcia i Tormo per optar

                                        al grau de Doctor en Ci\`encies F\'\i siques
                                        
                                        \vspace*{\stretch{1}}
                                        
                                        {Director de la tesi: Dr. Joan Soto i Riera}
                                        
                                        \vspace*{\stretch{4}}
                                        
                                        Departament d'Estructura i Constituents de la Mat\`eria

%                                       {\it ``F\'\i sica Te\`orica''}}

                                        Programa de doctorat de {\it ``F\'\i sica avan\c cada''} 

                                        Bienni 2001-2003

                                        {\bf Universitat de Barcelona}
                                        
                                        \vspace*{\stretch{2}}

                                        \end{center}

                                        \pagebreak

                                        \chapter*{   }
                                        \thispagestyle{empty}

                                        \begin{flushright}
                                        
                                        \vspace*{\stretch{1}}
 
                                        No diguis blat\\fins que no sigui al sac i ben lligat\\

\vspace{.2cm}

DITA POPULAR
           
%\vspace{1cm}
%                            
%                                        Luke, I am your father\\
%
%\vspace{.2cm}
%
%DARTH VADER                                  
                                        
                                        \vspace*{\stretch{4}}
                                        
                                        \end{flushright}

                                        \pagebreak

\chapter*{Acknowledgments}
\thispagestyle{empty}
%\begin{center}
%{\bf\Large{Acknowledgments}}
%\end{center}

%\vspace{3cm}

%Thanks for reading!
Finally I have finished my thesis. The process of preparing and writing this thesis has lasted for about four to five years. Several people has helped or encouraged me during that time. I would like to mention them here (hoping not to forget anybody!).
\selectlanguage{catalan}

En primer lloc (i de manera especial) agrair a en Joan Soto el seu ajut durant tot aquest temps. Òbviament sense la seva ajuda, dirigint-me la tesi, aquest document no hagués estat possible. També agrair a l'Antonio Pineda l'ajuda i els consells durant aquest temps (i gràcies per avisar la meva mare i el meu germà que ja sortia de l'avió provinent d'Itàlia ;-) ). Gràcies també a en Lluís Garrido per ajudar-nos en diverses ocasions amb el tractament de dades experimentals. També agrair a Ignazio Scimemi (amb qui més pàgines de càlcul he compartit) i, en definitiva, a tota aquella gent amb qui he estat interactuant en el departament durant aquest temps.

\selectlanguage{american}
I would also like to thank Nora and Antonio for their help and for the kind hospitality all the times I have been in Milano (and special thanks for the nice dinners!). Thanks also to Matthias Neubert for his help. This year I spent two months in Boston, I would like to thank specially Iain Stewart for the kind hospitality during the whole time I was there. I am also very grateful to the other people at the CTP who made my stay there so enjoyable. Thanks also to my roommates in Boston.  

\selectlanguage{catalan}
Aquesta tesi ha estat realitzada amb el suport del Departament d'Universitats Recerca i Societat de la Informació de la Generalitat de Catalunya i del Fons Social Europeu.

No puc deixar d'anomenar als companys i amics del departament, amb qui he estat compartint el dia a dia durant aquests anys. En primer lloc als meus actuals, Míriam, Carlos i Román (àlies Tomàs), i anteriors, Aleix, Toni i Ernest (gràcies per votar sempre Menjadors!), companys de despatx. I a tots aquells que he trobat en el departament (o prop d'ell), Quico, Luca, Àlex, Joan, Jaume, Jan, Enrique, Enric, Toni M., Lluís, Jorge, Jordi G., Majo, David, Jordi M., Chumi, Arnau, Mafalda, Otger, Dolors, David D., Julián, Dani, Diego, Pau, Laia, Sandro, Sofyan, Guillem, Valentí, Ester, Olga... i molts d'altres que em dec estar oblidant d'anomenar!

\selectlanguage{american}

And finally thanks to you for reading those acknowledgments, now you can start with the thesis!

\tableofcontents
\listoffigures
\listoftables

\chapter{General introduction}
\section{Preface}
This thesis deals with the study of the structure and the interactions of the
fundamental constituents of matter. We arrived at the end of the twentieth
century describing the known fundamental properties of nature in terms of
quantum field theories (for the electromagnetic, weak and strong nuclear
forces) and general relativity (for the gravitational force). The Standard
Model (SM) of the fundamental interactions of nature comprises quantum field
theories for the electromagnetic (quantum electrodynamics, QED) and weak
interactions (which are unified in the so-called electro-weak theory) and for
the strong interactions (quantum chromodynamics, QCD). This SM is supplemented
by the classical (not-quantum) theory of gravitation (general relativity). All
the experiments that has been performed in accelerators (to study the basic
constituents of nature), up to now, are consistent with this framework. This has led us to the beginning of the twenty-first century waiting for the next generation of experiments to unleash physics beyond that well established theories. Great expectations have been put on the Large Hadron Collider (LHC) actually being built at CERN and scheduled to start being operative in 2007. It is hoped that LHC will open us the way to new physics. Either by discovering the Higgs particle (the last yet-unseen particle in the SM) and triggering this way the discovery of new particles beyond the SM or by showing that there is not Higgs particle at all, which would demand a whole new theoretical framework for the explanation of the fundamental interactions in nature\footnote{Let us do not worry much about the scaring possibility that LHC finds the Higgs, closes the SM, and shows that there are no new physics effects at any scale we are capable to reach in accelerators. Although this is possible, it is, of course, extremely undesirable.}. Possible extensions of the SM have been widely studied during the last years. The expectation is that those effects will show  up in that new generation of experiments. Obviously accelerator Earth-based experiments are not the only source of information for new physics, looking at the sky and at the information that comes from it (highly energetic particles, cosmic backgrounds...) is also a widely studied and great option.

But in this way of finding the next up-to-now-most-fundamental theory of nature we do not want to lose the ability to use it to make concise predictions for as many processes as possible and we also want to be able to understand how the previous theory can be obtained from it, in an unambiguous way. It is obviously the dream of all physicists to obtain an unified framework for explaining the four known interactions of nature. But not at the price of having a theory that can explain everything but is so complicated that does not explain anything. In that sense, constructing a new theory is as important as developing appropriate tools for its use. As mentioned, this is true in a two-fold way, we should be able to understand how the previous theory can be derived from the new one and we also should be able to precisely predict as many observables as possible (to see if the observations can really be accommodated in our theory or if new effects are needed). Let us end this preface to the thesis with a little joke. According to what have been said here, the title of the seminar that would suppose the end of (theoretical) physics is not: \emph{M-theory: a unification of all the interactions in nature}, but rather: \emph{How to obtain the metabolism of a cow from M-theory}.

\section{Effective Field Theories}
In the thesis we will focus in the study of systems involving the strong interacting sector of the Standard Model (SM). The piece of the SM which describes the strong interactions is known as Quantum ChromoDynamics (QCD). QCD is a non-abelian $SU(3)$ quantum field theory, that describes the interactions between quarks and gluons. Its Lagrangian is extremely simple and it is given by
\begin{equation}
\mathcal{L}_{QCD}=\sum_{i=1}^{N_f}\bar{q}_i\left(iD\!\!\!\!\slash-m_i\right)q_i-\frac{1}{4}G^{\mu\nu\, a}G_{\mu\nu}^a
\end{equation}
In that equation $q_i$ are the quark fields, $igG_{\mu\nu}=[D_{\mu},D_{\nu}]$, with $D_{\mu}=\partial_{\mu}+igA_{\mu}$, $A_{\mu}$ are the gluon fields and $N_f$ is the total number of quark flavors. QCD enjoys the properties of confinement and asymptotic freedom. The strong coupling constant becomes large at small energies and tends to zero at large energies. At large energies quarks and gluons behave as free particles. Whereas at low energies they are confined inside color singlet hadrons. QCD develops an intrinsic scale $\Lambda_{QCD}$ at low energies, which gives the main contribution to the mass of most hadrons. $\Lambda_{QCD}$ can be thought of in several, slightly different, ways, but it is basically the scale where the strong coupling constant becomes order one (and perturbative calculations in $\als$ are no longer reliable). It can be thought as some scale around the mass of the proton. The presence of this intrinsic scale $\Lambda_{QCD}$ and the related fact that the spectrum of the theory consists of color singlet hadronic states, causes that direct QCD calculations may be very complicated (if not impossible) for many physical systems of interest. The techniques known as Effective Field Theories (EFT) will help us in this task. 

In general in quantum field theory, the study of any process which involve
more than one relevant physical scale is complicated. The calculations (and
integrals) that will appear can become very cumbersome if more than one scale
enters in them. The idea will be then to construct a new theory (the effective
theory) derived from the fundamental one, in such a way that it just involves
the relevant degrees of freedom for the particular energy regime we are
interested in. The general idea underlying the EFT techniques is simply the
following one: to describe physics in a particular energy region we do not
need to know the detailed dynamics of the other regions. Obviously this is a
very well known and commonly believed fact. For instance, to describe a
chemical reaction one does not need to know about the quantum electrodynamical
interaction between the photons and the electrons, but rather a model of the
atom with a nucleus and orbiting electrons is more adequate. And one does not
need to use this atomic model to describe a macroscopic biological process. The implementation of this, commonly known, idea in the framework of quantum field theories is what is known under the generic name of \emph{Effective Field Theories}. As mentioned before, those techniques are specially useful for processes involving the strong interacting sector of the SM, which is in what this thesis focus. The process of constructing an EFT comprises the following general steps. First one has to identify the relevant degrees of freedom for the process one is interested in. Then one should make use of the symmetries that are present for the problem at hand and finally any hierarchy of energy scales should be exploited. It is important to notice that the EFT is constructed in such a way that it gives equivalent physical results (equivalent to the fundamental theory) in its region of validity. We are not constructing a model for the process we want to study, but rigorously deriving the desired results from the fundamental theory, in a well controlled expansion.

More concretely, in this thesis we will focus in the study of systems
involving heavy quarks. As it is well known, there are six flavors of quarks
in QCD. Three of them have masses below the intrinsic scale of QCD
$\Lambda_{QCD}$, and are called \emph{light}. The other three have masses
larger than $\Lambda_{QCD}$ and are called \emph{heavy}. Therefore, rather than describing and classifying EFT in general we will describe heavy quark systems and the EFT that can be constructed for them (as it will be more adequate for our purposes here).

\section{Heavy quark and quarkonium systems}\label{secinthq}
Three of the six quarks present in QCD have masses larger than $\Lambda_{QCD}$
and are called heavy quarks. The three heavy quarks are the charm quark, the
bottom quark and the top quark. The EFT will take advantage of this large mass
of the quarks and construct an expansion in the heavy quark limit, of infinite
quark masses. The simpler systems that can be constructed involving heavy
quarks are hadrons composed of one heavy quark and one light (anti-)quark. The
suitable effective theory for describing this kind of systems is known as
\emph{Heavy Quark Effective Theory} (HQET), and it is nowadays (together with
chiral perturbation theory, which describes low energy interactions among
pions and kaons, and the Fermi theory of weak interactions, which describes
weak disintegrations below the mass of the $W$) a widely used example to show
how EFT work in a realistic case (see \cite{Neubert:1993mb} for a review of HQET). In brief, the relevant scales for this kind of systems are the heavy quark mass $m$ and $\Lambda_{QCD}$. The effective theory is then constructed as an expansion in $\Lambda_{QCD}/m$. The momentum of a heavy quark is decomposed as 
\begin{equation}
p=mv+k
\end{equation}
where $v$ is the velocity of the hadron (which is basically the velocity
of the heavy quark) and $k$ is a residual momentum of
order $\Lambda_{QCD}$. The dependence on the large scale $m$ is extracted from
the fields, according to
\begin{equation}
Q(x)=e^{-im_Qv\cdot x}\tilde{Q}_v(x)=e^{-im_Qv\cdot x}\left[h_v(x)+H_v(x)\right]
\end{equation}
 and a theory for the soft fluctuations around the heavy quark mass is
 constructed. The leading order Lagrangian of HQET is given by
\begin{equation}
\mathcal{L}_{HQET}=\bar{h}_viv\cdot Dh_v
\end{equation}
This leading order Lagrangian presents flavor and spin symmetries, which can
be exploited for phenomenology.

The systems in which this thesis mainly focus (although not exclusively) are
those known as heavy quarkonium. Heavy quarkonium is a bound state composed of
a heavy quark and a heavy antiquark. We can therefore have \emph{charmonium}
($c\bar{c}$) and \emph{bottomonium} ($b\bar{b}$) systems. The heaviest of the
quarks, the top, decays weakly before forming a bound state; nevertheless
$t-\bar{t}$ production in the non-relativistic regime (that is near threshold)
can also be studied with the same techniques. The relevant physical scales for
the heavy quarkonium systems are the heavy quark mass $m$, the typical three
momentum of the bound state $mv$ (where $v$ is the typical relative velocity
of the heavy quark-antiquark pair in the bound state) and the typical kinetic
energy $mv^2$. Apart from the intrinsic hadronic scale of QCD
$\Lambda_{QCD}$. The presence of all those scales shows us that heavy
quarkonium systems probe all the energy regimes of QCD. From the hard
perturbative region to the low energy non-perturbative one. Heavy quarkonium
systems are therefore an excellent place to improve our understanding of QCD
and to study the interplay of the perturbative and the non-perturbative
effects in QCD \cite{Brambilla:2004wf}. To achieve this goal, EFT for this
system will be constructed. Using the fact that the mass $m$ of the heavy
quark is much larger than any other scale present in the problem (a procedure
which is referred to as \emph{integrating out the scale $m$}) one arrives at
an effective theory known as \emph{Non-Relativistic QCD} (NRQCD) \cite{Bodwin:1994jh}. In that theory, which describes the dynamics of heavy quark-antiquark pairs at energy scales much smaller than their masses, the heavy quark (and antiquark) is treated non-relativistically by (2 component) Pauli spinors. Also gluons and light quarks with a four momentum of order $m$ are integrated out and not present any more in the effective theory. What we have achieved with the construction of this EFT is the systematic factorization of the effects at the hard scale $m$ from the effects coming from the rest of scales. NRQCD provides us with a rigorous framework to study spectroscopy, decay, production and many other heavy quarkonium processes. The leading order Lagrangian for this theory is given by
\begin{equation}
\mathcal{L}_{NRQCD}=
\psi^{\dagger} \left( i D_0 + {1\over 2 m} {\bf D}^2 \right)\psi  + 
\chi^{\dagger} \left( i D_0 - {1\over 2 m} {\bf D}^2 \right)\chi
\end{equation}
where $\psi$ is the field that annihilates a heavy quark and $\chi$ the field
that creates a heavy antiquark. Sub-leading terms (in the $1/m$ expansion) can then be derived. One might be
surprised, at first, that heavy quarkonium decay processes can be studied in
NRQCD. Since the annihilation of a $Q\bar{Q}$ pair will produce gluons and
light quarks with energies of order $m$, and those degrees of freedom are not
present in NRQCD. Nevertheless those annihilation processes can be explained
within NRQCD (in fact the theory is constructed to reproduce that kind of
physics). The answer is that annihilation processes are incorporated in NRQCD
through local four fermion operators. The $Q\bar{Q}$ annihilation rate is
represented in NRQCD by the imaginary parts of $Q\bar{Q}\to Q\bar{Q}$
scattering amplitudes. The coefficients of the four fermion operators in the
NRQCD Lagrangian, therefore, have imaginary parts, which reproduces the $Q\bar{Q}$ annihilation rates. In that way we can describe inclusive heavy quarkonium decay widths to light particles. 

NRQCD has factorized the effects at the hard scale $m$ from the rest of scales
in the problem. But if we want to describe heavy quarkonium physics at the
scale of the binding energy, we will face with the complication that the soft,
$mv$, and ultrasoft, $mv^2$, scales are still entangled in NRQCD. It would be
desirable to disentangle the effects of those two scales. To solve this
problem one can proceed in more than one way. One possibility is to introduce
separate fields for the soft and ultrasoft degrees of freedom at the NRQCD
level. This would lead us to the formalism now known as \emph{velocity NRQCD}
(vNRQCD) \cite{Luke:1999kz}. Another possibility is to exploit further the non-relativistic
hierarchy of scales in the system ($m\gg mv\gg mv^2$) and construct a new effective theory which just contains the relevant degrees of freedom to
describe heavy quarkonium physics at the scale of the binding energy. That
procedure lead us to the formalism known as \emph{potential NRQCD} (pNRQCD) \cite{Brambilla:1999xf}
and is the approach that we will take here, in this thesis. When going from
NRQCD to pNRQCD one is integrating out gluons and light quarks with energies
of order of the soft scale $mv$ and heavy quarks with energy fluctuations at
this soft scale. This procedure is sometimes referred to as \emph{integrating
  out the soft scale}, although the scale $mv$ is still active in the three
momentum of the heavy quarks. The resulting effective theory, pNRQCD, is
non-local in space (since the gluons are massless and the typical momentum
transfer is at the soft scale). The usual potentials in quantum mechanics appear as
Wilson coefficients of the effective theory. This effective theory will be
described in some more detail in section \ref{secpNRQCD}. 

The correct treatment of some heavy quark and quarkonium processes will
require additional degrees of freedom, apart from
those of HQET or NRQCD. When we want to describe regions of phase space where
the decay products have large energy, or exclusive decays of heavy particles,
for example,
collinear degrees of freedom would need to be present in the theory. The
interaction of collinear and soft degrees of freedom has been implemented in
an EFT framework in what now is known as \emph{Soft-Collinear Effective
  Theory} (SCET) \cite{Bauer:2000yr,Beneke:2002ph}. This effective theory will
also be described in a following section \ref{secSCET}. Just let us mention
here that, due to the peculiar nature of light cone interactions, this EFT
will be non-local in a light cone direction (collinear gluons can not interact
with soft fermions without taking them far off-shell). 

The study of heavy
quark and quarkonium systems has thus lead us to the construction of effective
quantum field theories of increasing richness and complexity. The full power
of the quantum field theory techniques (loop effects, matching procedures,
resummation of large logarithms...) is exploited to obtain systematic
improvements in our understanding of those systems.

\section{Structure of the thesis}
This thesis is structured in the following manner. Next chapter (chapter \ref{cat}) is
a summary of the whole thesis written in Catalan (it does not contain any
information which is not present in other chapters, except for the
translation). Chapter \ref{chapback} contains an introduction to potential
NRQCD and Soft-Collinear Effective Theory, the two effective theories that are
mainly used during the thesis. The three following chapters (chapters \ref{chappotest},
\ref{chapda} and \ref{chapraddec}) comprise the original contributions of this
thesis. Chapter \ref{chappotest} is devoted to the study of the (infrared
dependence of the) QCD static
potential, employing pNRQCD techniques. Chapter \ref{chapda} is devoted to the
calculation of an anomalous dimension in SCET (two loop $n_f$ terms are
obtained), which is relevant in many processes under recent study. And chapter
\ref{chapraddec} is devoted to the study of the semi-inclusive radiative decays of
heavy quarkonium to light hadrons, employing a combination of pNRQCD and
SCET. Chapter \ref{chapconcl} is devoted to the final conclusions. This chapter
is followed by
three appendices. The first appendix contains definitions of several factors
appearing throughout the thesis. The second appendix contains Feynman rules
for pNRQCD an SCET. And, finally, the third appendix contains the
factorization formulas for the NRQCD matrix elements in the strong coupling regime.

%%% Local Variables: 
%%% mode: latex
%%% TeX-master: t
%%% End: 

\chapter{Summary in Catalan}\label{cat}
\selectlanguage{catalan}
Per facilitar la lectura, i una eventual comparació amb d'altres referències escrites en anglès, incloem, en la taula \ref{tabtrad}, la traducció emprada per alguns dels termes presents en la tesi.
\section{Introducció general}
Aquesta tesi versa sobre l'estudi de l'estructura i les interaccions dels constituents fonamentals de la matèria. Vàrem arribar al final del segle XX descrivint les propietats més fonamentals conegudes de la matèria en termes de teories quàntiques de camps (pel que fa a les interaccions electromagnètiques, nuclear forta i nuclear feble) i de la relativitat general (pel que fa a la interacció gravitatòria). El Model Estàndard (ME) de les interaccions fonamentals en la natura engloba teories quàntiques de camps per descriure les interaccions electromagnètiques (l'anomenda ElectroDinàmica Quàntica, EDQ) i nuclears febles (que estan unificades en l'anomenada teoria electro-feble) i per descriure les interaccions fortes (l'anomedada CromoDinàmica Quàntica, CDQ). Aquest ME ve complementat per la teoria clàssica (no quàntica) de la gravitació, la relativitat general. Tots els experiments que s'han dut a terme en acceleradors de partícules (per tal d'estudiar els constituents bàsics de la matèria), fins el dia d'avui, són consistents amb aquest marc teòric. Això ens ha portat a començar el segle XXI esperant que la següent generació d'experiments destapi la física que hi pot haver més enllà d'aquestes teories, que han quedat ja ben establertes. Hi ha grans esperances posades en el gran accelerador hadrònic, anomenat \emph{Large Hadron Collider} (LHC), que s'està construint actualment al CERN. Està planificat que aquesta màquina comenci a ser operativa l'any 2007. El que s'espera és que l'LHC ens obri el camí cap a nous fenòmens físics no observats fins ara. Això es pot aconseguir de dues maneres. Una possibilitat és que l'LHC descobreixi la partícula de Higgs (l'única partícula del ME que encara no s'ha observat) i que això desencadeni la descoberta de noves partícules més enllà del ME. L'altra possibilitat és que l'LHC demostri que no hi ha tal partícula de Higgs; cosa que demanaria un marc teòric totalment nou i diferent l'actual (per explicar les interaccions fonamentals de la natura)\footnote{Intentarem no preocupar-nos gaire per la possibilitat que l'LHC descobreixi el Higgs, tanqui el ME i mostri que no hi ha efectes de nova física en cap escala d'energia que serem capaços d'assolir amb acceleradors de partícules. Tot i que això és possible no és, òbviament, gens desitjable.}. Les possibles extensions del ME han estat ja estudiades àmpliament i amb gran detall. El que s'espera és que tots aquests efectes es facin palesos en aquesta nova generació d'experiments. No cal dir que els experiments basats en acceleradors de partícules no són l'única opció que tenim, per tal de descobrir efectes associats a nova física. Una altra gran oportunitat (que també ha estat àmpliament estudiada) és la d'observar el cel i la informació que ens arriba d'ell (partícules altament energètiques, fons còsmics de radiació...).

Però en aquest camí a la recerca de la següent teoria més fonamental coneguda fins al moment, no volem perdre l'habilitat de fer sevir aquesta teoria per fer prediccions concises per un ampli ventall de processos físics, i també volem poder entendre (d'una forma no ambigua) com la teoria precedent es pot obtenir a partir de la nova. Òbviament el somni de qualsevol físic és trobar una descripició unificada de les quatre interaccions fonamentals conegudes de la matèria; però no al preu de tenir una teoria que pot explicar-ho tot però que és tant complicada que, de fet, no explica res. Acabarem aquests paràgrafs que fan de prefaci a la tesi amb un petit acudit. D'acord amb el que hem dit aquí, el títol de la conferència que suposaria el punt i final de la física (tèorica) no és: \emph{La teoria M: una unificació de totes les interaccions de la natura}, sinó més aviat: \emph{Com obtenir el metabolisme d'una vaca a partir de la teoria M}.

\subsection{Teories Efectives}
En aquesta tesi ens centrarem en l'estudi de sistemes que involucren el sector de les interaccions fortes en el ME. La part del ME que descriu les interaccions fortes és, com s'ha comentat abans, la CromoDinàmica Quàntica. CDQ és una teòrica quàntica de camps basada en el grup no abelià $SU(3)$ i descriu les  interaccions entre quarks i gluons. El seu Lagrangià és extremadament simple i ve donat per
\begin{equation}
\mathcal{L}_{QCD}=\sum_{i=1}^{N_f}\bar{q}_i\left(iD\!\!\!\!\slash-m_i\right)q_i-\frac{1}{4}G^{\mu\nu\, a}G_{\mu\nu}^a
\end{equation}
En aquesta equació $q_i$ són els camps associats als quarks, $igG_{\mu\nu}=[D_{\mu},D_{\nu}]$, amb $D_{\mu}=\partial_{\mu}+igA_{\mu}$, $A_{\mu}$ són els camps pels gluons i $N_f$ és el número total de sabors (tipus) de quarks. La CDQ presenta les propietats de llibertat asimptòtica i de confinament. La constant d'acoblament de les interaccions fortes esdevé gran a energies petites i tendeix a zero per energies grans. D'aquesta manera, per energies altes els quarks i els gluons es comporten com a partícules lliures, mentre que a baixes energies apareixen sempre confinats a l'interior d'hadrons (en una combinació singlet de color). La CDQ desenvolupa una escala intrínseca, $\Lambda_{QCD}$, a baixes energies; escala que dóna la contribució principal a la massa de la majoria dels hadrons. $\Lambda_{QCD}$ es pot interpretar de diferents maneres, però és bàsicament l'escala d'energia on la constant d'acoblament de les interaccions fortes esdevé d'ordre 1 (i la teoria de perturbacions en $\als$ ja no és fiable). Es pot pensar que és una escala de l'ordre de la massa del protó. La presència d'aquesta escala intrinseca i el fet, íntimament relacionat, que l'espectre de la teoria consisteixi en estats hadrònics singlets de color (i no dels quarks i gluons) provoca que els càlculs directes des de CDQ siguin extremadament complicats, sinó impossibles, per molts sistemes físics d'interès. Les tècniques conegudes amb el nom de teories efectives (TE) ens ajudaran en aquesta tasca.

Com a regla general, l'estudi de qualsevol procés, en teoria quàntica de camps, que involucri més d'una escala física rellevant és complicat. Els càlculs (i les integrals) que ens apareixeran poden resultar molt complicats si més d'una escala entra en ells. La idea serà doncs construir una nova teoria (la teoria efectiva), derivada de la teoria fonamental, de manera que només involucri els graus de llibertat rellevants per la regió que ens interessa. La idea general que hi ha sota les tècniques de TE és simplement la següent: per tal d'estudiar la física d'una determinada regió d'energies no necessitem conèixer la dinàmica de les altres regions de forma detallada. Aquest és, òbviament, un fet ben conegut i àmpliament acceptat. Per exemple, tothom entent que per descriure una reacció química no cal conèixer la interacció quàntica electrodinàmica entre els fotons i els electrons, per contra un model de l'àtom que consisteixi en un nucli i electrons orbitant al voltant és més convenient. I de la mateixa manera no cal usar aquest model atòmic per tal de descriure un procés biològic macroscòpic. La implementació d'aquesta ben coneguda idea en el marc de la teoria quàntica de camps és el que es coneix sota el nom genèric de \emph{Teories Efectives}. Tal i com ja s'ha dit abans, aquestes tècniques esdevenen especialment útils en l'estudi de processes que involucren les interaccions fortes. Per tal de construir una TE cal seguir els següents passos generals (a grans trets). En primer lloc és necessari identificar els graus de llibertat que són rellevants pel problema en que estem interessats. Després cal fer ús de les simetries  presents en el problema i, finalment, hem d'aprofitar qualsevol jerarquia d'escales que hi pugui haver. És important remarcar que el que estem fent  no és construir un model pel procés que volem estudiar. Per contra la TE està construida de manera de sigui equivalent a la teoria fonamental, en la regió on és vàlida; estem obtenint els resultats desitjats a partir d'una expansió ben controlada de la nostra teoria fonamental.

Més concretament, en aquesta tesi ens centrarem en l'estudi de sistemes que involucren els anomenats quarks pesats. Com és ben conegut hi ha sis sabors (tipus) de quarks en CDQ. Tres d'ells tene masses  per sota de l'escala $\lQ$ i s'anomenen \emph{lleugers}, mentre que els altres tres tenen masses per sobre d'aquesta escala $\lQ$ i s'anomenen \emph{pesats}. El que farem a continuació és descriure sistemes amb quarks pesats i les teories efectives que es poden construir per ells.

\subsection{Sistemes de quarks pesats i quarkoni}
El que faran les TE pels sistemes amb quarks pesats és aprofitar-se d'aquesta escala gran, la massa, i construir una expansió en el límit de quarks infinítament massius. Els sistemes més simples que es poden tenir involucrant quarks pesats són aquells composats d'un quark pesat i un (anti-)quark lleuger. La TE adequada  per descriure aquest tipus de sistemes rep el nom de \emph{Teoria Efectiva per Quarks Pesats} (TEQP). Aquesta teoria és avui en dia, i juntament amb la teoria de perturbacions quiral (que descriu les interaccions de baixa energia entre pions i kaons) i la teoria de Fermi per les interaccions febles (que decriu les desintegracions febles per a energies per sota de la massa del bosó $W$), un exemple àmpliament usat per mostrar com les TE funcionen en un cas realista. De manera molt breu, les escales físiques rellevants per aquest sistema són la massa $m$ del quark pesat i $\lQ$. La TE es construeix, per tant, com una expansió en $\lQ/m$. El moment del quak pesat es descomposa d'acord amb
\begin{equation}
p=mv+k
\end{equation}
on $v$ és la velocitat de l'hadró (que és bàsicament la velocitat del quark pesat) i $k$ és un moment residual d'ordre $\lQ$. La dependència en l'escala $m$ s'extreu dels camps de la TE d'acord amb
\begin{equation}
Q(x)=e^{-im_Qv\cdot x}\tilde{Q}_v(x)=e^{-im_Qv\cdot x}\left[h_v(x)+H_v(x)\right]
\end{equation}
i es construeix una teoria per les fluctuacions suaus al voltant de la massa del quark pesat. El Lagrangià de la TEQP a ordre dominant ve donat per
\begin{equation}
\mathcal{L}_{HQET}=\bar{h}_viv\cdot Dh_v
\end{equation}
Aquest Lagrangià presenta simetries de sabor i spin, que es poden aprofitar per a fer fenomenologia.

Els sistemes en que aquesta tesi se centrarà (encara que no de manera exclusiva) són aquells  coneguts amb el nom de quarkoni pesat. El quarkoni pesat és un estat lligat composat per un quark pesat i un antiquark pesat. Per tant podem tenir sistemes de \emph{charmoni} ($c\bar{c}$) i de  \emph{bottomoni} ($b\bar{b}$). El més pesat de tots els quarks, el quark top, es desintegra a través de les interaccions febles abans que pugui formar estats lligats; de tota manera la producció de parelles $t-\bar{t}$ prop del llindar de producció (per tant, en un règim no relativista) es pot estudiar amb les mateixes tècniques. Les escales físiques rellevants pels sistemes de quarkoni pesat són l'escala $m$ de la massa del quark pesat, el tri-moment típic de l'estat lligat $mv$ ($v$ és la velocitat relativa típica de la parella quark-antiquark en l'estat lligat) i l'energia cinètica típica $mv^2$. A part de l'escala intrínseca de la CDQ, $\lQ$, que sempre és present. La presència simultània de totes aquestes escales ens indica que els sistemes de quarkoni pesat involucren tots els rangs d'energia de CDQ, des de les regions perturbatives d'alta energia fins a les no-perturbatives de baixa energia. És per tant un bon sistema per estudiar la interacció entre els efectes perturbatius i els no perturbatius en CDQ i per millorar el nostre coneixement de CDQ en general. Per tal d'aconseguir aquest objectiu contruirem TE adequades per la descripció d'aquest sistema. Si fem servir el fet que la massa $m$ és molt més gran que qualsevol altra escala d'energia present el problema, arribem a una TE coneguda amb el nom de \emph{CDQ No Relativista} (CDQNR). En aquesta teoria, que descriu la dinàmica de  parelles de quark-antiquark per energies força menors a les seves masses, els quarks pesats vénen representats per spinors no relativistes de dues components. A més a més, gluons i quarks lleugers amb quadri-moment a l'escala $m$ són intergats de la teoria i ja no hi apareixen. El que hem aconseguit amb la construcció d'aquesta  teoria és factoritzar, de manera sistemàtica, els efectes que vénen de l'escala $m$ de la resta d'efectes provinents de les altres escales del problema. CDQNR ens proporciona un marc teòric  rigorós on estudiar processos de desintegració, producció i espectroscòpia de quarkoni pesat. El Lagrangià a ordre  dominant ve donat per
\begin{equation}
\mathcal{L}_{NRQCD}=
\psi^{\dagger} \left( i D_0 + {1\over 2 m} {\bf D}^2 \right)\psi  + 
\chi^{\dagger} \left( i D_0 - {1\over 2 m} {\bf D}^2 \right)\chi
\end{equation}
on $\psi$ és el camp que anihila el quark pesat i $\chi$ el camp que crea l'antiquark pesat. Termes sub-dominants, en l'expansió en $1/m$, poden ser derivats. D'entrada pot resultar sorprenent que els processos de desintegració puguin ser estudiats en el marc de la CDQNR. L'anihilació de la parella $Q\bar{Q}$ produirà gluons i quarks lleugers amb energies d'ordre $m$, i aquests graus de llibertat ja no són presents en CDQNR. Tot i això els processos de desintergació poden ser estudiats en el marc de la CDQNR, de fet la teoria està construida per tal de poder explicar aquests processos. La resposta és que els processos d'anihilació s'incorporen en CDQNR a través d'interaccions locals de quatre fermions. Les raons de desintegració vénen representades en CDQNR per les parts imaginàries de les amplituds de dispersió $Q\bar{Q}\to Q\bar{Q}$. Els coeficients dels operadors de quatre fermions tenen, per tant, parts imaginàries que codifiquen les raons de desintegració. D'aquesta manera podem estudiar les desintegracions inclusives de quakonium pesat en partícules lleugeres.

La CDQNR ens ha factoritzat els efectes a l'escala $m$ de la resta. Ara bé, si volem estudiar la física del quarkoni pesat a l'escala de l'energia de lligam del sistema, ens  trobarem amb el problema que les escales suau,  corresponent al tri-moment típic $mv$, i ultrasuau, corresponent a l'energia cinètica típica $mv^2$, estan encara entrellaçades en CDQNR. Seria desitjable separar els efectes d'aquestes dues escales. Per tal de solucionar aquest problema es pot procedir de més d'una manera. L'estrategia que emprarem en aquesta tesi és la d'aprofitar de manera més àmplia la jerarquia no relativista d'escales que presenta el sistema ($m\gg mv\gg mv^2$) i construir una nova teoria efectiva que només contingui els graus de llibertat rellevants per tal de descriure els sistemes de quarkoni pesat a l'escala de l'energia de lligam. La teoria que s'obté és coneguda amb el nom de \emph{CDQNR de potencial} (CDQNRp). Aquesta teoria serà descrita breument en la següent secció.

El tractament correcte d'alguns sistemes de quarks pesats i de quarkoni pesat demanarà la presència de graus de llibertat addicionals, a part dels presents en TEQP o en CDQNR. Quan volem descriure regions de l'espai fàsic on els productes de la desintegració tenen una energia gran, o quan volguem descriure desintegracions exclusives, per exemple, graus de llibertat col·lineals hauran de ser presents en la teoria. La interacció dels graus de llibertat col·lineals amb els graus de llibertat suaus ha estat implementada en el marc de les TE en el que avui es coneix com a Teoria Efectiva Col·lineal-Suau (TECS). Aquesta teoria la descriurem també breument en la següent secció.

En  definitiva, l'estudi de sistemes de quarks pesats i quarkoni ens ha portat a la construcció de teories efectives de camps de riquesa i complexitat creixents. Tota la potència de les tècniques de teoria quàntica de camps (efectes de bagues, resumació de logartimes...) és explotat per tal de millorar la nostra comprensió d'aquests sistemes.

\begin{table}[t]
\begin{center}
\begin{tabular}{|c|c|}
\hline
 Anglès & Català \\
\hline
 & \\
Quantum ChromoDynamics (QCD) & CromoDinàmica Quàntica (CDQ) \\
 & \\
Soft-Collinear Effective Theory (SCET) & Teoria Efectiva Col·lineal-Suau (TECS) \\
 & \\
loop & baga \\
 & \\
Standard Model (SM) & Model Estàndard (ME)\\
 & \\
Quantum ElectroDynamics (QED) & Electrodinàmica Quàntica (EDQ)\\
 & \\
 quarkonium & quarkoni\\
  & \\
Heavy Quark Effective Theory (HQET) & Teoria Efectiva per Quarks Pesats (TEQP)\\ & \\
Non-Relativistic QCD (NRQCD) & CDQ No Relativista (CDQNR)\\
 & \\
potential NRQCD (pNRQCD) & CDQNR de potencial (CDQNRp) \\
 & \\
matching coefficients & coeficients de coincidència \\
 & \\
label operators & operadors etiqueta \\
 & \\
jet & doll\\
 & \\
\hline
\end{tabular}
\caption[English-Catalan translations]{Traducció anglès-català d'alguns termes usats en la tesi.}\label{tabtrad}
\end{center}
\end{table}

\section{Rerefons}
\subsection{CDQNRp}
Com ja s'ha dit abans, les escales rellevants pels sistemes de quarkoni pesat són la massa $m$, l'escala suau $mv$ i l'escala ultrasuau $mv^2$. A part de l'escala $\lQ$. Quan aprofitem la jerarquia no relativista del sistema en la seva  totalitat arribem a la CDQNRp. Per tal d'identificar els graus de llibertat rellevants en la teoria final, cal especificar la importància relativa de $\lQ$ respecte les escales suau i ultrasuau. Dos règims rellevants han estat identificats. Són els anomenats \emph{règim d'acoblament feble} $mv^2\gtrsim\Lambda_{QCD}$ i \emph{règim d'acoblament fort} $mv\gtrsim\Lambda_{QCD}\gg mv^2$.
\subsubsection{Règim d'acoblament feble}
En aquest règim els graus de  llibertat de CDQNRp són semblants als de CDQNR, però amb les cotes superiors en energia i tri-moments abaixades. Els graus de llibertat de CDQNRp consisteixen en quarks i antiquarks pesats amb un tri-moment fitat superiorment per $\nu_p$ ($\vert\mathbf{p}\vert\ll\nu_p\ll m$) i una energia fitada per $\nu_{us}$
($\frac{\mathbf{p}^2}{m}\ll\nu_{us}\ll\vert\mathbf{p}\vert$), i en gluons i quarks lleugers amb un quadri-moment fitat per $\nu_{us}$. El Lagrangià es pot escriure com
\[
\mathcal{L}_{pNRQCD}=\int d^3{\bf r} \; {\rm Tr} \,  
\Biggl\{ {\rm S}^\dagger \left( i\partial_0 
- h_s({\bf r}, {\bf p}, {\bf P}_{\bf R}, {\bf S}_1,{\bf S}_2) \right) {\rm S} 
+ {\rm O}^\dagger \left( iD_0 
- h_o({\bf r}, {\bf p}, {\bf P}_{\bf R}, {\bf S}_1,{\bf S}_2) \right) {\rm O} \Biggr\}+
\]
\[
+V_A ( r) {\rm Tr} \left\{  {\rm O}^\dagger {\bf r} \cdot g{\bf E} \,{\rm S}
+ {\rm S}^\dagger {\bf r} \cdot g{\bf E} \,{\rm O} \right\} 
+ {V_B (r) \over 2} {\rm Tr} \left\{  {\rm O}^\dagger {\bf r} \cdot g{\bf E} \, {\rm O} 
+ {\rm O}^\dagger {\rm O} {\bf r} \cdot g{\bf E}  \right\}-
\]
\begin{equation}
- {1\over 4} G_{\mu \nu}^{a} G^{\mu \nu \, a} 
+  \sum_{i=1}^{n_f} \bar q_i \, i D\!\!\!\!\slash \, q_i
\end{equation}
amb
\begin{equation}
h_s({\bf r}, {\bf p}, {\bf P}_{\bf R}, {\bf S}_1,{\bf S}_2) = 
{{\bf p}^2 \over \, m_{\rm red}} 
+ {{\bf P}_{\bf R}^2 \over 2\, m_{\rm tot}} + 
V_s({\bf r}, {\bf p}, {\bf P}_{\bf R}, {\bf S}_1,{\bf S}_2)
\end{equation}
\begin{equation}
h_o({\bf r}, {\bf p}, {\bf P}_{\bf R}, {\bf S}_1,{\bf S}_2) = 
{{\bf p}^2 \over \, m_{\rm red}}  
+  {{\bf P}_{\bf R}^2 \over 2\, m_{\rm tot}} + 
V_o({\bf r}, {\bf p}, {\bf P}_{\bf R}, {\bf S}_1,{\bf S}_2)
\end{equation}
i
\begin{equation}
D_0 {\rm O} \equiv i \partial_0 {\rm O} - g [A_0({\bf R},t),{\rm O}]\quad {\bf
  P}_{\bf R} = -i{\bf D}_{\bf R}\quad m_{\rm red} =\frac{m_1m_2}{m_{\rm
  tot}}\quad m_{\rm tot} = m_1 + m_2
\end{equation}
$S$ és el camp singlet pel quarkoni i $O$ el camp octet per ell. $\mathbf{E}$ representa el camp cromoelèctric. Podem veure que els potencials usuals de mecànica quàntica apareixen com a coeficients de coincidència en la teoria efectiva.

\subsubsection{Règim d'acoblament fort}
En aquesta situació la física a l'escala de l'energia de lligam està per sota de l'escala $\lQ$. Per tant és millor discutir la teoria en termes de graus de llibertat hadrònics. Guiant-nos per algunes consideracions generals i per indicacions provinents CDQ en el reticle, podem suposar que el quarkoni ve descrit per un camp singlet. I si ignorem els bosons de Goldstone, aquests són tots els graus de llibertat en aques règim. El Lagrangià ve ara donat per 
\begin{equation}
L_{\rm pNRQCD} = \int d^3 {\bf R}  \int d^3 {\bf r}  \;
S^\dagger \big( i\partial_0 - h_s({\bf x}_1,{\bf x}_2, {\bf p}_1, {\bf p}_2,
{\bf S}_1,  {\bf S}_2) \big) S
\end{equation}
amb
\begin{equation}
h_s({\bf x}_1,{\bf x}_2, {\bf p}_1, {\bf p}_2, {\bf S}_1,  {\bf S}_2) =
{{\bf p}_1^2\over 2m_1} + {{\bf p}_2^2\over 2m_2}
+ V_s({\bf x}_1,{\bf x}_2, {\bf p}_1, {\bf p}_2, {\bf S}_1,  {\bf S}_2)
\end{equation}
El potencial $V_s$ és ara una quantitat no perturbativa. El procediment de coincidència de la teoria fonamental i la teoria efectiva ens donarà expressions pel potencial (en termes de les anomenades bagues de Wilson). 

\subsection{TECS}
L'objectiu d'aquesta teoria és descriure processos on graus de llibertat molt energètics (col·lineals) interactuen amb graus de llibertat suaus. Així la teoria es pot aplicar a un ampli ventall de processos, on aquesta situació cinemàtica és present. Qualsevol procés que contingui hadrons molt energètics, juntament amb una font per ells, contindrà partícules, anomenades col·lineals, que es mouen a prop d'una direcció del con de llum $n^{\mu}$. Com que aquestes partícules han de tenir una energia $E$ gran i alhora una massa invariant petita, el tamany de les components del seu quadri-moment (en coordenades del con de llum, $p^{\mu}=(\bar{n}p)n^{\mu}/2+p^{\mu}_{\perp}+(np)\bar{n}^{\mu}/2$) és molt diferent. Típicament $\bar{n}p\sim E$, $p_{\perp}\sim
E\lambda$ i $np\sim E\lambda^2$, amb $\lambda$ un paràmetre petit. És d'aquesta jerarquia d'escales que la TE treurà profit.

Els graus de llibertat que cal inlcoure en la teoria efectiva depenen de si un vol estudiar processos inclusius o exclusius. Les dues teories que en resulten es coneixen amb els noms de TECS$_{\rm I}$ i TECS$_{\rm II}$, respectivament.

\subsubsection{TECS$_{\rm I}$}
Aquesta és la teoria que conté graus de llibertat col·lineals $(p^{\mu}=(\bar{n}p,p_{\perp},np)\sim(1,\lambda,\lambda^2))$ i ultrasuaus $(p^{\mu}\sim(\lambda^2,\lambda^2,\lambda^2))$. Els graus de llibertat col·lineals tenen massa invariant d'ordre $E\Lambda_{QCD}$. Malauradament, tant en TECS$_{\rm I}$ com en TECS$_{\rm II}$, no hi ha una notació estàndard en la literatura. Dos formalismes (suposadament equivalents) han estat usats.

El Lagrangià a ordre dominant ve donat per 
\begin{equation}
\mathcal{L}_c=\bar{\xi}_{n,p'}\left\{inD+gnA_{n,q}+\left(\mathcal{P}\!\!\!\!\slash_{\perp}+gA\!\!\!\slash_{n,q}^{\perp}\right)W\frac{1}{\bar{\mathcal{P}}}W^{\dagger}\left(\mathcal{P}\!\!\!\!\slash_{\perp}+gA\!\!\!\slash_{n,q'}^{\perp}\right)\right\}\frac{\bar{n}\!\!\!\slash}{2}\xi_{n,p}
\end{equation}
$\xi_{n,p}$ és el camp pel quark col·lineal, $A_{n,p}$ el camp pel gluó col·lineal (la dependència en les escales grans ha estat extreta d'ells de manera semblant a en TEQP). Els $\mathcal{P}$ són els anomenats operdors etiqueta que donen les components grans (extretes) dels camps. Les $W$ són línies de Wilson.
\subsubsection{TECS$_{\rm II}$}
Aquesta és la teoria que descriu processos on els graus de llibertat col·lineals en l'estat final tenen massa invariant d'ordre $\Lambda_{QCD}^2$. Aquesta teoria és més complicada que l'anterior, ja que en el procés d'anar des de CDQ a TECS$_{\rm II}$ la presència de dos tipus de modes col·lineals s'ha de tenir en compte. En la tesi bàsicament no usarem aquesta teoria i, per tant, no en direm res més.

\section{El potencial estàtic singlet de CDQ}
El potencial estàtic entre un quark i un antiquark és un objecte clau per tal d'entendre la dinàmica de la CDQ. Aquí ens centrarem en estudiar la dependència infraroja del potencial estàtic singlet. Obtindrem la dependència infraroja sub-dominant del mateix fent servir la CDQNRp

L'expansió perturbativa del potencial estàtic singlet ve donada per 
\[
V_s^{(0)}(r)=-\frac{C_f\als(1/r)}{r}\left(1+\frac{\als(1/r)}{4\pi}\left(a_1+2\gamma_E\beta_0\right)+\left(\frac{\als(1/r)}{4\pi}\right)^2\bigg(
  a_2+\right.
\]
\begin{equation}\label{Vcat}
+\left.\left(\frac{\pi^2}{3}+4\gamma_E^2\right)\beta_0^2+\gamma_E\left(4a_1\beta_0+2\beta_1\right)\bigg)+\left(\frac{\als(1/r)}{4\pi}\right)^3\left(\tilde{a}_3+\frac{16\pi^2}{3}C_A^3\log r\mu\right)+\cdots\right)
\end{equation}
on
\begin{equation}
a_1=\frac{31}{9}C_A-\frac{20}{9}T_Fn_f
\end{equation}
i
\begin{eqnarray}
  a_2&=&
  \left[{4343\over162}+4\pi^2-{\pi^4\over4}+{22\over3}\zeta(3)\right]C_A^2
  -\left[{1798\over81}+{56\over3}\zeta(3)\right]C_AT_Fn_f-
  \nonumber\\
  &&{}-\left[{55\over3}-16\zeta(3)\right]C_fT_Fn_f
  +\left({20\over9}T_Fn_f\right)^2
\end{eqnarray}
el logaritme que veiem en l'expressió pel potencial és la dependència infraroja dominant. Aquí trobarem la dependència infraroja sub-dominant; és a dir una part de la correcció a quart ordre del potencial. Per fer-ho estudiarem el procés de fer coincidir CDQNR amb CDQNRp. El que cal fer és calcular la conicidència a ordre $r^2$ en l'expansió multipolar de CDQNRp. Per fer això cal evaluar el segon diagrama de la part dreta de la igualtat de la figura \ref{figmatchr2}. Quan calculem la primera correcció en $\als$ d'aquest diagrama (després del terme dominant) obtenim la dependència infraroja sub-dominant que busquem (el terme dominant del diagrama donava la dependència infraroja dominant). El resultat pels termes infrarojos sub-dominants del potencial és
\[
V_s^{(0)}(r)=(\mathrm{Eq.}\ref{Vcat})-
\]
\begin{equation}
-\frac{C_f\als(1/r)}{r}\left(\frac{\als(1/r)}{4\pi}\right)^4\frac{16\pi^2}{3}C_A^3\left(-\frac{11}{3}C_A+\frac{2}{3}n_f\right)\log^2r\mu-
\end{equation}
\begin{equation}
-\frac{C_f\als(1/r)}{r}\left(\frac{\als(1/r)}{4\pi}\right)^4\frac{16\pi^2}{3}C_A^3\left(a_1+2\gamma_E\beta_0-\frac{1}{27}\left(20
 n_f-C_A(12 \pi ^2+149)\right)\right)\log r\mu
\end{equation}

\section{Dimensió anòmala del corrent lleuger-a-pesat en TECS a dues bagues: termes $n_f$}
Els corrents hadrònics lleuger-a-pesat $J_{\mathrm{had}}=\bar{q}\Gamma b$ ($b$
representa el quark pesat i $q$ el quark lleuger), que apareixen en operadors de la teoria nuclear feble a una escala d'energia $\mu\sim m_b$, es poden fer coincidir amb els corrents de TECS$_{\mathrm{I}}$. A ordre més baix en el paràmetre d'expansió $\lambda$ el corrent en TECS ve donat per
\begin{equation}
J_{\mathrm{had}}^{SCET}=c_0\left(\bar{n}p,\mu\right)\bar{\xi}_{n,p}\Gamma h+c_1\left(\bar{n}p,\bar{n}q_1,\mu\right)\bar{\xi}_{n,p}\left(g\bar{n}A_{n,q_1}\right)\Gamma h+\cdots
\end{equation}
És a dir, un nombre arbitrari de gluons $\bar{n}A_{n,q}$ poden ser afegits, sense que això suposi supressió en el comptatge en el paràmetre $\lambda$. Els coeficients de Wilson poden ser evolucionats, en la teoria efectiva, a una escala d'energia més baixa. Com que tots els corrents estan relacionats per invariancia de galga col·lineal, és suficient estudiar el corrent $\bar{\xi}\Gamma
h$ (que és òbviament més simple). L'evolució del corrent a una baga ve determinada per la dimensió anòmala
\begin{equation}
\gamma=-\frac{\alpha_s}{4\pi}C_f\left(5+4\log\left(\frac{\mu}{\bar{n}P}\right)\right)
\end{equation}
$P$ és el moment total sortint del doll de partícules. Aquí volem trobar els termes $n_f$ de la correcció a dues bagues d'aquest resultat. Per tal de calcular-los cal evaluar els diagrames de la figura \ref{figdiagnf}. A part també necessitem la correcció a dues bagues dels propagadors del quark col·lineal i del quark pesat. La correcció del propagador del quark col·lineal coincideix amb la usual de CDQ (ja que en el seu càlcul només hi entren partícules col·lineals, i no ultrasuaus); mentre que la correcció al propagador del quark pesat és la usual de TEQP. Tenint en compte el resultat dels diagrames i aquestes correccions als propagadors, obtenim el resultat desitjat pels termes $n_f$ a dues bagues de la dimensió anòmala
\begin{equation}
\gamma^{(2bagues\;n_f)}=\left(\frac{\alpha_s}{4\pi}\right)^2\frac{4T_Fn_fC_f}{3}\left(\frac{125}{18}+\frac{\pi^2}{2}+\frac{20}{3}\log\left(\frac{\mu}{\bar{n}P}\right)\right)
\end{equation}

\section{Desintegracions radiatives de quarkoni pesat}
Les desintegracions semi-inclusives radiatives de quarkoni pesat a hadrons lleugers han estat estudiades des dels inicis de la CDQ. Aquests primers treballs tractaven el quarkoni pesat en analogia amb la desintegració de l'orto-positroni en EDQ. Diversos experiments, que es van fer posteriorment, van mostrar que la regió superior $z\to 1$ de l'espectre del fotó ($z$ és la fracció d'energia del fotó, respecte la màxima possible) no podia ser ben explicada amb aquests càlculs. Posteriors càlculs de correccions relativistes i de resumació de logaritmes, tot i que anaven en la bona direcció, no eren tampoc suficients per explicar les dades experimentals. Per contra, l'espectre podia ser ben explicat amb models que incorporaven una massa pel gluó. L'aparició de la CDQNR va permetre analitzar aquestes desintegracions de manera sistemàtica, però, tot i així, una massa finita pel gluó semblava necessària. Ben aviat, per això, es va notar que en aquesta regió superior la factorització de la CDQNR no funcionava. S'havien d'introduir les anomenades funcions d'estrucutra (en el canal octet de color), que integraven contribucions de diversos ordres en l'expansió de CDQNR. Alguns primers intents de modelitzar aquestes funcions d'estructura dugueren a resultats en fort desacord amb les dades. Més endavant es va reconèixer que per tractar correctament aquesta regió superior de l'espectre calia combinar la CDQNR amb la TECS (ja que els graus de llibertat col·lineals també eren importants en aquesta regió cinemàtica). D'aquesta manera les resumacions de logaritmes van ser estudiades en aquest marc (i es corregiren i ampliaren els resultats previs). Aquí farem servir una combinació de la CDQNRp amb la TECS per tal de calcular aquestes funcions d'estructura suposant que el quarkoni que es desintegra es pot tractar en el règim d'acoblament feble. Quan combinem de manera consistent aquests resultats amb els resultats previs coneguts, s'obté una bona descripció de l'espectre (sense que ja no calgui introduir una massa pel gluó) en tot el rang de $z$.

Per tal de calcular aquestes funcions d'estrucutra, el primer que cal fer es escriure els corrents en CDQNRp+TECS, que és on els calcularem. Un cop es té això ja es poden calcular els diagrames corresponents i aleshores, comparant amb les fórmules de factorització per aquest procés, es poden indentificar les funcions d'estrucutra desitjades. Els diagrames que cal calcular vénen representats a la figura \ref{figdos}. Del càlcul d'aquests diagrames s'obtenen les funcions d'estructura. El resultat que s'obté és divergent ultraviolat i ha de ser renormalitzat. Un cop s'ha fet això, si comparem el resultat teòric que tenim ara per l'espectre amb les dades experimentals en la regió superior, trobem un bon acord; tal i com es pot veure en la figura \ref{figepsubst} (les dues corbes en la figura representen diferents esquemes de renormalització). Fins ara hem pogut explicar, doncs, la regió superior de l'espectre. El que ara falta fer és veure si aquests resultats es poden combinar amb els càlculs anteriors, per la resta de l'espectre, i obtenir un bon acord amb les dades experimentals en tot el rang de $z$. Cal anar amb compte a l'hora de combinar aquests resultats, ja que en les diferents regions de l'espectre són necessàries diferents aproximacions teòriques (per tal de poder calcular). El procés emprat consisteix doncs en expandir (per $z$ en la regió central) les expressions que hem obtingut per la regió superior de l'espectre. Aleshores cal combinar les expressions d'acord amb la fórmula
\be
\frac{1}{\Gamma_0}\frac{d\Gamma^{dir}}{dz}=\frac{1}{\Gamma_0}\frac{d\Gamma^{c}}{dz}+\left(\frac{1}{\Gamma_0
}\frac{d\Gamma_{SC}^{e}}{dz}-\left.{\frac{1}{\Gamma_0
}\frac{d\Gamma_{SC}^{e}}{dz}}\right\vert_c\right)+\left(\frac{1}{\Gamma_0}\frac{d\Gamma_{OC}^{e}}{dz}-\left.{\frac{1}{\Gamma_0
}\frac{d\Gamma_{OC}^{e}}{dz}}\right\vert_c\right)
\ee
on $SC$ representa la contribució sinlget de color, $OC$ la contribució octet de color i els superíndexs $c$ i $e$ es refereixen a les expressions per la regió central i per l'extrem superior de l'espectre, respectivament. Quan usem aquesta fórmula aconseguim obtenir l'expressió vàlida per la regió central en la regió central i l'expressió vàlida per l'extrem superior de l'espectre en l'extrem superior, a part de termes que són d'ordre superior en el comptatge de la teoria efectiva en les respectives regions. I ho hem fet sense haver d'introduir talls o cotes arbitràries per tal de delimitar les diferents regions de l'espectre (cosa que hagués introduit incerteses teòriques bàsicament incontrolables en els nostres resultats). Quan comparem el resultat d'aquesta corba\footnote{També cal afegir les anomenades contribucions de fragmentació. A l'ordre en que estem treballant aquí són completament independents de les contribucions directes de la fórmula anterior.} (que ara ja conté tots els termes que, d'acord amb el nostre comptatge, han de ser presents) amb les dades experimentals, obtenim un molt bon acord. La comparació es pot veure a les figures \ref{figtotal} i \ref{figtotalnou} (la corba vermella (clara) contínua en aquestes figures és la predicció teòrica per l'espectre).

Un cop ja tenim l'espectre ben descrit des del punt de vista teòric, podem fer-lo servir per estudiar propietats del quarkoni pesat. En concret, és possible fer servir aquests espectres per tal de determinar en quin règim d'acoblament es troben els diferents quarkonis que es desintegren. Si calculem el quocient d'espectres de dos estats ($n$ i $r$) en el règim d'acoblament fort obtenim
\begin{equation}
\frac{\displaystyle\frac{d\Gamma_n}{dz}}{\displaystyle\frac{d\Gamma_r}{dz}} =\frac{\Gamma\left(V_Q(nS)\to e^+e^-\right)}{\Gamma\left(V_Q(rS)\to e^+e^-\right)}\left[\!1\!-\!\frac{\mathrm{Im}g_{ee}\left(\phantom{}^3S_1\right)}{\mathrm{Im}f_{ee}\left(\phantom{}^3S_1\right)}\frac{E_{n}-E_{r}}{m}\right]\left(1+\frac{C_1'\left[\phantom{}^3S_1\right](z)}{C_1\left[\phantom{}^3S_1\right](z)}\frac{1}{m}\left(E_{n}-E_{r}\right)\right)
\end{equation}
(totes les quantitats que apareixen en aquesta equació són conegudes), mentre que si un dels dos estats és en el règim d'acoblament feble la fórmula que obtenim presenta una dependència en $z$ diferent. Per tant si la fórmula anterior reprodueix bé el quocient d'espectres, això ens estarà indicant que els dos quarkonis estan en el règim d'acoblament fort, mentre que si no és així almenys un dels dos serà en el règim d'acoblament feble. Com que hi ha dades disponibles pels estats $\Upsilon(1S)$, $\Upsilon(2S)$ i $\Upsilon(3S)$ podem portar aquest procés a la pràctica. La comparació amb els resultats experimentals es pot veure a les figures \ref{fig1s2s}, \ref{fig1s3s} i \ref{fig2s3s} (l'estat $\Upsilon(1S)$ esperem que estigui en el règim d'acoblament feble, cosa que és compatible amb la gràfica \ref{fig1s2s}). Els errors són molt grans, però $\Upsilon(2S)$ i $\Upsilon(3S)$ semblen compatibles amb ser estats de règim d'acoblament fort (cal comparar la corba contínua amb els punts. Si coincideixen indica que els dos estats són en el règim d'acoblament fort).

\section{Conclusions}
En aquesta tesi hem fet servir les tècniques de teories efectives per tal d'estudiar el sector de quarks pesats del Model Estàndard. Ens hem centrat en l'estudi de tres temes. En primer lloc hem estudiat el potencial estàtic singlet de CDQ, fent servir la CDQ No Relativista de potencial. Amb l'ajuda d'aquesta teoria efectiva hem estat capaços de determinar la dependència infraroja sub-dominant d'aquest potencial estàtic. Entre altres possibles aplicacions, aquest resultat és rellevant en l'estudi de la producció de $t-\bar{t}$ prop del llindar de producció (a tercer ordre). Aquest és un procés que cal ser estudiat amb molt de detall amb vista a la possible futura construcció d'un gran accelerador lineal electró-positró. Després hem estudiat una dimensió anòmala en la TECS. Aquesta teoria té aplicacions molt importants en el camp de la física de mesons $B$. I aquest és un camp de gran importància per a la recerca indirecta de processos associats a nova física (mitjançant l'estudi de la violació de $CP$ i de la matriu de CKM). Finalment hem estudiat les desintegracions raditives semi-inlcusives de quarkoni pesat a hadrons lleugers. Per tal d'explicar bé aquest procés ha estat necessària una combinació de la CDQNRp amb la TECS. Mirant-s'ho des de la perspectiva actual, es pot veure aquest procés com un bonic exemple de com, una vegada s'incorporen tots els graus de llibertat rellevants en un problema (i es fa servir un comptatge ben definit per ells), aquest és ben descrit per la teoria. Un cop aquest procés està entès, es pot fer servir per estudiar algunes de les propietats del quarkoni pesat que es desintegra; com també hem mostrat en la tesi.

\selectlanguage{american}

%%% Local Variables: 
%%% mode: latex
%%% TeX-master: t
%%% End: 

\chapter{Background}\label{chapback}
In this chapter we describe the two effective field theories that will be
mainly used and studied in the thesis: \emph{potential Non-Relativistic QCD} and
\emph{Soft-Collinear Effective Theory}. It does not attempt to be a
comprehensive review but just provide the sufficient ingredients to follow the
subsequent chapters. %While review articles about
%non-relativistic theories (and about pNRQCD in particular) already exist, this
%is not the case for, the quite recent, SCET ei ei ei 

\section{potential Non Relativistic QCD}\label{secpNRQCD}
As it has already been explained in the introduction of the thesis, heavy quarkonium systems are
characterized by three intrinsic scales. Those are, the heavy quark mass $m$ (which is referred to
as the \emph{hard} scale and sets the mass of the quarkonium state), the
relative three-momentum of the heavy quark-antiquark pair
$\vert\mathbf{p}\vert\sim mv$ (which is
referred to as the \emph{soft} scale and sets the size of the bound state. $v$
is the typical relative velocity between the quark and the antiquark) and
the kinetic energy of the heavy quark and antiquark $E\sim mv^2$ (which is referred to as
the \emph{ultrasoft} scale and sets the binding energy of the quarkonium
state), and by the generic hadronic scale of QCD $\Lambda_{QCD}$. All those
scales are summarized in table \ref{tabpNsc}. The
interplay of $\Lambda_{QCD}$ with the other three scales determines the nature
of the different heavy quarkonium systems. By definition of heavy quark, $m$ is
always much larger than $\Lambda_{QCD}$; so the inequality $m\gg\Lambda_{QCD}$
always holds. Exploiting the inequality $m\gg\vert\mathbf{p}\vert,E$ one
arrives at Non-Relativistic QCD (NRQCD), as it has been described in the
previous chapter (note that at this level, after the
definition of heavy quark, one still
does not need to specify the interplay of $\Lambda_{QCD}$ with the remaining
scales, to identify the relevant degrees of freedom). Going one step further, using
the full non-relativistic hierarchy of the heavy quarkonium systems $m\gg mv\gg mv^2$, one
arrives at potential NRQCD (pNQRCD)\footnote{See \cite{Brambilla:2004jw} for a
review of pNRQCD.}. Now it is necessary to set the relative
importance of $\Lambda_{QCD}$ with the scales $\vert\mathbf{p}\vert$ and $E$
to fix the degrees of freedom of the resulting theory, the aim of which is to
study physics at the scale of the binding energy $E$. Two relevant regimes
have been identified so far; the so called \emph{weak coupling regime}, where
$mv^2\gtrsim\Lambda_{QCD}$, and the so called \emph{strong coupling regime},
where $mv\gtrsim\Lambda_{QCD}\gg mv^2$.
\begin{table}[t]
\begin{center}
\begin{tabular}{|c|c|}
\hline
 Scale & Value \\
\hline
 & \\
hard & $m$ \\
 & \\
soft & $mv$ \\
 & \\
ultrasoft & $mv^2$ \\
 & \\
generic hadronic QCD scale & $\Lambda_{QCD}$\\
 & \\
\hline
\end{tabular}
\caption{Relevant physical scales in heavy quarkonium}\label{tabpNsc}
\end{center}
\end{table}

\subsection{Weak coupling regime}

In this situation, the degrees of freedom of pNRQCD are not very different
from those of NRQCD. They are heavy quarks and antiquarks with a three
momentum cut-off $\nu_p$ ($\vert\mathbf{p}\vert\ll\nu_p\ll m$) and an energy
cut-off $\nu_{us}$
($\frac{\mathbf{p}^2}{m}\ll\nu_{us}\ll\vert\mathbf{p}\vert$), and gluons and
light quarks with a four momentum cut-off $\nu_{us}$. The most distinct feature
is that now non local terms in $r$, that is potentials, can appear (as it has
been discussed before, in the introductory section \ref{secinthq}). These degrees of freedom can
be arranged in several ways in the effective theory. One first way is to express them with the same
fields as in NRQCD. Then the pNRQCD Lagrangian has the following form
\begin{equation}
L_{pNRQCD}=L_{NRQCD}^{us}+L_{pot}
\end{equation}
where $L_{NRQCD}^{us}$ is the NRQCD Lagrangian but restricted to ultrasoft
gluons and $L_{pot}$ is given by
\begin{equation}
L_{pot}=-\int
d^3\mathbf{x}_1d^3\mathbf{x}_2\psi^{\dagger}\left(t,\mathbf{x}_1\right)\chi\left(t,\mathbf{x}_2\right)V\left(\mathbf{r},\mathbf{p}_1,\mathbf{p}_2,\mathbf{S}_1,\mathbf{S}_2\right)(us\textrm{
  gluon fields})\chi^{\dagger}\left(t,\mathbf{x}_2\right)\psi\left(t,\mathbf{x}_1\right)
\end{equation}
$\psi$ is the field that annihilates a quark and $\chi$ the one that creates
and antiquark; $\mathbf{p}_i=-i\bfnabla_{x_i}$ and $\mathbf{S}_i=\bfsigma_i/2$. Another option to express the
degrees of freedom is to represent the quark-antiquark pair by a wavefunction
field $\Psi$ (that is to project the theory to the one heavy quark-one heavy
antiquark sector)
\begin{equation}
\Psi (t,{\bf x}_1, {\bf x}_2)_{\alpha\beta}
 \sim
 \psi_{\alpha} (t,{\bf x}_1) \chi_{\beta}^\dagger (t,{\bf x}_2)
 \sim
 {1 \over N_c}\delta_{\alpha\beta}\psi_{\sigma} (t,{\bf x}_1)
 \chi_{\sigma}^\dagger (t,{\bf x}_2)
 +
 {1 \over T_F} T^a_{\alpha\beta}T^a_{\rho\sigma}\psi_{\sigma} (t,{\bf x}_1)
 \chi_{\rho}^\dagger (t,{\bf x}_2)
\end{equation}
Now the Lagrangian has the form ($m_1$ is the mass of the heavy quark and
$m_2$ the mass of the heavy antiquark, later on we will mainly focus
in the equal mass case $m_1=m_2\equiv m$)
\[
L_{NRQCD}^{us}=\int d^3{\bf x}_1 \, d^3{\bf x}_2 \; 
{\rm Tr}\, \left\{\Psi^{\dagger} (t,{\bf
  x}_1 ,{\bf x}_2 ) \left(
iD_0  +{{\bf D}_{{\bf x}_1 }^2\over 2\, m_1}+{{\bf D}_{{\bf x}_2 }^2\over 2\,
  m_2} + \cdots \right)\Psi (t,{\bf x}_1 ,{\bf x}_2 )\right\}-
\]
\begin{equation}
-\int d^3 x \; {1\over 4} G_{\mu \nu}^{a}(x) \,G^{\mu \nu \, a}(x) + 
\int d^3 x \;
\sum_{i=1}^{n_f} \bar q_i(x) \, i D\!\!\!\!\slash \,q_i(x)
+ \cdots
\end{equation}
\begin{equation}
L_{pot}=\int d^3{\bf x}_1 \, d^3{\bf x}_2 \; 
{\rm Tr} \left\{ \Psi^{\dagger} (t,  {\bf x}_1,{\bf x}_2)\,
V( {\bf r}, {\bf p}_1, {\bf p}_2, {\bf S}_1,{\bf S}_2)(us\hbox{ gluon fields}) \,
\Psi(t, {\bf x}_1, {\bf x}_2 )
\right\}
\end{equation}
%ei ei ei perque el canvi de signe? deu ser una tonteria pero mirar 
where the
dots represent higher order terms in the $1/m$ expansion and 
\begin{equation}
iD_0 \Psi (t,{\bf x}_1 ,{\bf x}_2)
= i\partial_0\Psi (t,{\bf x}_1 ,{\bf x}_2 ) -g A_0(t,{\bf x}_1)\,  
\Psi (t,{\bf x}_1 ,{\bf x}_2) + \Psi (t,{\bf x}_1 ,{\bf x}_2)\, g A_0(t,{\bf
  x}_2).
\end{equation}
The gluon fields
can be enforced to be ultrasoft by multipole expanding them in the relative
coordinate $\mathbf{r}$ (we define the center of mass coordinates by
$\mathbf{R}=(\mathbf{x}_1+\mathbf{x}_2)/2$ and $\mathbf{r}=\mathbf{x}_1-\mathbf{x}_2$), the
problem is that the multipole expansion spoils the manifest gauge invariance of
the Lagrangian. The gauge invariance can be restored by decomposing the
wavefunction field into (singlet and octet) components which have homogeneous
gauge transformations with respect to the center of mass coordinate
\[
\Psi (t,{\bf x}_1 ,{\bf x}_2)= 
P\bigl[e^{ig\int_{{\bf x}_2}^{{\bf x}_1} {\bf A} \cdot 
                   d{\bf x}} \bigr]\;{\rm S}({{\bf r}}, {{\bf R}}, t)
+P\bigl[e^{ig\int_{{\bf R}}^{{\bf x}_1} {\bf A} \cdot  d{\bf x}}
\bigr]\; {\rm O} ({\bf r} ,{\bf R} , t) \;
P\bigl[e^{ig\int^{{\bf R}}_{{\bf x}_2} {\bf A} \cdot d{\bf x}}\bigr]=
\]
\begin{equation}
=U_P\left(\mathbf{x}_1,\mathbf{R}\right)\left({\rm S}({{\bf r}}, {{\bf R}}, t)+{\rm O} ({\bf r} ,{\bf R} , t)\right)U_P\left(\mathbf{R},\mathbf{x}_2\right)
\end{equation}
with
\begin{equation}
U_P\left(\mathbf{x}_1,\mathbf{R}\right)=P\bigl[e^{ig\int_{{\bf R}}^{{\bf x}_1} {\bf A}(t,\mathbf{x}) \cdot  d{\bf x}}
\bigr]
\end{equation}
and the following color normalization for the singlet and octet fields
\begin{equation}
{\rm S} = { S 1\!\!{\rm l}_c / \sqrt{N_c}}  \quad\quad\quad 
{\rm O} = O^a { {\rm T}^a / \sqrt{T_F}}
\end{equation}
Arranging things that way, the lagrangian density (at order $p^3/m^2$) reads
\[
\mathcal{L}_{pNRQCD}=\int d^3{\bf r} \; {\rm Tr} \,  
\Biggl\{ {\rm S}^\dagger \left( i\partial_0 
- h_s({\bf r}, {\bf p}, {\bf P}_{\bf R}, {\bf S}_1,{\bf S}_2) \right) {\rm S} 
+ {\rm O}^\dagger \left( iD_0 
- h_o({\bf r}, {\bf p}, {\bf P}_{\bf R}, {\bf S}_1,{\bf S}_2) \right) {\rm O} \Biggr\}+
\]
\[
+V_A ( r) {\rm Tr} \left\{  {\rm O}^\dagger {\bf r} \cdot g{\bf E} \,{\rm S}
+ {\rm S}^\dagger {\bf r} \cdot g{\bf E} \,{\rm O} \right\} 
+ {V_B (r) \over 2} {\rm Tr} \left\{  {\rm O}^\dagger {\bf r} \cdot g{\bf E} \, {\rm O} 
+ {\rm O}^\dagger {\rm O} {\bf r} \cdot g{\bf E}  \right\}-
\]
\begin{equation}\label{pNRSO}
- {1\over 4} G_{\mu \nu}^{a} G^{\mu \nu \, a} 
+  \sum_{i=1}^{n_f} \bar q_i \, i D\!\!\!\!\slash \, q_i
\end{equation}
where
\begin{equation}
h_s({\bf r}, {\bf p}, {\bf P}_{\bf R}, {\bf S}_1,{\bf S}_2) = 
{{\bf p}^2 \over \, m_{\rm red}} 
+{{\bf P}_{\bf R}^2 \over 2\, m_{\rm tot}} + 
V_s({\bf r}, {\bf p}, {\bf P}_{\bf R}, {\bf S}_1,{\bf S}_2)
\end{equation}
\begin{equation}
h_o({\bf r}, {\bf p}, {\bf P}_{\bf R}, {\bf S}_1,{\bf S}_2) = 
{{\bf p}^2 \over \, m_{\rm red}} 
+{{\bf P}_{\bf R}^2 \over 2\, m_{\rm tot}} + 
V_o({\bf r}, {\bf p}, {\bf P}_{\bf R}, {\bf S}_1,{\bf S}_2)
\end{equation}
and
\begin{equation}
D_0 {\rm O} \equiv i \partial_0 {\rm O} - g [A_0({\bf R},t),{\rm O}]\quad {\bf
  P}_{\bf R} = -i{\bf D}_{\bf R}\quad m_{\rm red} =\frac{m_1m_2}{m_{\rm
  tot}}\quad m_{\rm tot} = m_1 + m_2
\end{equation}
%the $c_S$ and $c_O$ above are matching coefficients,
%which at leading order are equal to one\footnote{Poincar\'e invariance
%  constrains some of these matching coefficients \cite{Brambilla:2003nt}.}. 
$\mathbf{E}^i=G^{i0}$ and
$\mathbf{B}^i=-\epsilon_{ijk}G^{jk}/2$ are the chromoelectric and
chromomagnetic fields, respectively. Some of the Feynman rules arising from this Lagrangian are displayed in appendix \ref{appFR}. When written in terms of these singlet
and octet fields, the power counting of the pNRQCD Lagrangian is easy to
establish. Since the Lagrangian is bilinear in these fields we have just to
set the size of the terms multiplying those bilinears. The derivatives with
respect to the relative coordinate and $1/r$ factors must be counted as the
soft scale and the time derivatives, center of mass derivatives and fields for
the light degrees of freedom must be counted as the ultrasoft scale. The
$\alpha_s$ that come from the matching from NRQCD must be understood as
$\alpha_s(1/r)$ and the ones associated with light degrees of freedom must be
understood as $\alpha_s(E)$.

It is not that one form of the Lagrangian is preferred among the others, but
the different forms of writing the Lagrangian are convenient for different
purposes. In principle it is possible to go from one form of the Lagrangian to
the others; as an easy example consider the leading order Lagrangian (in
$\als$ and in the multipole expansion) in the static limit ($m\to\infty$)
written in terms of the wavefunction field
\[
L_{pNRQCD}=\int d^3{\bf x}_1 \, d^3{\bf x}_2 \; 
{\rm Tr}\, \left\{\Psi^{\dagger} (t,{\bf
  x}_1 ,{\bf x}_2 ) \left(
iD_0\right)\Psi (t,{\bf x}_1 ,{\bf x}_2 )\right\}+
\]
\[
+\int d^3{\bf x}_1 d^3{\bf x}_2 \;\frac{\als}{\vert\mathbf{x}_1-\mathbf{x}_2\vert}\mathrm{Tr}\left(T^a\Psi^{\dagger} (t,{\bf
  x}_1 ,{\bf x}_2 )T^a\Psi (t,{\bf x}_1 ,{\bf x}_2 )\right)-
\]
\begin{equation}
-\int d^3 x \; {1\over 4} G_{\mu \nu}^{a}(x) \,G^{\mu \nu \, a}(x) + 
\int d^3 x \;
\sum_{i=1}^{n_f} \bar q_i(x) \, i D\!\!\!\!\slash \,q_i(x)
\end{equation}
we will forget about the last line in the equation above, since it remains
the same. Now we introduce the singlet and octet fields, and take into account
that at leading order in the multipole expansion the Wilson lines are equal to
one, to obtain
\begin{equation}\label{wftoso}
\int d^3{\bf R} \, d^3{\bf r}\;\mathrm{Tr}\left\{\left(\mathrm{S}^{\dagger}+\mathrm{O}^{\dagger}\right)iD_0\left(\mathrm{S}+\mathrm{O}\right)\right\}+\int d^3{\bf R} \, d^3{\bf r}\;\frac{\als}{r}\mathrm{Tr}\left\{T^a\left(\mathrm{S}^{\dagger}+\mathrm{O}^{\dagger}\right)T^a\left(\mathrm{S}+\mathrm{O}\right)\right\}
\end{equation}
now, since
$iD_0(\mathrm{S}+\mathrm{O})=i\partial_0(\mathrm{S}+\mathrm{O})-g\left[A_0,\mathrm{O}\right]$
and taking into account that the trace of a single color matrix is
zero, we obtain from the first term in (\ref{wftoso})
\begin{equation}
\mathrm{Tr}\left\{\mathrm{S}^{\dagger}i\partial_0\mathrm{S}+\mathrm{O}^{\dagger}iD_0\mathrm{O}\right\}
\end{equation}
and from the second term
\begin{equation}
\frac{\als}{r}\mathrm{Tr}\left\{T^a\mathrm{S}^{\dagger}T^a\mathrm{S}+T^a\mathrm{O}^{\dagger}T^a\mathrm{O}\right\}=\frac{\als}{r}\mathrm{Tr}\left\{C_f\mathrm{S}^{\dagger}\mathrm{S}-\frac{1}{2N_c}\mathrm{O}^{\dagger}\mathrm{O}\right\}
\end{equation}
which gives us the static pNRQCD Lagrangian at leading order written in
terms of singlet and octet fields
\[
L_{pNRQCD}=\int d^3{\bf R} \, d^3{\bf r}\;\mathrm{Tr}\left\{\mathrm{S}^{\dagger}\left(i\partial_0+\frac{C_f\als}{r}\right)\mathrm{S}+\mathrm{O}^{\dagger}\left(iD_0-\frac{1}{2N_c}\frac{\als}{r}\right)\mathrm{O}\right\}-
\]
\begin{equation}
-\int d^3{\bf R}{1\over 4} G_{\mu \nu}^{a}G^{\mu \nu \, a} + 
\int d^3 {\bf R}
\sum_{i=1}^{n_f} \bar q_i \, i D\!\!\!\!\slash \,q_i
\end{equation}
While this procedure is relatively simple at leading order, in general it is
more convenient to construct each form of the pNRQCD Lagrangian independently
(by using the appropriate symmetry arguments and matching to NRQCD).

Note that, as mentioned before, the usual quantum mechanical potentials appear
as matching coefficients of the effective theory. Renormalization group improved expressions for the potentials can then be obtained \cite{Pineda:2000gz,Pineda:2001ra}.

\subsection{Strong coupling regime}
In this situation (where, remember, $\vert\mathbf{p}\vert\gtrsim\Lambda_{QCD}\gg E$) the
physics at the scale of the binding energy (which is in what we are
interested) is below the scale $\Lambda_{QCD}$. This implies that QCD is
strongly coupled, which in turn indicates that is better to formulate the
theory in terms of hadronic degrees of freedom. Hence we have, unavoidable, to
rely on some general considerations and indications from the lattice data to
identify the relevant degrees of freedom. Therefore we assume that a singlet
field describing the heavy quarkonium state together with Goldstone boson
fields, which are ultrasoft degrees of freedom, are the relevant degrees of
freedom for this theory. For this assumption to hold, we have to consider that there is
an energy gap of order $\Lambda_{QCD}$ from the ground state energy to the
higher hybrid excitations (that is states with excitations of the gluonic spin), which seems to be supported by lattice data, and
also that we are away from the energy threshold for the creation of a
heavy-light meson pair (in order to avoid mixing effects with these
states). If one forgets about the Goldstone boson fields (switch off light
fermions), as it is usually done, we are left with just the singlet field and the
theory takes the form of the potential models. In that case the pNRQCD
Lagrangian is given by
\begin{equation}
L_{\rm pNRQCD} = \int d^3 {\bf R}  \int d^3 {\bf r}  \;
S^\dagger \big( i\partial_0 - h_s({\bf x}_1,{\bf x}_2, {\bf p}_1, {\bf p}_2,
{\bf S}_1,  {\bf S}_2) \big) S
\end{equation}
with
\begin{equation}
h_s({\bf x}_1,{\bf x}_2, {\bf p}_1, {\bf p}_2, {\bf S}_1,  {\bf S}_2) =
{{\bf p}_1^2\over 2m_1} + {{\bf p}_2^2\over 2m_2}
+ V_s({\bf x}_1,{\bf x}_2, {\bf p}_1, {\bf p}_2, {\bf S}_1,  {\bf S}_2)
\end{equation}
The potential $V_s$ is now a
non-perturbative quantity (the different parts of which can be organized according to their scaling in $m$). The matching procedure will give us expressions for
the different parts of the potential in terms of Wilson loop amplitudes (which in
principle could be calculated on the lattice or with some vacuum model of QCD). When
considering annihilation processes (in which case, obviously, $m_1=m_2=m$), these expressions
translate into formulas for the NRQCD matrix elements. Hence, in the strong
coupling regime, the NRQCD matrix elements can be expressed in terms of wave functions at the origin and a few universal (that is bound state
independent) parameters. A list of some of the pNRQCD expressions for the
matrix elements can be found in appendix \ref{appME}. 

In the process of integrating out the degrees of freedom, from the scale $m$
to the ultrasoft scale, new momentum regions may appear (which were not
present in the weak coupling regime, since now we are also integrating
$\Lambda_{QCD}$). It turns out that the intermediate three momentum scale
$\sqrt{m\Lambda_{QCD}}$ is also relevant (it give contributions to loop
diagrams where gluons of energy $\Lambda_{QCD}$ are involved. Note that
$\sqrt{m\Lambda_{QCD}}$ is the three momentum scale that corresponds to the energy scale $\Lambda_{QCD}$). Hence, effects coming from this intermediate scale have also to be taken into account for the matching in the strong coupling regime \cite{Brambilla:2003mu}.

To establish the power counting of this Lagrangian we have to assign the soft scale to derivatives with respect to the relative coordinate and $1/m$ factors, and the ultrasoft scale $E$ to time derivatives and the static $V_s^{(0)}$. By definition of strong coupling regime $\als$ evaluated at the scale $E$ must be taken as order one. If we want to stay in the most conservative situation we should assume $\Lambda_{QCD}\sim mv$, in which case $\als(1/r)\sim 1$. Expectation values of fields for the light degrees of freedom should be counted as $\Lambda_{QCD}$ to the power of their dimension.

\section{Soft-Collinear Effective Theory}\label{secSCET}
The aim of this theory is to
describe processes in which very energetic (collinear) modes interact with
soft degrees of freedom. Soft-Collinear Effective Theory (SCET) can thus be applied
to a wide range of processes, in which this kinematic situation is
present. Those include exclusive and semi-inclusive $B$ meson decays, deep inelastic
scattering and Drell-Yan processes near the end-point, exclusive and
semi-inclusive quarkonium decays and many others. %ei ei ei cal posar
%referencies i fer la llista, potser, mes complerta...

Generally speaking, any process that contains highly energetic hadrons (that
is hadrons with energy much larger than its mass), together with a source for
them, will contain particles
(referred to as \emph{collinear}) which move close to a light cone direction
$n^{\mu}$. Since these particles are constrained to have large energy $E$ and
small invariant mass, the size of the different components (in light cone
coordinates, $p^{\mu}=(\bar{n}p)n^{\mu}/2+p^{\mu}_{\perp}+(np)\bar{n}^{\mu}/2$) of their momentum $p$ is very different; typically $\bar{n}p\sim E$, $p_{\perp}\sim
E\lambda$ and $np\sim E\lambda^2$, with $\lambda$ a small parameter. It is of
this hierarchy, $\bar{n}p\gg p_{\perp}\gg np$, that the effective theory takes
advantage. Due to the peculiar nature of the light cone interactions, the resulting
theory turns out to be non-local in one of the light cone directions (as it
has been mentioned in the introduction of the thesis). 

Unfortunately there is not a standard notation for the theory. Apart from
differences in the naming of the distinct modes, there are
basically two different formalisms (or notations). The one originally used in
\cite{Bauer:2000yr,Bauer:2001ct}, which uses the label operators (sometimes
referred to as the \emph{hybrid momentum-position space representation}) and the one
first employed in \cite{Beneke:2002ph}, which uses light-front multipole expansions to
ensure a well defined power counting (this is sometimes referred to as the
\emph{position space representation})\footnote{Note that the multipole expansions used in the
  position space representation are, to some extent, similar to the ones used
  in pNRQCD, while the hybrid representation is, not surprisingly, closer to
  the so called vNRQCD formalism. In any case the main difference between the
  vNRQCD and pNRQCD approaches is the way the soft and ultrasoft effects are
  disentangled. While vNRQCD introduces separate fields for the soft and
  ultrasoft degrees of freedom at the NRQCD level, the pNRQCD approach
  integrates out the soft scale producing thus a chain of effective theories
  QCD$\to$NRQCD$\to$pNRQCD, so that the final theory just contains the
  relevant degrees of freedom to study physics at the scale of the binding
  energy. In that sense any version of SCET is closer to vNRQCD than to
  pNRQCD, since separate (and overlapping) fields are introduced for the soft
  and collinear degrees of freedom (which probably is more adequate in this case) %ei ei ei %???? see [??????] for a discusion on how
%  the double counting should be solved???? ei ei ei).
.}. The two
formalisms are supposed to be completely equivalent (although precise comparisons are, many times, difficult).

The modes one need to include in the effective theory depend on whether one
want to study inclusive or exclusive processes. The resulting theories are
usually called SCET$_{\rm{I}}$ and SCET$_{\rm{II}}$, respectively. When one is
studying an inclusive process, collinear degrees of freedom with a typical
offshellness of order $\sqrt{E\Lambda_{QCD}}$ are needed. While in an
exclusive process the collinear degrees of freedom in the final state have
typical offshellness of order $\Lambda_{QCD}$; the simultaneous presence of
two type of collinear modes must then be taken into account in the matching
procedure from QCD to SCET$_{\rm{II}}$. We will briefly describe these two theories in turn, in the following subsections. In this
thesis we will be mainly using the SCET$_{\rm I}$ framework (consequently the
peculiarities and subtleties of SCET$_{\rm II}$ will just be very briefly mentioned).

\subsection{SCET$_{\rm I}$}
This is the theory containing collinear
$(p^{\mu}=(\bar{n}p,p_{\perp},np)\sim(1,\lambda,\lambda^2))$ and ultrasoft
$(p^{\mu}\sim(\lambda^2,\lambda^2,\lambda^2))$ modes\footnote{Be aware that
  the terminology for the different modes varies a lot in the literature. One
  should check the terminology used in each case to avoid unnecessary
  confusions (this is also true for SCET$_{\rm II}$).} (for some applications collinear fields in more than one direction could be needed), where the final collinear states have virtualities of order $E\Lambda_{QCD}$. The theory was first written in the (sometimes called) label or hybrid formalism \cite{Bauer:2000yr,Bauer:2001ct}. Within that approach the large component of the momentum $p$ is extracted from the fields (it becomes a label for them) according to
\begin{equation}
\phi(x)=\sum_{\tilde{p}\neq 0}e^{-i\tilde{p}x}\phi_{n,p}
\end{equation}
where $p=\tilde{p}+k$ and $\tilde{p}$ contains the large components of the momentum. In that way $\bar{n}p$ and $p_{\perp}$ have become labels for the field. Derivatives acting on $\phi_{n,p}$ will just give contributions of order $\lambda^ 2$. Then the so called label operators $\mathcal{P}$ are introduced. Those operators, when acting on the effective theory fields, give the sum of large labels in the fields minus the sum of large labels in the conjugate fields. We have, therefore
\begin{equation}
f\left(\bar{\mathcal{P}}\right)\left(\phi_{n,q_1}^{\dagger}\cdots\phi_{n,p_1}\cdots\right)=f\left(\bar{n}p_1+\cdots-\bar{n}q_1\cdots\right)\left(\phi_{n,q_1}^{\dagger}\cdots\phi_{n,p_1}\cdots\right)
\end{equation}
an analogous operator is defined for the transverse label
$\mathcal{P}_{\perp}^{\mu}$. With that technology, building blocks to form
invariant operators (under collinear and ultrasoft gauge transformations) can
be constructed. A scaling in $\lambda$ is assigned to the fields in the effective theory, such that the action for the kinetic terms counts as $\lambda^0$. The scaling for the various fields is summarized in table \ref{scaling}.
\begin{table}[t]
\begin{center}
\begin{tabular}{|c|c|}
\hline
Fields & Scaling \\
\hline
 & \\
collinear quark $\xi$ & $\lambda$ \\
 & \\
ultrasoft quark $q$ & $\lambda^3$ \\
 & \\
ultrasoft gluon $A_{us}^{\mu}$ & $\lambda^2$ \\
 & \\
collinear gluon $(\bar{n}A_{n,q},A_{n,q}^{\perp},nA_{n,q})$ & $(1,\lambda ,\lambda^2)$\\
 & \\
\hline
\end{tabular}
\caption{$\lambda$ scaling of the fields in SCET$_{\mathrm{I}}$.}\label{scaling}
\end{center}
\end{table}
The leading order Lagrangian for the SCET is then derived. This leading order (in the power counting in $\lambda$) Lagrangian is given by
\begin{equation}\label{LSCET}
\mathcal{L}_c=\bar{\xi}_{n,p'}\left\{inD+gnA_{n,q}+\left(\mathcal{P}\!\!\!\!\slash_{\perp}+gA\!\!\!\slash_{n,q}^{\perp}\right)W\frac{1}{\bar{\mathcal{P}}}W^{\dagger}\left(\mathcal{P}\!\!\!\!\slash_{\perp}+gA\!\!\!\slash_{n,q'}^{\perp}\right)\right\}\frac{\bar{n}\!\!\!\slash}{2}\xi_{n,p}
\end{equation}
in that equation $\xi$ are the fields for the collinear quarks, $A$ are the gluon fields, the covariant derivative $D$ contains ultrasoft gluon fields and $W$ are collinear Wilson lines given by
\begin{equation}
W=\left[\sum_{\mathrm{perm.}}e^{-g\frac{1}{\bar{\mathcal{P}}}\bar{n}A_{n,q}}\right]
\end{equation}
where the label operator acts only inside the square brackets. We can see that couplings to an arbitrary number of $\bar{n}A_{n,q}$ gluons are present at leading order in $\lambda$. The Feynman rules arising from this Lagrangian are given in appendix \ref{appFR}. 

Subsequently power suppressed (in $\lambda$) corrections to that Lagrangian were derived. This was first done in \cite{Beneke:2002ph,Beneke:2002ni}, where the position space formalism for SCET was introduced. In the position space formalism, the different modes present in the theory are also defined by the scaling properties of their momentum. But the strategy to construct the theory consists now of three steps. First one performs a field redefinition on the QCD fields, to introduce the fields with the desired scaling properties. Then the resulting Lagrangian is expanded, in order to achieve an homogeneous scaling in $\lambda$ of all the terms in it. This step involves multipole expanding the ultrasoft fields in one light cone direction, according to
\begin{equation}
\phi_{us}(x)=\phi_{us}(x_-)+\left[x_{\perp}\partial\phi_{us}\right](x_-)+\frac{1}{2}nx\left[\bar{n}\partial\phi_{us}\right](x_-)+\frac{1}{2}\left[x_{\mu\perp}x_{\nu\perp}\partial^{\mu}\partial^{\nu}\phi_{us}\right](x_-)+\mathcal{O}\left(\lambda^3\phi_{us}\right)
\end{equation}
where $x_-=1/2(\bar{n}x)n$. And finally the last step consists in a further field redefinition which restores the explicit (collinear and ultrasoft) gauge  invariance of the Lagrangian (which was lost by the multipole expansions). With that procedure the Lagrangian for SCET up to corrections of order $\lambda^2$ (with respect to the leading term (\ref{LSCET})) was obtained. Later on this power suppressed terms were also derived in the label formalism \cite{Pirjol:2002km}.

Note that the purely collinear part of the Lagrangian is equivalent to full
QCD (in a particular reference frame). The notion of collinear particle
acquires a useful meaning when, in a particular reference frame, we have a source that create such particles.

\subsection{SCET$_{\rm II}$}
This is the theory that describe processes in which the collinear particles in the final state have virtualities of order $\Lambda_{QCD}^2$. The simultaneous presence of two kinds of collinear modes must be taken into account in this case. We will have hard-collinear modes, with a typical scaling $p^{\mu}\sim(1,\lambda,\lambda^2)$ and virtuality of order $E\Lambda_{QCD}$ (these correspond to the collinear modes of the previous section, in SCET$_{\rm{I}}$) and collinear modes, with a typical scaling $p^{\mu}\sim(1,\lambda^2,\lambda^4)$ and virtuality of order $\Lambda_{QCD}^2$; together with ultrasoft modes with scaling $p^{\mu}\sim(\lambda^2,\lambda^2,\lambda^2)$. 

In the final effective theory (SCET$_{\rm II}$) only modes with virtuality
$\mathcal{O}(\Lambda_{QCD}^2)$ must be present. The contributions from the
intermediate hard-collinear scale must then be integrated out in this
case. This can be done with a two step process, where first the hard scale
$E$ is integrated and one ends up with SCET$_{\rm I}$. Then the hard-collinear
modes are integrated and one is left with an effective theory containing only
modes with virtuality of order $\Lambda_{QCD}^2$. SCET$_{\rm II}$ is therefore
much more complex than SCET$_{\rm I}$. In particular one of the most
controversial issues is how one should deal with end-point singularities that
may appear in convolutions for the soft-collinear factorization
formulas. Those can be treated, or regulated, in several different ways. If
one works in dimensional regularization in the limit of vanishing quark masses
a new mode, called soft-collinear messenger \cite{Becher:2003qh}, must be introduced in the theory. It provides a systematic way to discuss factorization and end-point singularities. Alternative regulators avoid the introduction of such a mode. Although this is clear now, to what extent the messenger should be considered as fundamental in the definition of the effective theory or not is still under debate.

%%% Local Variables: 
%%% mode: latex
%%% TeX-master: t
%%% End: 

\chapter{The singlet static QCD potential}\label{chappotest}
In this chapter we will calculate the logarithmic fourth order perturbative
correction to the static quark-antiquark potential for a color singlet
state (that is the sub-leading infrared dependence). This work appears here for the first time. It will later be reported in \cite{potestinprep}.

\section{Introduction}
The static potential between a quark and an antiquark is a key object for
understanding the dynamics of QCD. The first thing almost every student learns
about QCD is that a linear growing potential at long distances is a signal for
confinement. Apart from that, it is also a basic ingredient of the
Schrödinger-like formulation of heavy quarkonium. What is more, precise lattice data for the short distance part of the potential is nowadays available, allowing for a comparison between lattice and perturbation theory. Therefore the static potential is an ideal place to study the interplay of the perturbative and the non-perturbative aspects of QCD. 

The quark-antiquark system can be in a color singlet or in a color octet configuration. Which will give rise to the singlet and octet potentials, respectively. Both of them are relevant for the modern effective field theory calculations in the heavy quarkonium system. Here we will focus in the singlet potential.

The perturbative expansion of the singlet static potential (in position space) reads
\[
V_s^{(0)}(r)=-\frac{C_f\als(1/r)}{r}\left(1+\frac{\als(1/r)}{4\pi}\left(a_1+2\gamma_E\beta_0\right)+\left(\frac{\als(1/r)}{4\pi}\right)^2\bigg(
  a_2+\right.
\]
\begin{equation}\label{pot}
+\left.\left(\frac{\pi^2}{3}+4\gamma_E^2\right)\beta_0^2+\gamma_E\left(4a_1\beta_0+2\beta_1\right)\bigg)+\left(\frac{\als(1/r)}{4\pi}\right)^3\left(\tilde{a}_3+\frac{16\pi^2}{3}C_A^3\log r\mu\right)+\cdots\right)
\end{equation}
the one-loop coefficient $a_1$ is given by \cite{Fischler:1977yf,Billoire:1979ih} 
\begin{equation}
a_1=\frac{31}{9}C_A-\frac{20}{9}T_Fn_f
\end{equation}
and the two loop coefficient $a_2$ by \cite{Peter:1996ig,Schroder:1998vy}
\begin{eqnarray}
  a_2&=&
  \left[{4343\over162}+4\pi^2-{\pi^4\over4}+{22\over3}\zeta(3)\right]C_A^2
  -\left[{1798\over81}+{56\over3}\zeta(3)\right]C_AT_Fn_f-
  \nonumber\\
  &&{}-\left[{55\over3}-16\zeta(3)\right]C_fT_Fn_f
  +\left({20\over9}T_Fn_f\right)^2
\end{eqnarray}
the non-logarithmic third order correction $\tilde{a}_3$ is still unknown. The form of the logarithmic
term in (\ref{pot}) corresponds to using dimensional regularization for the
ultrasoft loop (which is the natural scheme when calculating from
pNRQCD).\footnote{Note that this is not the natural scheme when calculating
from NRQCD. In that case one would regulate also the potentials in
$d$-dimensions.}

We will calculate here the logarithmic fourth order correction to the potential. Since this calculation follow the same lines as that of the third order logarithmic terms, we will briefly review it in the next section.

\section{Review of the third order logarithmic correction}
The leading infrared (IR) logarithmic dependence of the singlet static potential was obtained in \cite{Brambilla:1999xf} by matching NRQCD to pNRQCD perturbatively. The matching is performed by comparing Green functions in NRQCD and pNRQCD (in coordinate space), order by order in $1/m$ and in the multipole expansion. 

\begin{figure}
\centering
\includegraphics{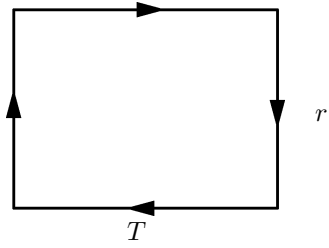}
\caption[Rectangular Wilson loop]{Rectangular Wilson loop. The corners are $x_1 = (T/2,{\bf r}/2)$, $x_2 = (T/2,-{\bf r}/2)$, 
$y_1 = (-T/2,{\bf r}/2)$ and  $y_2 = (-T/2,-{\bf r}/2)$}\label{figWL}
\end{figure}

To perform that matching, first of all one need to identify interpolating fields in NRQCD with the same quantum numbers and transformation properties as the singlet and octet fields in pNRQCD. The chosen fields are
\begin{equation}
\chi^\dagger({\bf x}_2,t) \phi({\bf x}_2,{\bf x}_1;t) \psi({\bf x}_1,t) 
\rightarrow  \sqrt{Z^{(0)}_s(r)} S({\bf r},{\bf R},t) 
+ \sqrt{Z_{E,s}(r)} \, r \, {\bf r}\cdot g{\bf E}^a({\bf R},t) O^a({\bf r},{\bf R},t) + \dots
\end{equation}
for the singlet, and
\begin{eqnarray}
\chi^\dagger({\bf x}_2,t) \phi({\bf x}_2,{\bf R};t) T^a \phi({\bf R},{\bf x}_1;t) \psi({\bf x}_1,t) 
&\rightarrow& \sqrt{Z^{(0)}_o(r)} O^a({\bf r},{\bf R},t)+
\nonumber\\
&& \hspace{-5mm}
+ \sqrt{Z_{E,o}(r)} \, r \, {\bf r}\cdot g{\bf E}^a({\bf R},t) S({\bf r},{\bf R},t) + \dots
\end{eqnarray}
for the octet, where
\begin{equation}
\phi({\bf y},{\bf x};t)\equiv P \, \exp \left\{ i \displaystyle 
\int_0^1 \!\! ds \, ({\bf y} - {\bf x}) \cdot g{\bf A}({\bf x} - s({\bf x} - {\bf y}),t) \right\}
\end{equation}
The $Z$ in the above expressions are normalization factors. The different combinations of fields in the pNRQCD side are organized according to the multipole expansion, just the first term of this expansion is needed for our purposes here. Then the matching is done using the Green function
\begin{equation}
G=\langle {\rm vac} \vert  \chi^\dagger(x_2) \phi(x_2,x_1) \psi(x_1) 
\psi^\dagger(y_1)\phi(y_1,y_2) \chi(y_2) \vert {\rm vac} \rangle
\end{equation}
In the NRQCD side we obtain
\begin{equation}\label{GNRQCD}
G_{\mathrm{NRQCD}}=\delta^3({\bf x}_1 - {\bf y}_1) \delta^3({\bf x}_2 - {\bf y}_2) 
\langle W_\Box \rangle
\end{equation}
where $W_\Box$ represents the rectangular Wilson loop of figure \ref{figWL}. Explicitly it is given by 
\begin{equation}
W_\Box \equiv P \,\exp\left\{{\displaystyle - i g \oint_{r\times T} \!\!dz^\mu A_{\mu}(z)}\right\}.
\end{equation}
The brackets around it in (\ref{GNRQCD}) represent an average over gauge fields and light quarks.
We are interested only in the large $T$ limit of the Wilson loop (to single out the soft scale), therefore we define the following expansion for $T\to\infty$
\begin{equation}
{i\over T}\ln  \langle W_\Box \rangle = u_0(r) + i {u_1(r)\over T} + {\cal
  O}\left( {1\over T^2}\right)
\end{equation}
In the pNRQCD side we obtain, at order $r^2$ in the multipole expansion\footnote{The superscripts $(0)$ in all those expressions are reminding us that we are in the static limit $m\to\infty$. Since in this chapter we are always in the static limit, we will omit them after (\ref{GpNRQCD}), to simplify the notation.}
\begin{eqnarray}\label{GpNRQCD}
& &G_{\rm pNRQCD}
= Z^{(0)}_s(r) \delta^3({\bf x}_1 - {\bf y}_1) \delta^3({\bf x}_2 - {\bf y}_2) 
e^{-iTV^{(0)}_s(r)}\cdot
\\
& &\cdot \left(1 
- {T_F\over N_c} V_A^2 (r)
\int_{-T/2}^{T/2} \! dt \int_{-T/2}^{t} \! dt^\prime \, 
e^{-i(t-t^\prime)(V^{(0)}_o-V^{(0)}_s)} 
\langle {\bf r}\cdot g{\bf E}^a(t) \phi^{\rm adj}_{ab}(t,t^\prime){\bf r}
\cdot g{\bf E}^b(t^\prime)\rangle
\right)
\nonumber
\end{eqnarray}
where the Wilson line
\begin{equation}
\phi(t,t')=P\, \exp \left\{ - ig \displaystyle \int_{t'}^{t} \!\! d\tilde{t} \, A_0(\tilde{t}) \right\}
\end{equation}
which comes from the octet propagator, is evaluated in the adjoint representation.

\begin{figure}
\centering
\includegraphics{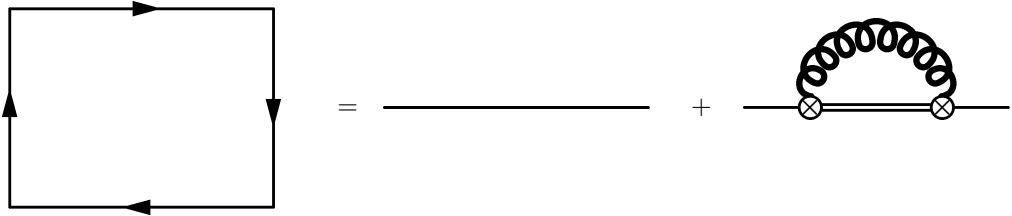}
\caption[NRQCD$\to$pNRQCD matching of the potential at $\mathcal{O}(r^2)$]{Schematic representation of the NRQCD$\to$pNRQCD matching for the static potential and the normalization factors, at order $r^2$ in the multipole expansion. On the pNQRCD side (right) the single line represents the singlet field, the double line the octet field, the circled cross the $\mathcal{O}(r)$ chromoelectric vertex and the thick springy line represents the correlator of chromoelectric fields. Remember that those diagrams are representing an expansion in $r$ (and in $1/m$) and not a perturbative expansion in $\alpha_s$.}\label{figmatchr2}
\end{figure}

Then comparing $G_{\rm NRQCD}$ with $G_{\rm pNRQCD}$ one obtains the matching conditions for the potential and the normalization factors. That matching is schematically represented in figure \ref{figmatchr2}.
Note that up to this point no perturbative expansion in $\als$ have been used yet. We will now evaluate the pNRQCD diagram perturbatively in $\als$. The dependence in $\als$ enters through the $V_A$, $V_s$ and $V_o$ potentials and through the field strength correlator of chromoelectric fields. To regulate IR divergences we will keep the $\als$ dependence in the exponential on the second line of (\ref{GpNRQCD}) unexpanded. $V_o-V_s$ will then act as our IR regulator. The tree level expression for $V_A$ is given by the NRQCD$\to$pNRQCD matching at order $r$ in the multipole expansion. It simply states that $V_A=1$ at tree level. We then need the tree level expression for the correlator of chromoelectric fields. Note that, since afterwards we want to integrate over $t$ and $t'$, we need this gluonic correlator in $d$ dimensions (see the following subsections for more details). Inserting the $d$-dimensional ($d=4-2\epsilon$) tree level expressions we obtain (in the $T\to\infty$ limit)
\[
G_{\rm pNRQCD}=Z_s(r) \delta^3({\bf x}_1 - {\bf y}_1) \delta^3({\bf x}_2 - {\bf y}_2) 
e^{-iTV_s(r)}\cdot
\]
\[
\cdot\left(1-iT\frac{N_c^2-1}{2N_c}\frac{\alpha_s(\mu)}{\pi}\frac{r^2}{3}\left(V_o-V_s\right)^3\left(\frac{1}{\epsilon}-\log\left(\frac{V_o-V_s}{\mu}\right)^2+const.\right)-\right.
\]
\begin{equation}\label{restl}
\left.-\frac{N_c^2-1}{2N_c}\frac{\alpha_s(\mu)}{\pi}r^2\left(V_o-V_s\right)^2\left(\frac{1}{\epsilon}-\log\left(\frac{V_o-V_s}{\mu}\right)^2+const.\right)\right)
\end{equation}
where we can explicitly see that $V_o-V_s$ acts as our IR regulator in the logarithms. The last line in (\ref{restl}) only affects the normalization factor (and not the potential) and will be omitted in the following. The $V_o-V_s$ that appears outside the logarithms must now be expanded in $\als$ (since are no longing acting as IR regulators). Therefore we have obtained the result
\[
G_{\rm pNRQCD}=Z_s(r) \delta^3({\bf x}_1 - {\bf y}_1) \delta^3({\bf x}_2 - {\bf y}_2) 
e^{-iTV_s(r)}\cdot
\]
\begin{equation}\label{restl2}
\cdot\left(1-iT\frac{N_c^2-1}{2N_c}\frac{\alpha_s(\mu)}{\pi}\frac{C_A^3}{24}\frac{1}{r}\alpha_s^3(1/r)\left(\frac{1}{\epsilon}-\log\left(\frac{V_o-V_s}{\mu}\right)^2+const.\right)+\mathcal{O}(T^0)\right)
\end{equation}
Note that the $\als$ coming from the potentials are evaluated at the soft scale $1/r$, while the $\als$ coming from the ultrasoft couplings is evaluated at the scale $\mu$\footnote{Obviously the distinction is only relevant at the next order, that is next section.}. The ultraviolet divergences in that expression can be re-absorbed by a renormalization of the potential. Comparing now the perturbative evaluations of $G_{\rm NRQCD}$ and $G_{\rm pNRQCD}$ (obviously the same renormalization scheme one has used in the evaluation of the pNRQCD diagram must be used in the calculation of the Wilson loop in NRQCD) we obtain
\begin{equation}
V_s(r,\mu)=(u_0(r))_{\rm two-loops} 
-\frac{N_c^2-1}{2N_c}{C_A^3\over 12} {\als\over r}{\als^3\over \pi}\ln ({r\mu})
\end{equation}
when $(u_0(r))_{\rm two-loops}$ is substituted by its known value we obtain the result (\ref{pot}), quoted in the introduction of this chapter. The calculation explained in this section has thus provided us the leading IR logarithmic dependence of the static potential. In \cite{Brambilla:1999xf} the cancellation of the IR cut-off dependence between the NRQCD and pNRQCD expressions was checked explicitly, by calculating the relevant graphs for the Wilson loop. That was an explicit check that the effective theory (pNRQCD) is correctly reproducing the IR.

In the following section we will use the same procedure employed here to obtain the next-to-leading IR logarithmic dependence of the static potential. That is the logarithmic $\als^5$ contribution to the potential (which is part of the N$^4$LO, in $\als$, correction to the potential).

\section{Fourth order logarithmic correction}
In the preceding section no perturbative expansion in $\als$ was used until
the paragraph following equation (\ref{GpNRQCD}). Therefore (\ref{GpNRQCD}) is
still valid for us here and will be our starting point (note that contributions from higher order operators in the multipole expansion have a
suppression of order $\als^2$ with respect to the second term of
(\ref{GpNRQCD}), and therefore are unimportant for us here). We need to calculate the $\als$ correction to the evaluation of the diagram in the preceding section. Remember that the dependence in $\als$ enters through $V_A$, $V_s$, $V_o$ and the correlator of gluonic fields, therefore we need the $\mathcal{O}(\alpha_s)$ correction to all this quantities. That terms will be discussed in the following subsections in turn. Then in subsection \ref{subseccal4o} we will obtain the fourth order logarithmic correction to the potential. Again we will regulate IR divergence by keeping the exponential of $V_o-V_s$ unexpanded.

\subsection{$\mathcal{O}(\alpha_s)$  correction of $V_A$, $V_s$ and $V_o$}
\begin{figure}
\centering
\includegraphics{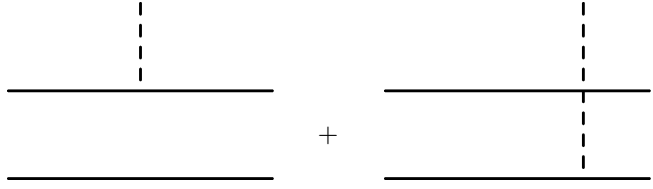}
\caption[Leading order matching of $V_A$]{NRQCD diagrams for the leading order matching of $V_A$. The solid lines represent the quark and the antiquark, the dashed line represents an $A_0$ gluon.}\label{figVALO}
\end{figure}

\begin{figure}
\centering
\includegraphics{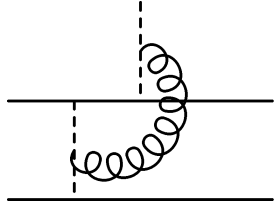}
\caption[Next-to-leading order matching of $V_A$]{Sample NRQCD diagram for the next-to-leading order matching of $V_A$.}\label{figVANLO}
\end{figure}

The $\mathcal{O}(\alpha_s)$ corrections to $V_s$ and $V_o$ are well known. They are given by
\begin{eqnarray}
V_s & = & -\frac{C_f}{r}\alpha_s(1/r)\left(1+\left(a_1+2\gamma_E\beta_0\right)\frac{\alpha_s(1/r)}{4\pi}\right)\\
V_o & = & \left(\frac{C_A}{2}-C_f\right)\frac{1}{r}\alpha_s(1/r)\left(1+\left(a_1+2\gamma_E\beta_0\right)\frac{\alpha_s(1/r)}{4\pi}\right)
\end{eqnarray}
as it has already been reported in the introduction of this chapter. The mixed
potential $V_A$ can be obtained by matching NRQCD to pNRQCD at order $r$ in
the multipole expansion (we are thinking in performing this matching for $V_A$
in pure
dimensional regularization). At leading order in $\alpha_s$ we have to calculate the diagrams shown in figure \ref{figVALO}. They give the tree level result $V_A=1$. The first corrections to this result are given by diagrams like that of figure \ref{figVANLO}. We can clearly see that the first corrections are $\mathcal{O}(\alpha_s^2)$
\begin{equation}
V_A=1+\mathcal{O}(\alpha_s^2)
\end{equation}
and therefore unimportant for us here.

\subsection{$\mathcal{O}(\alpha_s)$ correction of the field strength correlator}
The $\mathcal{O}(\alpha_s)$ correction to the QCD field strength correlator was calculated in \cite{Eidemuller:1997bb}. Let us review here that result and explain how we need to use it.

\begin{figure}
\centering
\includegraphics{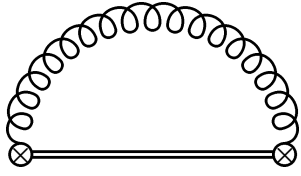}
\caption[Field strength correlator at leading order]{Leading order contribution to the field strength correlator. The gluonic string is represented by a double line.}\label{figfscLO}
\end{figure}

\begin{figure}
\centering
\includegraphics{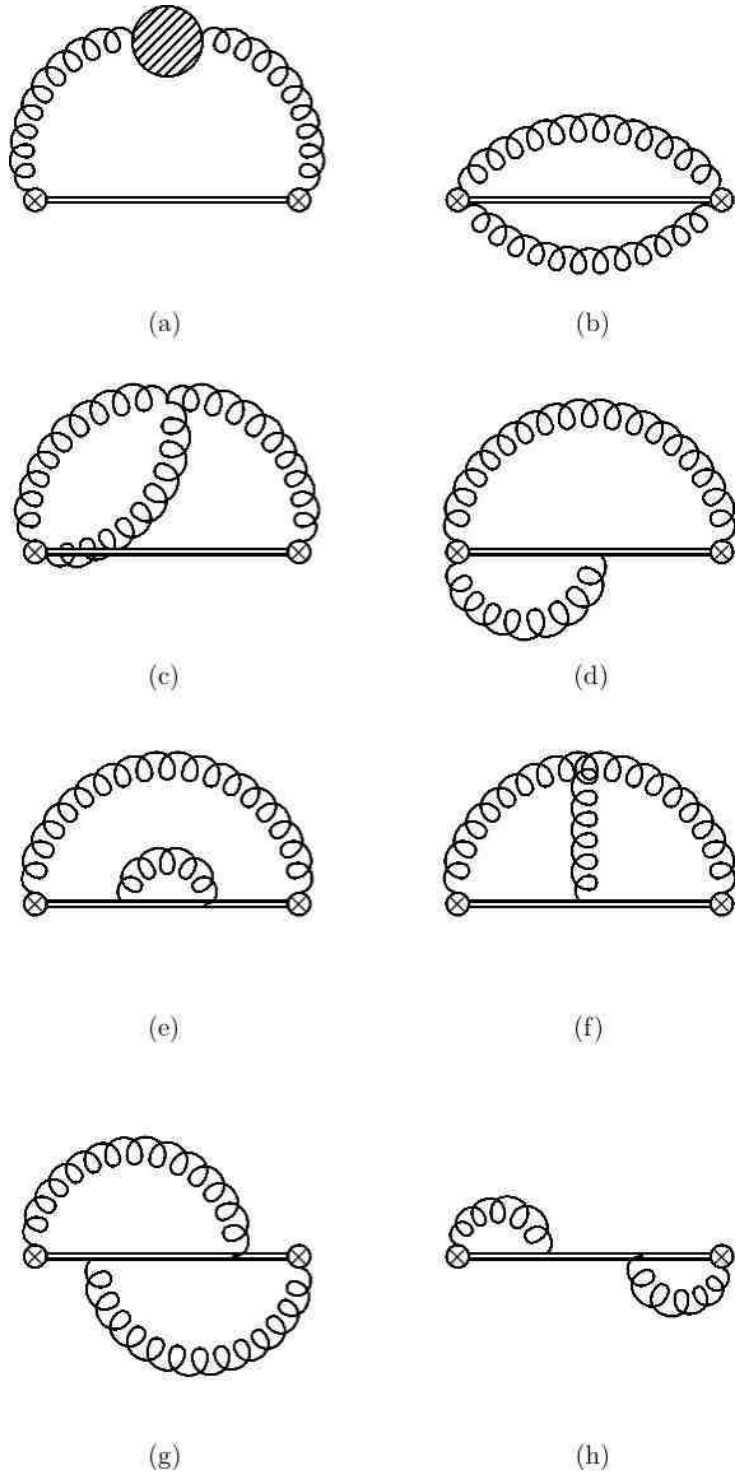}
\caption[Field strength correlator at next-to-leading order]{Next-to-leading order contributions to the field strength correlator. The gluonic string is represented by a double line. The shaded blob represents the insertion of the 1-loop gluon self-energy. Symmetric graphs are understood for (c) and (d).}\label{figfscNLO}
\end{figure}

\begin{figure}
\centering
\includegraphics{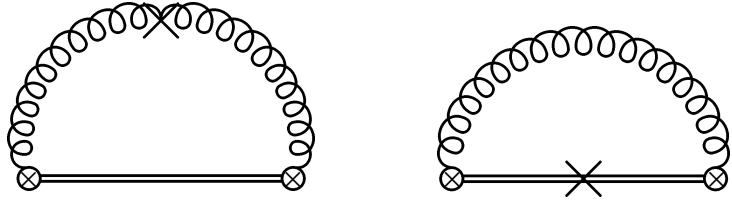}
\caption[$\mathcal{O}(\als)$ counterterm diagrams for the chromoelectric correlator]{$\mathcal{O}(\als)$ counterterm diagrams for the chromoelectric correlator. The gluonic string (which comes from the octet propagator) is represented by a double line.}\label{figctEE}
\end{figure}

The two-point field strength correlator
\begin{equation}
\mathcal{D}_{\mu\nu\lambda\omega}(z)\equiv\left<\mathrm{vac}\right\vert T\left\{G_{\mu\nu}^a(y)\mathcal{P}e^{gf^{abc}z^{\tau}\int_0^1d\sigma A_{\tau}^c(x+\sigma z)}G_{\lambda\omega}^b(x)\right\}\left\vert\mathrm{vac}\right>
\end{equation}
can be parametrised in terms of two scalar functions $\mathcal{D}(z^2)$ and $\mathcal{D}_1(z^2)$ according to
\[
\mathcal{D}_{\mu\nu\lambda\omega}(z)=\left(g_{\mu\lambda}g_{\nu\omega}-
g_{\mu\omega}g_{\nu\lambda}\right)\left(\mathcal{D}(z^2)+\mathcal{D}_1(z^2)\right)+
\]
\begin{equation}
+\left(g_{\mu\lambda}z_\nu z_\omega-g_{\mu\omega}
z_\nu z_\lambda-g_{\nu\lambda} z_\mu z_\omega+g_{\nu\omega}z_\mu z_\lambda\right)\frac{\partial\mathcal{D}_1(z^2)}{\partial z^2}
\end{equation}
where $z=y-x$. In (\ref{GpNRQCD}) $x$ and $y$ just differ in the time component, so $z=t-t'$ for us. Furthermore, we are interested in the chromoelectric components, so we need the contraction
\begin{equation}
\mathcal{D}_{i0i0}(z)=-(d-1)\left(\mathcal{D}(z^2)+\mathcal{D}_1(z^2)+z^2\frac{\partial\mathcal{D}_1(z^2)}{\partial z^2}\right)
\end{equation}
The tree level contribution is given by the diagram shown in figure \ref{figfscLO}\footnote{Note a slight change in notation in the diagrams with respect to \cite{Eidemuller:1997bb}. We represent the gluonic string by a double (not dashed) line and we always display it (also when it reduces to $\delta^{ab}$).}, the result is 
\begin{equation}
\mathcal{D}_1^{(0)}(z^2)=\mu^{2\epsilon}(N_c^2-1)\frac{\Gamma(2-\epsilon)}{\pi^{2-\epsilon}z^{4-2\epsilon}}\qquad \mathcal{D}^{(0)}(z^2)=0
\end{equation} 
The next-to-leading ($\mathcal{O}(\als)$) contribution is given by the
diagrams in figure \ref{figfscNLO}. Here we need the expression in $d$
dimensions. The $d$-dimensional result for the $\als$ correction is
\cite{pcJamin}\footnote{We are indebted to Matthias Jamin for sharing the $d$-dimensional results for the field strength correlator with us.}
\begin{eqnarray}
\mathcal{D}^{(1)}(z^2) & = & N_c(N_c^2-1)\frac{\als}{\pi}\frac{\mu^{4\epsilon}}{4\pi^{2-2\epsilon}}\Gamma^2(1-\epsilon)\left(\frac{1}{z^2}\right)^{2-2\epsilon}g(\epsilon)\\
\mathcal{D}_1^{(1)}(z^2) & = & N_c(N_c^2-1)\frac{\als}{\pi}\frac{\mu^{4\epsilon}}{4\pi^{2-2\epsilon}}\Gamma^2(1-\epsilon)\left(\frac{1}{z^2}\right)^{2-2\epsilon}g_1(\epsilon)
\end{eqnarray}
with
\begin{eqnarray}
g(\epsilon) & = & \frac{2 \epsilon ^3+2 (1-\epsilon) B(2 \epsilon -1,2 \epsilon -2) \epsilon ^2-6 \epsilon ^2+8
   \epsilon -3}{\epsilon  \left(2 \epsilon ^2-5 \epsilon +3\right)}\\
g_1(\epsilon) & = & \frac{-6 \epsilon ^3+17 \epsilon ^2-18 \epsilon +6}{\epsilon ^2 \left(2 \epsilon ^2-5 \epsilon
   +3\right)}+\frac{2 (1-\epsilon+\epsilon^2) B(2 \epsilon -1,2 \epsilon -2)+\frac{2
   (1-\epsilon) n_f}{N_c}}{\epsilon  (2 \epsilon -3)}
\end{eqnarray}
Since the external points, $x$ and $y$, are fixed in this calculation, the divergences we will encounter in $\mathcal{D}_{i0i0}$ (coming from the expressions above) should be canceled by the vertex and (gluon and octet field) propagator counterterms. The counterterm for the vertex is zero, since as we have seen in the previous subsection the first correction to $V_A$ is of order $\als^2$. The counterterm for the gluon propagator is the usual one in QCD. The counterterm for the octet propagator coincides with the counterterm for the quark propagator in Heavy Quark Effective Theory but with the quark in the adjoint representation. We can represent the counterterm contributions by the diagrams of figure \ref{figctEE}. We have checked that when we compute $\mathcal{D}_{i0i0}$ the divergence coming from the first diagram in figure \ref{figfscNLO} is canceled by the counterterm of the gluon propagator. That diagram \ref{figfscNLO}b does not give a divergent contribution, as it should. And that when we add the remaining diagrams the divergence we obtain is exactly canceled by the counterterm of the octet propagator.

The contributions of the counterterms are given by
\begin{eqnarray}
\mathcal{D}^{\mathrm{c.t.}}(z^2) & = & 0\\
\mathcal{D}^{\mathrm{c.t.}}_1(z^2) & = & N_c(N_c^2-1)\frac{\als}{\pi}\frac{\mu^{2\epsilon}}{4\pi^{2-\epsilon}}\Gamma(2-\epsilon)\frac{1}{z^{4-2\epsilon}}\left[\frac{-2}{\epsilon}+\frac{1}{\epsilon}\left(-\frac{5}{3}+\frac{4}{3}T_F\frac{n_f}{N_c}\right)\right]
\end{eqnarray}
where the first $1/\epsilon$ in the square bracket corresponds to the octet propagator and the second one to the gluon propagator. Then the total $d$-dimensional result (including the contributions from the counterterms) for the $\als$ correction to the chromoelectric correlator is
\begin{equation}\label{EENLO}
\mathcal{D}_{i0i0}^{(1)}=-(d-1)\left(\mathcal{D}^{(1)}(z^2)+(-1+2\epsilon)\mathcal{D}_1^{(1)}(z^2)+\mathcal{D}^{\mathrm{c.t.}}(z^2)+(-1+\epsilon)\mathcal{D}_1^{\mathrm{c.t.}}(z^2)\right)
\end{equation}
which no longer have $1/\epsilon$ poles.

\subsection{Calculation of the fourth order logarithmic correction}\label{subseccal4o}
The results of the two preceding subsections provides us all the necessary ingredients to compute the next-to-leading IR dependence of the static potential. We have to evaluate the second term in the parenthesis of (\ref{GpNRQCD}) at next-to-leading order. Let us define
\begin{equation}
\mathcal{G}^{(r^2)}\equiv - {T_F\over N_c} V_A^2 (r)
\int_{-T/2}^{T/2} \! dt \int_{-T/2}^{t} \! dt^\prime \, 
e^{-i(t-t^\prime)(V_o-V_s)} 
\langle {\bf r}\cdot g{\bf E}^a(t) \phi^{\rm adj}_{ab}(t,t^\prime){\bf r}
\cdot g{\bf E}^b(t^\prime)\rangle
\end{equation}
First we will consider the contribution we obtain when we insert the $\als$ correction (\ref{EENLO}) to the field strength correlator. We have just to perform the integrations over $t$ and $t'$. To do that we change the integration variables to $t+t'$ and $t-t'\equiv t_-$. The integral over the sum just gives us a factor $T-t_-$. The $T$ term will give a contribution to the potential and the $t_-$ term a contribution to the normalization factor (that is unimportant for us here, it will be omitted in the following). The remaining integral over $t_-$ can be done by using
\begin{equation}
\int_0^{\infty}dx\,x^ne^{-ax}=\frac{\Gamma(n+1)}{a^{n+1}}
\end{equation}
The result we obtain is then (in the $T\to\infty$ limit)
\[
\mathcal{G}^{(r^2)}_{<EE>\vert_{\mathcal{O}(\als)}}=-iT\left(\frac{\alpha_s(\mu)}{\pi}\right)^2\alpha_s^3(1/r)\frac{N_c^2-1}{2}\frac{C_A^3}{8}\frac{1}{r}\cdot
\]
\begin{equation}\label{coEE}
\cdot\left(\frac{A}{\epsilon^2}+\frac{B}{\epsilon}+C_1\log^2\frac{V_o-V_s}{\mu}+C_2\log\frac{V_o-V_s}{\mu}+const.\right)
\end{equation}
with
\begin{eqnarray}
A & = & \frac{1}{24} \left(\frac{2 n_f}{3N_c}-\frac{11}{3}\right)\nonumber\\
B & = & \frac{1}{108} \left(-\frac{5 n_f}{N_c}+6 \pi ^2+47\right)\nonumber\\
C_1 & = & \frac{1}{6} \left(-\frac{2 n_f}{3N_c}+\frac{11}{3}\right)\nonumber\\
C_2 & = & \frac{1}{54} \left(\frac{20 n_f}{N_c}-12 \pi ^2-149\right)
\end{eqnarray}
We get another contribution to $\mathcal{G}^{(r^2)}$ when we use the leading order expression for the chromoelectric correlator (then we arrive at (\ref{restl})) and then insert the next-to-leading order correction to $V_o-V_s$. This contribution is given by
\begin{equation}\label{coprop}
\mathcal{G}^{(r^2)}_{V_o-V_s\vert_{\mathcal{O}(\als)}}=-iT\frac{\als(\mu)}{\pi}\frac{\als^4(1/r)}{4\pi}\frac{N_c^2-1}{2N_c}\frac{C_A^3}{8}\frac{1}{r}\left(a_1+2\gamma_E\beta_0\right)\left(\frac{1}{\epsilon}-\log\left(\frac{V_o-V_s}{\mu}\right)^2+const.\right)
\end{equation}
The ultraviolet divergences we encounter in expressions (\ref{coEE}) and (\ref{coprop}) can again be
re-absorbed by a renormalization of the potential. Finally we get another
contribution that comes from changing $\als(\mu)$ to $\als(1/r)$ in equation
(\ref{restl2}), after renormalization (we want all the $\als$ evaluated at the scale
$1/r$ in the potential). It is given by
\begin{equation}
  \mathcal{G}^{(r^2)}_{\mu\to 1/r}=-iT\frac{\als^5(1/r)C_A^3\beta_0}{48\pi^2}\frac{N_c^2-1}{2N_c}2\log(r\mu)\log\left(\frac{V_o-V_s}{\mu}\right)
\end{equation}

We see that the $\log^2((V_o-V_s)/\mu)$ and the
$\log(r\mu)\log((V_o-V_s)/\mu)$ terms appear with the right coefficients to
form, together with the double IR logarithms that would come from the NRQCD
calculation of the Wilson loop, an IR cut-off independent quantity for the
matching coefficient (as it should).
Moreover the coefficient for the double logarithm we have obtained here (which remember came from the correction to the gluonic correlator) coincides with what one obtains expanding the renormalization group improved static potential of \cite{Pineda:2000gz}. These two facts are checks of our calculation. 

We have therefore obtained the $\als^5\log r\mu$ (and $\als^5\log^2r\mu$) terms
of the singlet static potential\footnote{\emph{Note added}: It is understood that (when renormalizing the potential) we have used the scheme where $1/\varepsilon-\gamma_E+\log\pi$ is subtracted. And this has been implemented by redefining $\mu^2\to\mu^2e^{\gamma_E}/\pi$ where applicable. Therefore one of the $\als$ in the third order correction in (\ref{pot}) is understood to be in this scheme, whereas the remaining $\als$ are understood to be in the $\overline{MS}$ scheme. Also we have chosen the scheme where only the $\log r\mu$ terms that compensate an infrared $\log((V_o-V_s)/\mu)$ are displayed in the potential.}.
\[
V_s^{(0)}(r)=(\mathrm{Eq.}\ref{pot})-
\]
\begin{equation}
-\frac{C_f\als(1/r)}{r}\left(\frac{\als(1/r)}{4\pi}\right)^4\frac{16\pi^2}{3}C_A^3\left(-\frac{11}{3}C_A+\frac{2}{3}n_f\right)\log^2r\mu-
\end{equation}
\begin{equation}
-\frac{C_f\als(1/r)}{r}\left(\frac{\als(1/r)}{4\pi}\right)^4 16\pi^2 C_A^3\left(a_1+2\gamma_E\beta_0-\frac{1}{27}\left(20
 n_f-C_A(12 \pi ^2+149)\right)\right)\log r\mu
\end{equation}

\section{Discussion}
In view of the possible future construction of an $e^+-e^-$ linear collider
(the aim of which will be the study of the possible new particles LHC will discover),
much theoretical effort is being put in the calculation of $t-\bar{t}$
production near threshold. The complete second order (N$^2$LO) corrections are
already computed. The second order renormalization group improved
expressions (N$^2$LL) are under study (several contributions are already
known) \cite{Hoang:2001mm,Pineda:2001et,Hoang:2002yy,Hoang:2003ns,Penin:2004ay}. Given the extremely good precision that such
a new machine could achieve, the third order corrections are also needed. These
third order corrections (N$^3$LO terms) are being computed at present, by
several different people. This is a gigantic project that requires the use of
state-of-the-art calculational and computational techniques \cite{Penin:2002zv,Beneke:2005hg,Penin:2005eu,Eiras:2005yt}. Once those third
order corrections are completed, the corresponding third order renormalization
group improved expressions (N$^3$LL) will also be needed (to achieve the desired
theoretical precision in the calculation). Just let us mention that the
results presented in this chapter will be a piece of these N$^3$LL computations.

%%% Local Variables: 
%%% mode: latex
%%% TeX-master: t
%%% End: 

\chapter[Two loop SCET heavy-to-light current a.d. : $n_f$ terms]{Two loop SCET heavy-to-light current anomalous dimension: $n_f$ terms}\label{chapda}

In this chapter we will calculate the two loop $n_f$ terms of the anomalous
dimension of the leading order (in $\lambda$) heavy-to-light
current in Soft-Collinear Effective Theory (SCET). The work presented in this
chapter, although mentioned in \cite{Neubert:2004dd,Neubert:2005nt}, appears
here for the first time. The calculation of the complete two loop anomalous
dimension will appear elsewhere \cite{dainprep}.

\section{Introduction}
The heavy-to-light hadronic currents $J_{\mathrm{had}}=\bar{q}\Gamma b$ ($b$
represents a heavy quark and $q$ a light quark) appearing in
operators of the weak theory at a scale $\mu\sim m_b$ can be matched into
SCET$_{\mathrm{I}}$ \cite{Bauer:2000yr}. The lowest order SCET hadronic current is not
$J_{\mathrm{had}}^{SCET}=C(\mu)\bar{\xi}\Gamma h$, but rather
\begin{equation}
J_{\mathrm{had}}^{SCET}=c_0\left(\bar{n}p,\mu\right)\bar{\xi}_{n,p}\Gamma h+c_1\left(\bar{n}p,\bar{n}q_1,\mu\right)\bar{\xi}_{n,p}\left(g\bar{n}A_{n,q_1}\right)\Gamma h+\cdots
\end{equation}
That is, an arbitrary number of $\bar{n}A_{n,q}$ gluons can be added without
suppression in the power counting. Here $\xi$ and $A$ are the fields for the
collinear quarks and gluons in the effective theory, respectively; $h$ is the
field for the heavy quark in HQET. Collinear gauge invariance relates all
these operators and organize the current into the (collinear gauge invariant)
form
\begin{equation}
J_{\mathrm{had}}^{SCET}=C_i\left(\mu,\bar{n}P\right)\bar{\chi}_{n,P}\Gamma h
\end{equation}
where
\begin{equation}
\bar{\chi}=\bar{\xi}W
\end{equation}
and $W$ is a collinear Wilson line (see section \ref{secSCET}). We can then run the Wilson coefficients down in
SCET. Note that it is enough to consider the simpler current $\bar{\xi}\Gamma
h$, because collinear gauge invariance relates them all. This was done at one
loop in \cite{Bauer:2000yr}. The result obtained there was\footnote{The
  coefficients for the different Dirac structures mix into themselves. There
  is no operator mixing at this order.}
\begin{equation}
Z=1+\frac{\alpha_sC_f}{4\pi}\left(\frac{1}{\epsilon^2}+\frac{2}{\epsilon}\log\left(\frac{\mu}{\bar{n}P}\right)+\frac{5}{2\epsilon}\right)
\end{equation}
\begin{equation}
\gamma=-\frac{\alpha_s}{4\pi}C_f\left(5+4\log\left(\frac{\mu}{\bar{n}P}\right)\right)
\end{equation}
$Z$ is the current counterterm in the effective theory, $\gamma$ is the
anomalous dimension ($P$ is the total outgoing jet momentum). Here we will calculate
the 2 loop $n_f$ corrections to this result.

\section{Calculation of the $n_f$ terms}
The effective theory diagrams that
are needed to compute the $n_f$ terms of the two loop anomalous
dimension are depicted in figure \ref{figdiagnf}. We will perform the
calculation in $d=4-2\epsilon$ dimensions. To distinguish infrared (IR)
divergences from the ultraviolet (UV) ones, we will take the collinear quark
off-shell by setting $p_{\perp}=0$ and the heavy quark with residual momentum
$\omega$. This will regulate the IR divergences of all the diagrams. We will
work in Feynman gauge. The gluon self-energy is the same as in QCD (for both
the collinear and the ultrasoft gluons), it is given in figure
\ref{figgse}. The Feynman rules which are needed to compute the diagrams are
given in figure \ref{figcurr} (for the  current), figure \ref{figHQET} (vertex and propagator rules for HQET) and appendix \ref{appFR} (vertex and propagator rules for SCET). 
\begin{figure}
\centering
\includegraphics{currlo.epsi}
\caption[Feynman rule for the $\mathcal{O}(\lambda^0)$ SCET heavy-to-light
current]{\label{figcurr} Feynman rule for the $\mathcal{O}(\lambda^0)$ SCET
  heavy-to-light current. The double line is the heavy quark. The dashed line
  is the collinear quark. Springy lines with a line inside are collinear gluons.}
\end{figure}
\begin{figure}
\centering
\includegraphics{HQET.epsi}
\caption[HQET Feynman rules]{\label{figHQET} HQET Feynman rules. The heavy quark is represented by a double line. The springy line represents the gluon.}
\end{figure}
For the ultrasoft diagrams the collinear quark propagator simplifies to
($s$ is an ultrasoft loop momentum, $p$ is the external collinear quark
momentum)
\begin{equation}
\frac{\bar{n}(p+s)}{(p+s)^2+i\eta}=\frac{\bar{n}(p)}{\bar{n}(p)n(p+s)+i\eta}=\frac{1}{ns+\frac{\bar{n}pnp}{\bar{n}p}+i\eta}=\frac{1}{ns+\frac{p^2}{\bar{n}p}+i\eta}\equiv\frac{1}{ns+\alpha+i\eta}
\end{equation}
To further simplify the integrals we will choose $\omega$ to be $\omega=\alpha/2$.

For the evaluation of the ultrasoft graphs we will just need the
integrals 
\[
\int\frac{d^ds}{(2\pi)^d}\frac{1}{ns+\alpha+i\eta}\frac{1}{vs+\omega+i\eta}\left(\frac{1}{s^2+i\eta}\right)^{\beta}=\frac{2i}{(4\pi)^{2-\epsilon}}(-1)^{2-\beta}\frac{\Gamma(2-\beta-\epsilon)\Gamma(-2+2\beta+2\epsilon)}{\Gamma(\beta)}\cdot
\]
\begin{equation}\label{ints1}
\cdot\int_0^1dy\left(2\omega y+\alpha(1-y)\right)^{2-2\beta-2\epsilon}y^{-2+\beta+\epsilon}=\frac{2i(-1)^{2-\beta}}{(4\pi)^{2-\epsilon}}\alpha^{2-2\beta-2\epsilon}\frac{\Gamma(2-\beta-\epsilon)\Gamma(-2+2\beta+2\epsilon)}{\Gamma(\beta)(-1+\beta+\epsilon)}
\end{equation}
\begin{equation}
\int\frac{d^ds}{(2\pi)^d}\frac{1}{vs+\omega+i\eta}\left(\frac{1}{s^2+i\eta}\right)^{\beta}=\frac{2i(-1)^{2-\beta}}{(4\pi)^{2-\epsilon}}(2\omega)^{3-2\beta-2\epsilon}\frac{\Gamma(2-\beta-\epsilon)\Gamma(-3+2\beta+2\epsilon)}{\Gamma(\beta)}
\end{equation}
which can be calculated with Feynman and Georgi parameterizations, we have used that $2\omega=\alpha$ in the last step of
(\ref{ints1}). Using these results we obtain
\[
\mathrm{Fig.}\;\ref{figdiagnf}a=\frac{\alpha_s^2}{(4\pi)^2}\left(\frac{p^2}{\mu\bar{n}p}\right)^{-4\epsilon}C_f\left(C_A\left(\frac{-5}{12\epsilon^3}-\frac{31}{36\epsilon^2}+\frac{1}{\epsilon}\left(-\frac{2}{27}-\frac{5\pi^2}{8}\right)\right)-\right.
\]
\begin{equation}
\left.-T_Fn_f\left(-\frac{1}{3\epsilon^3}-\frac{5}{9\epsilon^2}+\frac{1}{\epsilon}\left(\frac{8}{27}-\frac{\pi^2}{2}\right)\right)\right)
\end{equation}
\[
\mathrm{Fig.}\;\ref{figdiagnf}b=\frac{\alpha_s^2}{(4\pi)^2}\left(\frac{p^2}{\mu\bar{n}p}\right)^{-2\epsilon}C_f\left(C_A\left(\frac{5}{3\epsilon^3}+\frac{1}{\epsilon}\left(-\frac{5}{3}+\frac{25\pi^2}{36}\right)\right)+\right.
\]
\begin{equation}
\left.+T_Fn_f\left(-\frac{4}{3\epsilon^3}+\frac{1}{\epsilon}\left(\frac{4}{3}-\frac{5\pi^2}{9}\right)\right)\right)
\end{equation}
where we have redefined $\mu^2\to\mu^2e^{\gamma_E}/(4\pi)$ (from now on, we
will always use this redefinition).

The evaluation of the collinear graphs requires the integral (which again can
be calculated with Feynman and Georgi parameterizations)
\[
\int\frac{d^ds}{(2\pi)^d}\frac{\bar{n}(p-s)}{\bar{n}s}\frac{1}{(s-p)^2+i\epsilon}\left(\frac{1}{s^2+i\epsilon}\right)^{\beta}=\frac{i(-p^2)^{-\epsilon}}{(4\pi)^{2-\epsilon}}\left(p^2\right)^{-\beta+1}\cdot
\]
\begin{equation}
\cdot\frac{\Gamma(1-\epsilon)\Gamma(\beta-1+\epsilon)\Gamma(1-\beta-\epsilon)}{\Gamma(\beta)\Gamma(2-\beta-2\epsilon)}\left(\frac{1-\epsilon}{2-\beta-2\epsilon}\right)
\end{equation}
plus other integrals which do not involve the $\bar{n}$ vector and can thus be
found in standard QCD books (see for instance \cite{Pascual:1984zb}). Using
these results we obtain
\[
\mathrm{Fig.}\;\ref{figdiagnf}c=\frac{\alpha_s^2}{(4\pi)^2}\left(\frac{p^2}{\mu^2}\right)^{-2\epsilon}C_f\left(C_A\left(\frac{5}{6\epsilon^3}+\frac{23}{9\epsilon^2}-\frac{1}{\epsilon}\left(-\frac{253}{27}+\frac{5\pi^2}{36}\right)\right)+\right.
\]
\begin{equation}
\left.+T_Fn_f\left(-\frac{2}{3\epsilon^3}-\frac{16}{9\epsilon^2}+\frac{1}{\epsilon}\left(-\frac{176}{27}+\frac{\pi^2}{9}\right)\right)\right)
\end{equation}
\[
\mathrm{Fig.}\;\ref{figdiagnf}d=\frac{\alpha_s^2}{(4\pi)^2}\left(\frac{p^2}{\mu^2}\right)^{-\epsilon}C_f\left(C_A\left(-\frac{10}{3\epsilon^3}-\frac{5}{3\epsilon^2}-\frac{1}{\epsilon}\left(5-\frac{5\pi^2}{18}\right)\right)-\right.
\]
\begin{equation}
\left.-T_Fn_f\left(-\frac{4}{3\epsilon^3}-\frac{2}{3\epsilon^2}+\frac{1}{\epsilon}\left(-2+\frac{\pi^2}{9}\right)\right)\right)
\end{equation}

\begin{figure}
\centering
\includegraphics{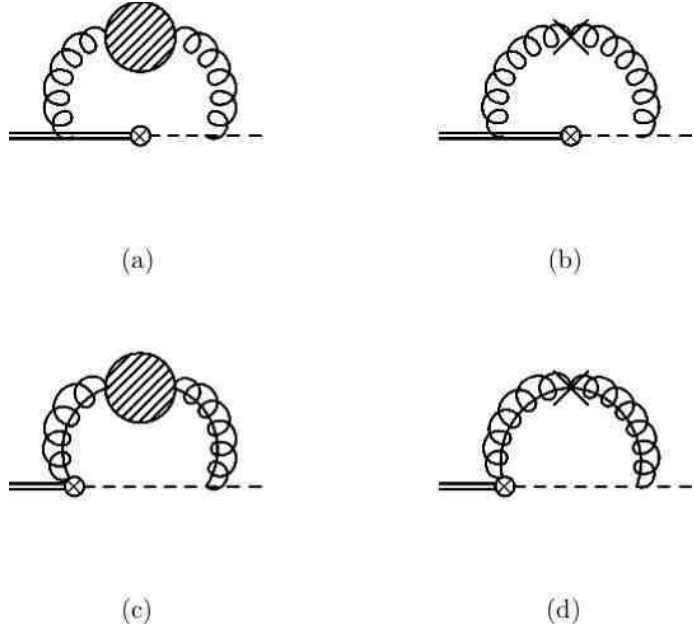}
\caption[2 loop $n_f$ diagrams]{\label{figdiagnf} Effective theory diagrams
  contributing to the 2 loop $n_f$ terms of the $\mathcal{O}(\lambda^0)$
  heavy-to-light current. The double line represents a heavy quark, the dashed
  line represents a collinear quark and the springy lines are gluons (collinear
  if they have a line inside ultrasoft if not). The circled cross is the
  insertion of the current. The shaded blob represents the
  one loop insertion of the gluon self-energy and the cross the corresponding counterterm.}
\end{figure}

\begin{figure}
\centering
\includegraphics{gse.epsi}
\caption[Gluon self-energy]{\label{figgse} Gluon self-energy graph. The gluons
in this graph can be understood either as ultrasoft or collinear.}
\end{figure}

To compute the anomalous dimension we also need the two loop correction to the
collinear and heavy quark propagators. Since the correction to the collinear
quark propagator just involves collinear particles (and not ultrasoft ones),
this is the same as in usual QCD. While the correction for the heavy quark
propagator is that of HQET. The corresponding counterterms are \cite{Egorian:1978zx,Broadhurst:1991fz}
\[
Z_{\xi}=1+\frac{\alpha_sC_f}{4\pi}\frac{1}{\epsilon}+\left(\frac{\alpha_s}{4\pi}\right)^2C_f\left(C_A\left(\frac{-1}{\epsilon^2}+\frac{34}{8\epsilon}\right)+C_f\left(\frac{-1}{2\epsilon^2}-\frac{3}{4\epsilon}\right)-T_Fn_f\frac{1}{\epsilon}\right)
\]
\begin{equation}
Z_{h}=1-\frac{\alpha_sC_f}{4\pi}\frac{2}{\epsilon}+\left(\frac{\alpha_s}{4\pi}\right)^2C_f\left(C_A\left(\frac{9}{2\epsilon^2}-\frac{19}{3\epsilon}\right)-C_f\frac{2}{\epsilon^2}-T_Fn_f\left(\frac{2}{\epsilon^2}-\frac{8}{3\epsilon}\right)\right)
\end{equation}

With this we obtain the two loop $n_f$ part of the counterterm in the
effective theory
\begin{equation}
Z^{(2loop\;n_f)}=\left(\frac{\alpha_s}{4\pi}\right)^2\frac{4}{3}C_fT_Fn_f\left(\frac{3}{4\epsilon^3}+\frac{5}{6\epsilon^2}+\frac{1}{\epsilon^2}\log\left(\frac{\mu}{\bar{n}P}\right)-\frac{125}{72}\frac{1}{\epsilon}-\frac{\pi^2}{8\epsilon}-\frac{5}{3\epsilon}\log\left(\frac{\mu}{\bar{n}P}\right)\right)
\end{equation}
where $P$ is the total outgoing jet momentum. We can then obtain the two loop
$n_f$ terms of the anomalous dimension by using the formula
\begin{equation}
\gamma=\frac{2}{Z}\frac{d}{d\mu}Z=\frac{2}{Z}\left(\left(-\epsilon\alpha_s-\beta_0\frac{\alpha_s^2}{4\pi}\right)\frac{\partial
  Z}{\partial\alpha_s}+\frac{\mu}{2}\frac{\partial
  Z}{\partial\mu}\right)
\end{equation}
The result is
\begin{equation}
\gamma^{(2loop\;n_f)}=\left(\frac{\alpha_s}{4\pi}\right)^2\frac{4T_Fn_fC_f}{3}\left(\frac{125}{18}+\frac{\pi^2}{2}+\frac{20}{3}\log\left(\frac{\mu}{\bar{n}P}\right)\right)
\end{equation}

\section{Discussion}
There is a lot of theoretical interest in obtaining the
$\mathcal{O}\left(\alpha_s^2\right)$ corrections to the $\bar{B}\to X_s\gamma$
decay rate. To measure this decay rate it is experimentally necessary to
put a cut on the energy of the observed photon. This cut introduces a new scale in the
problem and, consequently, induces a possible new source of corrections that
must be taken into account in the evaluation of the decay rate. The effects of
this scale can be systematically treated in an effective field
theory framework using SCET \cite{Neubert:2004dd}. A factorization formula for
the decay rate with the cut in the photon energy, which disentangle the
effects of all these scales, can then be derived in a systematic way. The
anomalous dimension calculated in this chapter enters in this expression. This
formula involves, among many other things, a jet function, which describes the
physics of the hadronic final state, and a soft function, which governs the
soft physics inside the $B$ meson. Expressions for the evolution equations of
these jet and shape functions can be derived. The evolution equation for the
jet function involves an anomalous dimension $\gamma^J$ which can be related
to the anomalous dimension for the jet function appearing in deep-inelastic
scattering (DIS) \cite{Neubert:2004dd,Becher:2006qw}. The two loop anomalous
dimension entering the evolution of the shape function was calculated in
\cite{Korchemsky:1992xv}\footnote{The original result in
  \cite{Korchemsky:1992xv} (from 1992) have been recently corrected in the
  revised version of the paper (from 2005). The revised result agrees with an
  independent calculation \cite{Gardi:2005yi}. The discovery of some of the
  mistakes in \cite{Korchemsky:1992xv} was triggered by the calculations described in this chapter.}. The current anomalous dimension
that we are dealing with in this chapter can be related to the anomalous
dimensions for these jet and shape functions \cite{Neubert:2004dd}. Recently
the soft and jet functions have been evaluated at two loops
\cite{Becher:2005pd,Becher:2006qw}. This calculation confirms the previous
results for the two loop anomalous dimension of the soft function
\cite{Korchemsky:1992xv,Gardi:2005yi} and provides the first direct
calculation (that is, not using the relation with DIS) for the two loop
anomalous dimension of the jet function. Therefore the last thing that remains
to be done is the direct two loop calculation of the leading order
heavy-to-light current in SCET (in this chapter we have presented the $n_f$
part of this calculation). This is important since it will ensure that we
correctly understand the renormalization properties of SCET (at two
loops). Given the peculiar structure of SCET (much different from other known effective field theories) some subtleties may arise here.

%%% Local Variables: 
%%% mode: latex
%%% TeX-master: t
%%% End: 

\chapter{Radiative heavy quarkonium decays}\label{chapraddec}

In this chapter we study the semi-inclusive radiative decay of heavy
quarkonium to light hadrons from an effective field theory point of view. As we will see below,
the correct treatment of the upper end-point region of the spectrum requires
the combined use of Non-Relativistic QCD and Soft-Collinear Effective
Theory. When these two effective theories are consistently combined a very
good description of the experimental data is achieved. The photon spectrum can
then be used to uncover some properties of the decaying quarkonia. The
contents of this chapter are basically based on work published in
\cite{GarciaiTormo:2004jw,GarciaiTormo:2004kb,GarciaiTormo:2005ch,GarciaiTormo:2005bs},
although some comparisons with new data (not available when some of the
preceding articles were published) are also presented.

\section{Introduction}

Although we will focus on the study of the semi-inclusive radiative decay, the
exclusive radiative decays have also been addressed in an effective field
theory approach. Exclusive decays will be very briefly commented in subsection \ref{subsecexcl}.

\subsection{Semi-inclusive radiative decays}
Semi-inclusive radiative decays of heavy quarkonium systems (see
\cite{Brambilla:2004wf} for a review) to light hadrons have been a subject of
investigation since the early days of QCD
\cite{Brodsky:1977du,Koller:1978qg}. In these references, the decay of the
heavy quarkonium state to $gg\gamma$ (and to $ggg$) is treated in lowest order
QCD, in analogy with the QED decays of orthopositronium to three photons. This
lowest order QCD calculation predicted a, basically, linear rise with $z$ ($z$
being the fraction of the maximum energy the photon may have) of the photon
spectrum. The angular distribution has also been studied in
\cite{Koller:1978qg}; it should be mentioned that this angular
  distribution is still assumed to be correct, and it is used for the
  comparison of the experimental results with theory and the subsequent
  extraction of QCD parameters\footnote{The recent data in
    \cite{Besson:2005jv} has allowed for the first time for a check of this
    prediction for the angular distribution. There it is found that data
    agrees adequately with \cite{Koller:1978qg}}. The upper end-point region of the photon spectrum (that was obtained by several later experiments \cite{Csorna:1985iv,Bizzeti:1991ze,Albrecht:1987hz}) appeared to be poorly described by this linear rise; a much softer spectrum, with a peak about $z\sim 0.6-0.7$, was observed instead. A subsequent resummation of the leading Sudakov ($\log (1-z)$) logarithms \cite{Photiadis:1985hn}, as well as a calculation of the leading relativistic corrections \cite{Keung:1982jb} (see also \cite{Yusuf:1996av}), although produced a softening of the spectrum in the upper end-point (namely $z\to 1$) region, were neither able to reproduce the observed spectrum. Instead, the data was well described by the model in ref. \cite{Field:1983cy}, where a parton-shower Monte Carlo technique was used to incorporate the effects of gluon radiation by the outgoing gluons in the decay.

\begin{figure}
\centering
\hspace{-2.5cm}\includegraphics[width=14cm]{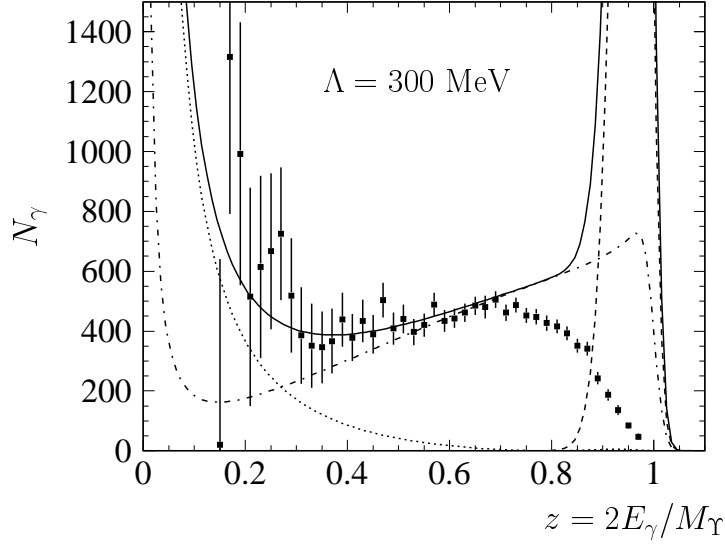}
\caption[Early attempts to modeling the color octet shape
functions]{Comparison of the spectrum obtained by the modeling
of the octet shape functions with data. The dot-dashed line is the (leading order) color singlet
contribution alone (color singlet model). The dashed line is the direct octet
contribution (where the shape functions enter). The
solid line is the total result. We can clearly see that color singlet model
alone is unable to reproduce data and that the modeling of the octet shape
function has produced a result in complete disagreement with data. $\Lambda$
is a parameter in the model for the shape functions. Plot from
\texttt{hep-ph/0010217} \cite{Wolf:2000pm}.}\label{figWolf}
\end{figure}

This led to some authors to claim that a non-vanishing gluon mass was
necessary in order to describe the data \cite{Consoli:1993ew}. With the advent
of Non-Relativistic QCD (NRQCD) \cite{Bodwin:1994jh}, these decays could be
analyzed in a framework where short distance effects, at the scale of the
heavy quark mass $m$ or larger, could be separated in a systematic manner
\cite{Maltoni:1998nh}. These short distance effects are calculated
perturbatively in $\als (m)$ and encoded in matching coefficients whereas long
distance effects are parameterized by matrix elements of local NRQCD
operators. Even within this framework, a finite gluon mass seemed to be
necessary to describe data \cite{Consoli:1997ts}. However, about the same time
it was pointed out that in the upper end-point region the NRQCD factorization
approach breaks down and shape functions, namely matrix elements of non-local
operators, rather than NRQCD matrix elements, must be introduced
\cite{Rothstein:1997ac}. Early attempts to modeling color octet shape
functions produced results in complete disagreement with data
\cite{Wolf:2000pm} (as shown in figure \ref{figWolf}), and hence later authors did not include them in their phenomenological analysis.

Notwithstanding this upper end-point region has received considerable attention lately, as it was recognized that Soft-Collinear Effective Theory (SCET) \cite{Bauer:2000ew,Bauer:2000yr} may help in organizing the calculation and in performing resummations of large (Sudakov) logs \cite{Bauer:2001rh,Fleming:2002rv,Fleming:2002sr,Fleming:2004rk}. In fact, the early resummation of Sudakov logarithms \cite{Photiadis:1985hn} has been recently corrected \cite{Fleming:2004rk} within this framework, and statements about the absence of Sudakov suppression in the color singlet channel \cite{Hautmann:2001yz} have been clarified \cite{Fleming:2002sr}. These SCET calculations will be explained in the following sections.

For the $\Upsilon (1S)$ state, the bound state dynamics is amenable of a weak
coupling analysis, at least as far as the soft scale ($mv$, $v\sim \als (mv)
\ll 1$, the typical velocity of the heavy quark in the quarkonium rest frame)
is concerned
\cite{Titard:1993nn,Titard:1994id,Titard:1994ry,Pineda:1997hz,Pineda:1998ja,Pineda:2001zq,Brambilla:2001fw,Brambilla:2001qk,Recksiegel:2002za,Kniehl:2002br,Penin:2002zv,Kniehl:2003ap,Kniehl:2002yv,Kniehl:1999mx}.
These calculations can most conveniently be done in the framework of potential
NRQCD (pNRQCD), a further effective theory where the contributions due to the
soft and ultrasoft ($\sim mv^2$) scales are factorized
\cite{Pineda:1997bj,Kniehl:1999ud,Brambilla:1999xf} (see section
\ref{secpNRQCD}). The color octet shape functions can then be calculated
combining pNRQCD and SCET. This calculation of the octet shape functions in the weak coupling regime will be the subject of subsection \ref{subseccalshpfct}.

Parallel to all that, shortly after \cite{Bodwin:1994jh}, in \cite{Catani:1994iz} it was pointed out that a parametrically leading contribution had been ignored so far. This was the contribution where the photon is emitted from the decay products of the heavy quark (light quarks), and not directly from the heavy quark itself (remember that we are always dealing with prompt photons, that is photons that do not come from hadronic decays). These type of contributions, called fragmentation contributions, completely dominate the spectrum in the lower end-point (namely $z\to 0$) region. At first it was thought that only the gluon to photon fragmentation function appeared in the process; so the radiative decays seemed a good place to determine this (yet unknown) gluon to photon fragmentation function; but a subsequent investigation \cite{Maltoni:1998nh} showed that this was not the case. When considering also the color octet contributions, the quark to photon fragmentation function also appeared, and their contributions can not be disentangled. 

When all the known contributions to the photon spectrum are taken into account
and are consistently combined, a very good description of the data is now
achieved (with no longer need for the introduction of a finite gluon
mass). This will be explained in detail in section \ref{secmerg}.

\subsection{Exclusive radiative decays}\label{subsecexcl}
Exclusive radiative decays of heavy quarkonium have been analyzed in an
effective field theory framework in \cite{Fleming:2004hc}. A combination of
NRQCD and SCET is also needed in this case. However, since in this case we are
dealing with an exclusive process, the existence (and effects) of two
different collinear scales have to be taken into account. Moreover, the fact
that the hadronic final states must be composed of collinear fields in a color
singlet configuration, causes that only color singlet contributions, and not
color octet ones, enter in the NRQCD-SCET analysis at leading order (in
contrast with the situation in the inclusive case, as we will see below). In
this case the final result of this effective theory analysis agrees with
the leading-twist order of previous known results \cite{Baier:1985wv,Ma:2001tt}.

\phantom{}

We will move now to the study of the semi-inclusive radiative decays, starting
in the next section.

\section{Effective Field Theory approach to the upper end-point region}
The NRQCD framework organizes the radiative decay in the following factorized form
\begin{equation}\label{NRQCDfact}
\frac{d\Gamma}{dz}=\sum_{i}C_i\left(M,z\right)\left<\mathrm{H}\right|\mathcal{O}_i\left|\mathrm{H}\right>
\end{equation}
where H represents a generic heavy quarkonium state and $M$ represents its mass. In that formula $C_i$ are the hard matching coefficients, which incorporates short distance effects, and the $\left<\mathrm{H}\right|\mathcal{O}_i\left|\mathrm{H}\right>$ are the NRQCD matrix elements, which parameterize the long distance effects.

However, as was already mentioned in the previous section, in the upper end-point region of the photon spectrum that standard NRQCD factorization is not applicable \cite{Rothstein:1997ac}. This is due to the fact that small scales induced by the kinematics enter the problem and have an interplay with the bound state dynamics. In order to study this region, one has to take into account collinear degrees of freedom in addition to those of NRQCD. This can be done using SCET as it has been described in \cite{Bauer:2001rh,Fleming:2002sr}. Using the SCET framework, the decay rate has been expressed in the factorized form \cite{Fleming:2002sr} 
\begin{equation}\label{SCETfact}
\frac{d\Gamma}{dz}=\sum_{\omega}H(M,\omega,\mu)\int dk^+S(k^+,\mu)\mathrm{Im}J_{\omega}(k^++M(1-z),\mu)
\end{equation}
where $H$ encodes the short distance effects, $J$ is the so called jet function, which incorporates effects at the collinear scale, and $S$ are the ultrasoft shape functions.

Using the combined counting in SCET (counting in $\lambda\sim\sqrt{\Lambda_{QCD}/(2m)}$) plus NRQCD (counting in $v$), one can see that we have color singlet and color octet operators contributing at the same order. More concretely \cite{Fleming:2004hc,Lee:2005gj}, at $\mathcal{O}(\lambda)$ in the SCET counting we have the $\phantom{}^1S_0$ and $\phantom{}^3P_J$ octet operators
\begin{equation}
\sum_iC_i^{(8,\phantom{}^1S_0)}\Gamma^i_{\alpha\mu}\chi^{\dagger}_{-\mathbf{p}}B_{\perp}^{\alpha}\psi_{\mathbf{p}}
\end{equation}  
\begin{equation}
\sum_i C^{(8,\phantom{}^3P_J)}_i \Gamma^i_{\alpha \mu \sigma \delta}  
   \chi^{\dagger}_{-{\bf p}}    B^\alpha_\perp  \Lambda \cdot \frac{{\bf p}^\sigma}{2m} 
   \Lambda \cdot {\mbox{\boldmath $\sigma$}}^\delta \psi_{\bf p}
\end{equation}
When considering also the $v$ counting (and taking into account the overlap with the $\phantom{}^3S_1$ quarkonium state) the two of these operators become $\mathcal{O}(v^5\lambda)$. Their matching coefficients start at order $\sqrt{\als(\mu_h)}$. At $\mathcal{O}(\lambda^2)$ in the SCET counting and with a matching coefficient starting at order $\als(\mu_h)$, we have the color singlet operator
\begin{equation}
\sum_i \Gamma^i_{\alpha \beta \delta \mu}
 \chi^\dagger_{-{\bf p}} \Lambda\cdot{\mbox{\boldmath $\sigma$}}^\delta \psi_{\bf p}
{\rm Tr} \big\{ B^\alpha_\perp \, 
C^{(1,{}^3S_1)}_i \, 
B^\beta_\perp \big\}
\end{equation}
when considering also the $v$ counting this operator becomes $\mathcal{O}(v^3\lambda^2)$. Then the octet-to-singlet ratio (considering $\lambda\sim v$) becomes $\frac{v}{\sqrt{\als(\mu_h)}}$; hence the color octet contributions become as important as the color singlet ones if we count $\als(\mu_h)\sim v^2\sim 1-z$.

Before going on, and explaining the calculations in the Effective Field Theory (EFT) approach, let us comment on the relation of these EFT calculations with the, phenomenologically very successful, model in \cite{Field:1983cy}. Recall that, as mentioned in the introduction of this chapter, in that reference a parton-shower Monte Carlo technique was used to incorporate the effects of gluon radiation by the outgoing gluons in the decay. When we are at the end-point region of the spectrum, the EFT analysis tells us that the decay is organized according to eq. (\ref{SCETfact}), then (as was already explained in ref. \cite{Fleming:2002sr}) the approach in \cite{Field:1983cy} is equivalent to consider that the collinear scale is non-perturbative and introduce a model with a gluon mass for the jet function $J$. When we are away of the upper end-point region, the EFT approach tells us that the decay is organized according to (\ref{NRQCDfact}). Then the effects of the gluon radiation modeled in \cite{Field:1983cy} should be incorporated in higher order NRQCD local matrix elements (this interpretation is consistent with the fact that the analysis in \cite{Field:1983cy} produced a not very large correction to the spectrum for all $z$, except in the upper end-point region, where the effect becomes $\mathcal{O}(1)$); in any case it is not justified, from an EFT point of view, why one should take into account the subset of corrections that are incorporated in \cite{Field:1983cy} and not other ones, which in principle could contribute with equal importance.

\subsection{Resummation of the color singlet contributions}\label{subsecressing}
The resummation of the Sudakov logarithms in the color singlet channel has
been performed in ref. \cite{Fleming:2002rv,Fleming:2002sr} and in
ref. \cite{Fleming:2004rk} (where the (small) effect of the operator mixing
was taken into account). It is found that all the logarithms come from
collinear physics, that is only collinear gluons appear in the diagrams for the running of the singlet operator.

The resummed rate is given by
\[
\frac{1}{\Gamma_0}\frac{d\Gamma^{e}_{CS}}{dz} = \Theta(M-2mz) \frac{8z}9 
\sum_{n \rm{\ odd}} \left\{\frac{1}{f_{5/2}^{(n)}}
\left[ \gamma_+^{(n)} r(\mu_c)^{2 \lambda^{(n)}_+ / \beta_0}  - 
\gamma_-^{(n)} r(\mu_c)^{2 \lambda^{(n)}_- / \beta_0} \right]^2+\right.
\]
\begin{equation}\label{singres}
\left.+
\frac{3 f_{3/2}^{(n)}}{8[f_{5/2}^{(n)}]^2}\frac{{\gamma^{(n)}_{gq}}^2}{\Delta^2}
\left[ r(\mu_c)^{2 \lambda^{(n)}_+ / \beta_0}  -  
r(\mu_c)^{2 \lambda^{(n)}_- / \beta_0} \right]^2\right\}
\end{equation}
where
\begin{equation}
f_{5/2}^{(n)} = \frac{n(n+1)(n+2)(n+3)}{9(n+3/2)}\quad;\quad f_{3/2}^{(n)} = \frac{(n+1)(n+2)}{n+3/2}
\end{equation}
\begin{equation}
r(\mu) = \frac{\alpha_s(\mu)}{\alpha_s(2m)}
\end{equation}
\begin{equation}
\gamma_\pm^{(n)} = \frac{\gamma_{gg}^{(n)} - \lambda^{(n)}_\mp}{\Delta}\quad;\quad \lambda^{(n)}_\pm = \frac{1}{2} \big[
\gamma^{(n)}_{gg} +  \gamma^{(n)}_{q\bar{q}} 
\pm \Delta \big]\quad;\quad\Delta = 
\sqrt{ (\gamma^{(n)}_{gg} -  \gamma^{(n)}_{q\bar{q}})^2 + 
                 4 \gamma^{(n)}_{gq} \gamma^{(n)}_{qg} }
\end{equation}
\begin{eqnarray}
\gamma^{(n)}_{q\bar{q}} &=& C_f \bigg[ \frac{1}{(n+1)(n+2)} -\frac{1}{2} - 2 \sum^{n+1}_{i=2} \frac{1}{i}
\bigg]\,\nonumber
\\
\gamma^{(n)}_{gq} &=&
\frac{1}{3}C_f \frac{n^2 + 3n +4}{(n+1)(n+2)}\,
\nonumber \\
\gamma^{(n)}_{qg} &=& 3 n_f \frac{n^2 + 3n +4}{n(n+1)(n+2)(n+3)}\,
\nonumber \\
 \gamma^{(n)}_{gg} &=&
 C_A \bigg[ \frac{2}{n(n+1)} + \frac{2}{(n+2)(n+3)}- \frac{1}{6} - 
 2 \sum^{n+1}_{i = 2} \frac{1}{i} \bigg] - \frac{1}{3}n_f
 \end{eqnarray}
This result corrects a previous calculation \cite{Photiadis:1985hn} performed
several years ago.

\subsection{Resummation of the color octet contributions}\label{subsecresoc}
The resummation of the Sudakov logarithms in the color octet channel was performed in ref. \cite{Bauer:2001rh}. In contrast with the color singlet channel, both ultrasoft and collinear gluons contribute to the running of the octet operators. The expression for the resummed Wilson coefficients is
\begin{equation}\label{ocres} 
C(x-z)=-\frac{d}{dz} \left\{
\theta(x-z) \; \frac{\exp [ \ell g_1[\alpha_s \beta_0 \ell/(4\pi)] + g_2[\alpha_s \beta_0
\ell/(4\pi)]]}{\Gamma[1-g_1[\alpha_s \beta_0
\ell/(4\pi)] - \alpha_s \beta_0 \ell/(4\pi) g_1^\prime[\alpha_s \beta_0
\ell/(4\pi)]]}\right\}
\end{equation}
where
\begin{equation}
\ell\approx-\log(x-z)
\end{equation}
\begin{eqnarray}
g_1(\chi) &=& 
-\frac{2 \Gamma^{\rm adj}_1}{\beta_0\chi}\left[(1-2\chi)\log(1-2\chi) -2(1-\chi)\log(1-\chi)\right] \nonumber \\
g_2(\chi) &=& -\frac{8 \Gamma^{\rm adj}_2}{\beta_0^2}
  \left[-\log(1-2\chi)+2\log(1-\chi)\right] \nonumber\\
 && - \frac{2\Gamma^{\rm adj}_1\beta_1}{\beta_0^3}
   \left[\log(1-2\chi)-2\log(1-\chi)
  +\frac12\log^2(1-2\chi)-\log^2(1-\chi)\right] \nonumber\\
 &&+\frac{4\gamma_1}{\beta_0} \log(1-\chi) + 
 \frac{2B_1}{\beta_0} \log(1-2\chi) \nonumber\\
&&-\frac{4\Gamma^{\rm adj}_1}{\beta_0}\log n_0
 \left[\log(1-2\chi)-\log(1-\chi)\right]
\end{eqnarray}
\begin{equation}
\Gamma^{\rm adj}_1 =  C_A \quad ;
\Gamma^{\rm adj}_2 =  
   C_A \left[ C_A \left( \frac{67}{36} - \frac{\pi^2}{12} \right) 
    - \frac{5n_f}{18} \right]  \quad; %\nonumber \\
B_1 = -C_A\quad; \gamma_1 = -\frac{\beta_0}{4}\quad; n_0=e^{-\gamma_E}
\end{equation}

\subsection{Calculation of the octet shape functions in the weak coupling regime}\label{subseccalshpfct}
In this subsection we will explain in detail the calculation of the octet shape functions. We will start by rewriting some of the expressions in the preceding sections in the pNRQCD language, which is the convenient language for the subsequent calculation of the shape functions. We begin from the formula given in \cite{Fleming:2002sr}
\begin{equation}\label{QCDexpr}
{d \Gamma\over dz}=z{M\over 16\pi^2} {\rm Im} T(z)\quad \quad T(z)=-i\int d^4 x e^{-iq\cdot x}\left< V_Q(nS)\vert T\{ J_{\mu} (x) J_{\nu} (0)\} \vert V_Q(nS)\right> \eta^{\mu\nu}_{\perp}
\end{equation}
where $J_{\mu} (x)$ is the electromagnetic current for heavy quarks in QCD and we have restricted ourselves to $\phantom{}^3S_1$ states. The formula above holds for states fulfilling relativistic normalization. In the case that non-relativistic normalization is used, as we shall do below, the right hand side of either the first or second formulas in (\ref{QCDexpr}) must be multiplied by $2M$. At the end-point region the photon momentum (in light cone coordinates, $q_\pm=q^0\pm q^ 3$) in the rest frame of the heavy quarkonium is $q=\left(q_{+},q_{-}, q_{\perp}\right)=(zM,0,0)$ with $z \sim 1$ ($M\sqrt{1-z} \ll M$). This together with the fact that the heavy quarkonium is a non-relativistic system fixes the relevant kinematic situation. It is precisely in this situation when the standard NRQCD factorization (operator product expansion) breaks down \cite{Rothstein:1997ac}. The quark (antiquark) momentum in the $Q\bar Q$ rest frame can be written as $p=(p_0, {\bf p})$, $p_0=m+l_0, {\bf p}={\bf l}$; $l_0 , {\bf l} \ll m$. Momentum conservation implies that if a few gluons are produced in the short distance annihilation process at least one of them has momentum $r=( r_{+},r_{-}, r_{\perp})$, $r_{-} \sim M/2$ ; $r_{+}, r_{\perp}\ll M$, which we will call collinear. At short distances, the emission of hard gluons is penalized by $\als (m)$ and the emission of softer ones by powers of  soft scale over $M $. Hence, the leading contribution at short distances consists of the emission of a single collinear gluon. This implies that the $Q\bar Q$ pair must be in a color octet configuration, which means that the full process will have an extra long distance suppression related to the emission of (ultra)soft gluons. The next-to-leading contribution at short distances already allows for a singlet $Q\bar Q$ configuration. Hence, the relative weight of color-singlet and color-octet configurations depends not only on $z$ but also on the bound state dynamics, and it is difficult to establish a priori. In order to do so, it is advisable to implement the constraints above by introducing suitable  EFTs. In a first stage we need NRQCD \cite{Bodwin:1994jh}, which factors out the scale $m$ in the $Q\bar Q$ system, supplemented with collinear gluons, namely gluons for which the scale $m$ has been factored out from the components $r_{+}, r_{\perp} $ (but is still active in the component $r_{-}$). For the purposes of this section it is enough to take for the Lagrangian of the collinear gluons the full QCD Lagrangian and enforce $r_{+}, r_{\perp} \ll m$ when necessary.

\subsubsection{Matching QCD to NRQCD+SCET}\label{subsubsecmQCD}
At tree level, the electromagnetic current in (\ref{QCDexpr}) can be matched to the following currents in this EFT \cite{Fleming:2002sr}

\begin{equation}\label{effcurr}
J_{\mu} (x)= e^{-i2mx_0}\left( \Gamma_{\alpha\beta i\mu}^{(1,\phantom{}^3S_1)}J^{i\alpha\beta}_{(1,\phantom{}^3S_1)} (x)
+\Gamma_{\alpha\mu}^{(8,\phantom{}^1S_0)}J^{\alpha}_{(8,\phantom{}^1S_0)} (x)+
\Gamma_{\alpha\mu ij}^{(8,\phantom{}^3P_J)}J^{\alpha ij}_{(8,\phantom{}^3P_J)} (x) + \dots
\right) + h.c.
\end{equation}

\bea
 & \Gamma_{\alpha\beta i\mu}^{(1,\phantom{}^3S_1)}={g_s^2 e e_Q\over 3 m^2}\eta^{\perp}_{\alpha\beta}\eta_{\mu i} &J^{i\alpha\beta}_{(1,\phantom{}^3S_1)} (x)= \chi^{\dagger}\bfsigma^{i} \psi Tr\{ B^{\alpha}_{\perp}  B^{\beta}_{\perp}\}(x) \cr & & \cr&\Gamma_{\alpha\mu}^{(8,\phantom{}^1S_0)}= {g_s e e_Q\over m} \epsilon_{\alpha\mu}^{\perp}&J^{\alpha}_{(8,\phantom{}^1S_0)} (x)=  \chi^{\dagger}B^{\alpha}_{\perp} \psi (x)\cr & & \cr
&\Gamma_{\alpha\mu ij}^{(8,\phantom{}^3P_J)}= {g_s e e_Q\over m^2}\left(\eta_{\alpha j}^{\perp} \eta_{\mu i}^{\perp} +\eta_{\alpha i}^{\perp} \eta_{\mu j}^{\perp}-\eta_{\alpha\mu}^{\perp} n^{j} n^{i}\right) & J^{\alpha i j}_{(8,\phantom{}^3P_J)} (x)= -i \chi^{\dagger}B^{\alpha}_{\perp} \bfnabla^{i}\bfsigma^{j}\psi (x)
\eea
where $n=(n_+,n_-,n_{\perp})=(1,0,0)$ and $\epsilon_{\alpha\mu}^{\perp}=\epsilon_{\alpha\mu\rho 0}n^\rho $. These effective currents can be identified with the leading order in $\als$ of the currents introduced in \cite{Fleming:2002sr} (which has already appeared at the beginning of this section). We use both Latin ($1$ to $3$) and Greek ($0$ to $3$) indices, $B^{\alpha}_{\perp}$ is a single collinear gluon field here, and $ee_Q$ is the charge of the heavy quark. Note, however, that in order to arrive at (\ref{effcurr}) one need not specify the scaling of collinear fields as $M(\lambda^2, 1, \lambda)$ but only the cut-offs mentioned above, namely $r_{+}, r_{\perp}\ll M$. Even though the $P$-wave octet piece appears to be $1/m$ suppressed with respect to the $S$-wave octet piece, it will eventually give rise to contributions of the same order once the bound state effects are taken into account. This is due to the fact that the $^3S_1$ initial state needs a chromomagnetic transition to become an octet $^1S_0$, which is $\als$ suppressed with respect to the chromoelectric transition required to become an octet $^3P_J$.

$T(z)$ can then be written as
\bea\label{T}
T(z)&=&H^{(1,\phantom{}^3S_1)}_{ii'\alpha\alpha'\beta\beta'}T_{(1,\phantom{}^3S_1)}^{ii'\alpha\alpha'\beta\beta'}+H^{(8,\phantom{}^1S_0)}_{\alpha\alpha^{\prime}}T_{(8,\phantom{}^1S_0)}^{\alpha\alpha^{\prime}}+
H^{(8,\phantom{}^3P_J)}_{\alpha ij\alpha^{\prime}i'j'}T_{(8,\phantom{}^3P_J)}^{\alpha ij\alpha^{\prime}i'j'}
+\cdots
\eea
where
\bea
H^{(1,\phantom{}^3S_1)}_{ii'\alpha\alpha'\beta\beta'}&=& \eta_{\perp}^{\mu\nu}\Gamma_{\alpha\beta i\mu}^{(1,\phantom{}^3S_1)}\Gamma_{\alpha'\beta'i'\nu}^{(1,\phantom{}^3S_1)}\nonumber\\
H^{(8,\phantom{}^1S_0)}_{\alpha\alpha^{\prime}}&=& \eta_{\perp}^{\mu\nu}\Gamma_{\alpha\mu}^{(8,\phantom{}^1S_0)}\Gamma_{\alpha'\nu}^{(8,\phantom{}^1S_0)}\nonumber\\
H^{(8,\phantom{}^3P_J)}_{\alpha ij\alpha^{\prime}i'j'}&=& \eta_{\perp}^{\mu\nu}\Gamma_{\alpha\mu ij}^{(8,\phantom{}^3P_J)}\Gamma_{\alpha'\nu i'j'}^{(8,\phantom{}^3P_J)}
\eea
and
\bea
T_{(1,\phantom{}^3 S_1)}^{ii'\alpha\alpha'\beta\beta'}(z)&= &-i\int d^4 x e^{-iq\cdot x-2mx_0}\left<V_Q(nS)\vert T\{ {J^{i\alpha\beta}_{(\phantom{}1,^3 S_1)} (x)}^{\dagger} J^{i'\alpha^{\prime}\beta^{\prime}}_{(1,\phantom{}^3 S_1)} (0)\}
 \vert V_Q(nS)\right> & \cr
 T_{(8,\phantom{}^1 S_0)}^{\alpha\alpha^{\prime}}(z)&= &-i\int d^4 x e^{-iq\cdot x-2mx_0}\left<V_Q(nS)\vert T\{ {J^{\alpha}_{(8,^1 S_0)} (x)}^{\dagger} J^{\alpha^{\prime}}_{(8,\phantom{}^1 S_0)} (0)\}
 \vert V_Q(nS)\right> & \cr
T_{(8,\phantom{}^3 P_J)}^{\alpha ij\alpha^{\prime}i'j'}(z)&= &-i\int d^4 x e^{-iq\cdot x-2mx_0}\left<V_Q(nS)\vert T\{ {J^{\alpha ij}_{(8,\phantom{}^3 P_J)} (x)}^{\dagger} J^{\alpha^{\prime}i'j'}_{(8,\phantom{}^3 P_J)} (0)\}
 \vert V_Q(nS)\right>
\eea
In (\ref{T}) we have not written a crossed term $(8,\phantom{}^1S_0$-$\phantom{}^3P_J)$ since it eventually vanishes at the order we will be calculating.

\subsubsection{Matching NRQCD+SCET to pNRQCD+SCET}
Thanks to the fact that in the end-point region ($M\gg M\sqrt{1-z}\gg M(1-z)$) the typical three momentum of the heavy quarks is given by
\begin{equation}\label{trim}
p\sim\sqrt{m\left(\frac{M}{2}(1-z)-E_1\right)}
\end{equation}
we can proceed one step further in the EFT hierarchy. NRQCD still contains quarks and gluons with energies $\sim m\als$, which in the kinematical situation of the end-point (where the typical three momentum is always much greater than the typical energy) can be integrated out. This leads to potential NRQCD (pNRQCD).

For the color singlet contributions we have
\begin{displaymath}
\left<V_Q(nS)\vert T\{ {J^{i\alpha\beta}_{(\phantom{}1,^3 S_1)} (x)}^{\dagger} J^{i'\alpha^{\prime}\beta^{\prime}}_{(1,\phantom{}^3 S_1)} (0)\}
 \vert V_Q(nS)\right> \longrightarrow
\end{displaymath}
\begin{equation}\label{hardcoll}
\longrightarrow 2N_c
{S_{V}^{i}}^{\dagger}({\bf x}, {\bf 0}, x_0)S_{V}^{i'}({\bf 0}, {\bf 0}, 0)
\left< \textrm{{\small VAC}}\vert Tr\{ B^{\alpha}_{\perp}  B^{\beta}_{\perp}\}(x)
Tr\{ B^{\alpha^{\prime}}_{\perp}  B^{\beta^{\prime}}_{\perp}\}(0)\vert \textrm{{\small VAC}}\right>
\end{equation}
The calculation of the vacuum correlator for collinear gluons above has been carried out in \cite{Fleming:2002sr}, and the final result, which is obtained by sandwiching (\ref{hardcoll}) between the quarkonium states, reduces to the one put forward in that reference.

For the color octet currents, the leading contribution arises from a tree level matching of the currents (\ref{effcurr}),

\bea\label{effcurr2}
J^{\alpha}_{(8,\phantom{}^1S_0)} (x) &\longrightarrow &\sqrt{2T_F} O_{P}^a ({\bf x}, {\bf 0}, x_0)B^{a \alpha}_{\perp}(x)\cr & & \cr
J^{\alpha ij}_{(8,\phantom{}^3P_J)} (x)& \longrightarrow &\sqrt{2T_F}\left.\left( i\bfnabla^i_{\bf y} O_{V}^{a j}({\bf x}, {\bf y}, x_0)\right)\right\vert_{{\bf y}={\bf 0}} B^{a \alpha}_{\perp}(x)
\eea
$S_{V}^{i}$, $O_{V}^{a i}$ and $O_{P}^a$ are the projection of the singlet and octet wave function fields introduced in \cite{Pineda:1997bj,Brambilla:1999xf} to their vector and pseudoscalar components, namely $S=(S_{P}+S_{V}^{i}\sigma^i)/\sqrt{2}$ and $O^a =(O_{P}^{a}+O_{V}^{a i}\sigma^i)/\sqrt{2}$. $T_F=1/2$ and $N_c=3$ is the number of colors.

\subsubsection{Calculation in pNRQCD+SCET}\label{subsubseccalc}
\begin{figure}
\centering
\includegraphics{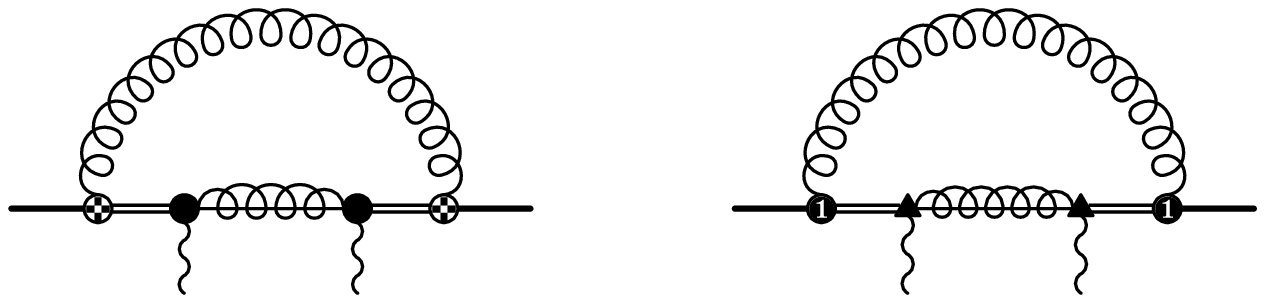}
\caption[Color octet contributions]{\label{figdos}Color octet contributions. $\bullet$ represents the color octet S-wave current, $\blacktriangle$ represents the color octet P-wave current. The notation for the other vertices is that of ref. \cite{Brambilla:2002nu}, namely \ding{60}:= ${ig c_F \over \sqrt{N_c T_F}} { \left ( {\bfsigma}_1 - {\bfsigma}_2 \right ) \over 2m } \, {\rm Tr} \left [ T^b {\bf B} \right ]$ and \ding{182}:= ${ig\over \sqrt{N_c T_F}} {\bf x} \, {\rm Tr} \left [ T^b {\bf E} \right ]$. The solid line represents the singlet field, the double line represents the octet field and the gluon with a line inside represents a collinear gluon.}
\end{figure}

We shall then calculate the contributions of the color octet currents in pNRQCD coupled to collinear gluons. They are depicted in figure \ref{figdos}. For the contribution of the $P$-wave current, it is enough to have the pNRQCD Lagrangian at leading (non-trivial) order in the multipole expansion given in \cite{Pineda:1997bj,Brambilla:1999xf}. For the contribution of the $S$-wave current, one needs a $1/m$ chromomagnetic term given in \cite{Brambilla:2002nu}.

Let us consider the contribution of the $S$-wave color octet current in some detail. We have, from the first diagram of fig. \ref{figdos},
\begin{displaymath}
 T_{(8,\phantom{}^1S_0)}^{\alpha\alpha^{\prime}}(z)=
-i\eta^{\alpha\alpha^{\prime}}_{\perp}(4\pi){32\over 3}T_F^2 \left({ c_F\over 2m}\right)^2 \als (\mu_u)C_f\int d^3 {\bf x}  \int d^3 {\bf x}^{\prime}
\psi^{\ast}_{n0}( {\bf x}^{\prime}) \psi_{n0}( {\bf x})\int\!\!\! {d^4 k\over (2\pi)^4} {{\bf k}^2\over k^2+i\epsilon}\times
\end{displaymath}
\begin{equation}\label{eqonas}
\times\left(1\over -k_0+E_n-h_o+i\epsilon\right)_{{\bf x}^{\prime},{\bf 0}}{1\over (M(1-z)-k_{+})M-{\bf k}_{\perp}^2+i\epsilon}
 \left(1\over -k_0+E_n-h_o+i\epsilon\right)_{{\bf 0}, {\bf x}}
\end{equation}
where we have used the Coulomb gauge (both for ultrasoft and collinear gluons). $E_n < 0$ is the binding energy ($M=2m+E_n$) of the heavy quarkonium, $\psi_{n0}( {\bf x})$ its wave function, and $h_o$ the color-octet Hamiltonian at leading order, which contains the kinetic term and a repulsive Coulomb potential \cite{Pineda:1997bj,Brambilla:1999xf}. $c_F$ is the hard matching coefficient of the chromomagnetic interaction in NRQCD \cite{Bodwin:1994jh}, which will eventually be taken to $1$. We have also enforced that $k$ is ultrasoft by neglecting it in front of $M$ in the collinear gluon propagator. We shall evaluate (\ref{eqonas}) in light cone coordinates.  If we carry out first the integration over $k_{-}$, only the pole $k_{-}={\bf k}_{\perp}^2/k_{+}$ contributes. Then the only remaining singularities in the integrand are in the collinear gluon propagator. Hence, the absorptive piece can only come from its pole $M^2(1-z)-M k_{+}= {\bf k}_{\perp}^2$. If $k_{+} \siml M(1-z)$, then  ${\bf k}_{\perp}^2 \sim M^2(1-z)$ which implies $k_{-}\sim M$. This contradicts the assumption that $k$ is ultrasoft. Hence, ${\bf k}_{\perp}^2$ must be expanded in the collinear gluon propagator. We then have
\begin{displaymath}
{\rm Im}\left(T_{(8,\phantom{}^1S_0)}^{\alpha\alpha^{\prime}}(z)\right)=-\eta^{\alpha\alpha^{\prime}}_{\perp}(4\pi){32\over 3}T_F^2 \left({ c_F\over 2m}\right)^2 \als(\mu_u)C_f\times
\end{displaymath}
\begin{displaymath}
\times\int d^3 {\bf x} \int d^3 {\bf x}^{\prime}
\psi^{\ast}_{n0}( {\bf x}^{\prime}) \psi_{n0}( {\bf x}){1\over 8\pi M}\int_0^{\infty} dk_+ \delta \left(M(1-z) -k_+\right)\times
\end{displaymath}
\begin{equation}\label{imts}
\times\int_0^{\infty} dx \left( \left\{ \delta ({\bf \hat x}) , {h_o-E_n \over h_o -E_n +{k_+ \over 2}+x} \right\} - {h_o-E_n \over h_o -E_n +{k_+ \over 2}+x}\delta ({\bf \hat x}){h_o-E_n \over h_o -E_n +{k_+ \over 2}+x} \right)_{{\bf x},{\bf x}^{\prime}}
\end{equation}
where we have introduced the change of variables $\vert {\bf k}_{\perp}\vert=\sqrt{2k_+x}$. Restricting ourselves to the ground state ($n=1$) and using the techniques of reference \cite{Beneke:1999gq} we obtain
\begin{displaymath}
{\rm Im}\left(T_{(8,\phantom{}^1S_0)}^{\alpha\alpha^{\prime}}(z)\right)=-\eta^{\alpha\alpha^{\prime}}_{\perp}{16\over 3}T_F^2 \left({ c_F\over 2m}\right)^2 \als(\mu_u)C_f{1\over M}\int_0^{\infty} dk_+ \delta (M(1-z) -k_+)\times
\end{displaymath}
\begin{displaymath}
\times\int_0^{\infty} dx \left( 2 \psi_{10}( {\bf 0})I_{S}({k_+\over 2} +x)- I_{S}^2({k_+\over 2} +x) \right)
\end{displaymath}
\begin{displaymath}
I_{S}({k_+\over 2} +x):=\int d^3 {\bf x} \psi_{10}( {\bf x})\left( {h_o-E_1 \over h_o -E_1 +{k_+ \over 2}+x}\right)_{{\bf x},{\bf 0}}=
\end{displaymath}
\begin{equation}
=m\sqrt{\gamma\over \pi}{\als N_c \over 2}{1\over 1-z'}\left( 1-{2z'\over 1+z'} \;\phantom{}_2F_1\left(-\frac{\lambda}{z'},1,1-\frac{\lambda}{z'},\frac{1-z'}{1+z'}\right)\right)
\end{equation}
where
\begin{equation}
\gamma=\frac{mC_f\als}{2}\quad z'=\frac{\kappa}{\gamma}\quad-\frac{\kappa^2}{m}=E_1-\frac{k_+}{2}-x\quad\lambda=-\frac{1}{2N_cC_f}\quad E_1=-{\gamma^2\over m}
\end{equation}
This result can be recast in the factorized form given in \cite{Fleming:2002sr} (equation \ref{SCETfact}).
\begin{equation}\label{ImTS}
{\rm Im}\left(T_{(8,\phantom{}^1S_0)}^{\alpha \alpha^{\prime}}(z)\right)=-\eta^{\alpha \alpha^{\prime}}_{\perp}\int dl_+ S_{S}(l_+)
{\rm Im} J_M (l_+ - M(1-z))
\end{equation}
\begin{equation}
{\rm Im} J_M(l_+ - M(1-z))=
T_F^2\left(N_c^2-1\right){2\pi\over M}\delta (M(1-z) -l_+)
\end{equation}
\begin{equation}\label{shpfctS}
S_{S}(l_+)={4\als (\mu_u)\over 3 \pi N_c} \left({ c_F\over 2m}\right)^2
\int_0^{\infty} dx \left( 2 \psi_{10}( {\bf 0})I_{S}({l_+\over 2} +x)- I_{S}^2({l_+\over 2} +x) \right)
\end{equation}
We have thus obtained the $S$-wave color octet shape function $S_{S}(l_+)$.
Analogously, for the $P$-wave color octet shape functions, we obtain from the second diagram of fig. \ref{figdos}
\begin{equation}\label{ImTP}
{\rm Im}\left(T_{(8,\phantom{}^3P_J)}^{\alpha i j \alpha^{\prime} i' j'}(z)\right)\!\!\!=\!\!-\eta_{\perp}^{\alpha \alpha^{\prime}}\delta^{j j'}\!\!\!\int\!\!\! dl_+\!\! \left( \delta_{\perp}^{i i'}S_{P1}(l_+)+\left(n^i n^{i'}\!\!-{1\over 2}\delta_{\perp}^{i i'}\right)S_{P2}(l_+)\right)
{\rm Im} J_M(l_+ - M(1-z))
\end{equation}
\begin{equation}\label{shpfctP1}
S_{P1}(l_+):=
{\als (\mu_u)\over 6 \pi N_c}
\int_0^{\infty}\!\!\!dx\left( 2\psi_{10}( {\bf 0})I_P(\frac{l_+}{2}+x)-I_P^2(\frac{l_+}{2}+x) \right)
\end{equation}
\begin{equation}\label{shpfctP2}
S_{P2}(l_+):=
{\als (\mu_u)\over 6 \pi N_c}
\int_0^{\infty}\!\!\!dx \frac{8l_+x}{\left(l_++2x\right)^2}\left(
\psi^2_{10}( {\bf 0})-2\psi_{10}( {\bf 0})I_P(\frac{l_+}{2}+x)+I_P^2(\frac{l_+}{2}+x)\right)
\end{equation}
where
\begin{displaymath}
I_{P}({k_+\over 2} +x):=-\frac{1}{3}\int d^3 {\bf x} {\bf x}^i \psi_{10}( {\bf x})\left( {h_o-E_1 \over h_o -E_1 +{k_+ \over 2}+x} \bfnabla^i \right)_{{\bf x},{\bf 0}}=
\end{displaymath}
\begin{displaymath}
=\sqrt{\frac{\gamma^3}{\pi}}
{8\over 3}\left(
2-\lambda \right)\!\!\frac{1}{4(1+z')^3}\Bigg( 2(1+z')(2+z')+(5+3z')(-1+\lambda)+2(-1+\lambda)^2+
\end{displaymath}
\begin{equation}
\left.+\frac{1}{(1-z')^2}\left(4z'(1+z')(z'^2-\lambda^2)\left(\!\!-1+\frac{\lambda(1-z')}{(1+z')(z'-\lambda)}+\phantom{}_2F_1\left(-\frac{\lambda}{z'},1,1-\frac{\lambda}{z'},\frac{1-z'}{1+z'}\right)\right)\right)\right)
\end{equation}
Note that two shape functions are necessary for the $P$-wave case. 

The shape functions (\ref{shpfctS}), (\ref{shpfctP1}) and (\ref{shpfctP2}) are ultraviolet (UV) divergent and require regularization and renormalization. In order to regulate them at this order it is enough to calculate the ultrasoft (US) loop (the integral over $k$ in (\ref{eqonas})) in $D$-dimensions, leaving the bound state dynamics in $3$ space dimensions ($D=4-2\varepsilon$). In fact, the expressions (\ref{shpfctS}) and (\ref{shpfctP1}) implicitly assume that dimensional regularization (DR) is used, otherwise linearly divergent terms proportional to $\psi^2_{10}({\bf 0})$ would appear (which make (\ref{shpfctS}) and (\ref{shpfctP1}) formally positive definite quantities). This procedure, to use DR for the US loop only, was the one initially employed in \cite{GarciaiTormo:2004jw}. There the following steps were performed:
\begin{itemize}
\item In order to isolate the $1/\varepsilon$ poles,  $I_S$ and $I_P$ were expanded up to $\mathcal{O}(1/{z'}^2)$ (the expansion of these functions up to $\mathcal{O}(1/{z'}^4)$ can be found in equations (\ref{IStaylor}) and (\ref{IPtaylor})). 
\item The result was subtracted and added to the integrand of (\ref{shpfctS})-(\ref{shpfctP1}) (for (\ref{shpfctP2}) this is not necessary since the only divergent piece is independent of $I_P$). The subtracted part makes the shape functions finite. The added part contains linear and logarithmic UV divergencies. 
\item The remaining divergent integrals were dimensionally regularized by making the substitution $dx\rightarrow dx (x/ \mu)^{-\varepsilon}$. That produced the $1/\varepsilon$ poles displayed in formulas (16) of ref. \cite{GarciaiTormo:2004jw}, which were eventually subtracted (linear divergences are set to zero as usual in DR).
\end{itemize}
That last point was motivated by the fact that $x \sim {\bf k}_\perp^2$ (${\bf k}_\perp$ being the transverse momentum of the US gluon) but differs from a standard $MS$ scheme. 

As was already mentioned in \cite{GarciaiTormo:2004jw}, this regularization and renormalization scheme is not the standard one in pNRQCD calculations. Later, in \cite{GarciaiTormo:2005ch}, a regularization-renormalization procedure closer to the standard one in pNRQCD was used; which is the one we will use here. That latter procedure consists in regularizing both the US loop and the potential loops (entering in the bound state dynamics) in DR; then US divergences are identified by taking the limit $D\rightarrow 4$ in the US loops while leaving the potential loops in $D$ dimensions \cite{Pineda:1997ie}; potential divergencies are identified by taken $D\rightarrow 4$ in the potential loops once the US divergencies have been subtracted. It turns out that all divergencies in $S_{P2}$ are US and all divergencies in $S_S$ are potential. $S_{P1}$ contains both US and potential divergencies. The potential divergences related with the bound state dynamics can be isolated using the methods of ref. \cite{Czarnecki:1999mw}. Following this procedure we obtain the following expressions for the singular pieces
\begin{displaymath}
\left.S_{S}(k_+)\right\vert_{\varepsilon\rightarrow 0}\simeq
{c_F^2\als (\mu_u )\gamma^3 C_f^2 \als^2 (\mu_p)\over 3\pi^2 N_c m}(1-\lambda)\left( -2+\lambda (2\ln2 + 1)\right)\cdot
\end{displaymath}
\begin{equation}\label{singularSS}
\cdot\left(\frac{1}{\varepsilon}+\ln\left(\frac{\mu_{pc}^2}{m\left(\frac{k_+}{2}+\frac{\gamma^2}{m}\right)}\right)+\cdots\right)
\end{equation}
\[
\left.S_{P1}(k_+)\right\vert_{\varepsilon\rightarrow 0} \simeq
{\als (\mu_u )\gamma^3 m C_f^2 \als^2 (\mu_p)\over 9\pi^2 N_c }
\left(\!\!-\frac{31}{6}+\lambda (4\ln2+\frac{19}{6})-\lambda^2 (2\ln 2 + {1\over 6})\right)\cdot
\]
\begin{equation}\label{singularSP1}
\cdot\left(\frac{1}{2\varepsilon}+\ln\left(\frac{\mu_{pc}^2}{m\left(\frac{k_+}{2}+\frac{\gamma^2}{m}\right)}\right)
+\cdots\right)+{2\als (\mu_u )\gamma^5  \over 9\pi^2 N_c m}\left(-{1\over\varepsilon}-\ln\left(\frac{\mu_c^2}{k_+^2}\right)+\cdots\right)
\end{equation}
\begin{equation}\label{singularP2}
\left.
S_{P2}(k_+)\right\vert_{\varepsilon\rightarrow 0}\simeq
{\als (\mu_u ) k_+ \gamma^3\over 3\pi^2 N_c}
\left(\frac{1}{\varepsilon}
+\ln\left(\frac{\mu_c^2}{k_+^2}\right)+\cdots
\right)
\end{equation}
For simplicity, we have set $D=4$ everywhere except in the momentum integrals. $\mu_p$, according to (\ref{trim}), is given by $\mu_p=\sqrt{m(M(1-z)/2-E_1)}$. $\mu_c$ and $\mu_{pc}$ are the subtraction points of the US and potential divergencies respectively. For comparison, let us mention that when we set $\mu_c=M\sqrt{1-z}$ and $\mu_{pc}=\sqrt{m\mu_c}$, as we will do, we obtain exactly the same result as in the procedure used in ref. \cite{GarciaiTormo:2004jw} for what the potential divergences is concerned\footnote{We assume that the correlation of scales advocated in \cite{Luke:1999kz} (see \cite{Pineda:2001et} for the implementation in our framework) must also be taken into account here.}; for the US divergences there is a factor $\ln\left(\frac{\mu_c}{2k_+}\right)$ of difference with respect to that former scheme.

\begin{figure}
\centering
\includegraphics{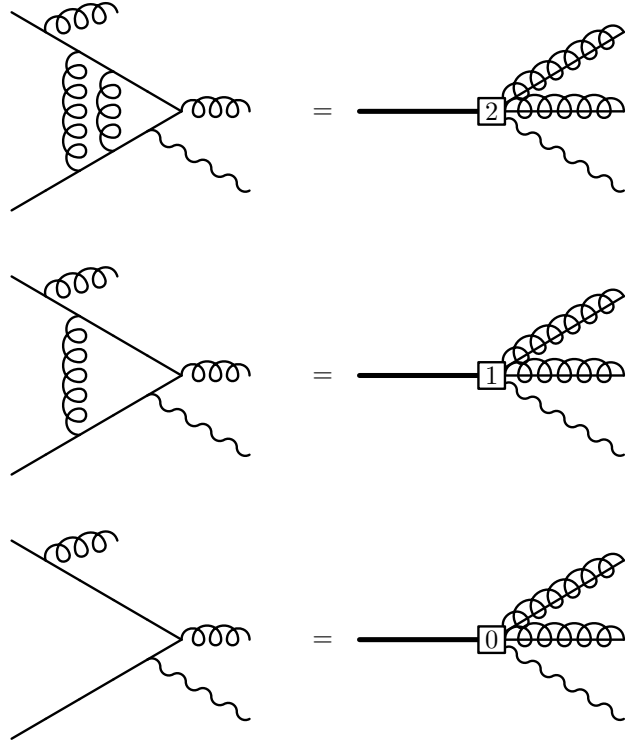}
\caption{Relevant diagrams in the matching calculation QCD $\rightarrow$ pNRQCD+SCET.}\label{figmatch}
\end{figure}

\begin{figure}
\centering
\includegraphics{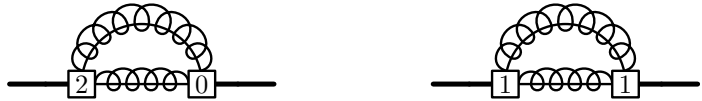}
\caption{Diagrams which induce an IR scale dependence which cancels against the UV one of the octet shape functions.
}\label{figdiv}
\end{figure}

The renormalization of that expressions is not straightforward. We will assume that suitable operators exists which may absorb the $1/\varepsilon$ poles so that an $MS$-like scheme makes sense to define the above expressions, and discuss in the following the origin of such operators. In order to understand the scale dependence of (\ref{singularSS})-(\ref{singularP2}) it is  important to notice that it appears because the term ${\bf k}_{\perp}^2$ in the collinear gluon propagator is neglected in (\ref{eqonas}). It should then cancel with an IR divergence induced by keeping the term ${\bf k}_{\perp}^2$, which implies assuming a size $M^2(1-z)$ for it, and expanding the ultrasoft scales accordingly. We have checked that it does. However, this contribution cannot be computed reliably within pNRQCD (neither within NRQCD) because it implies that the $k_{-}$ component of the ultrasoft gluon is of order $M$, and hence it becomes collinear. A reliable calculation involves (at least) two steps within the EFT strategy. The first one is the matching calculation of the singlet electromagnetic current at higher orders both in $\als$ and in $({\bf k}_{\perp}/M)^2$ and $k_+/M$. The second is a one loop calculation with collinear gluons involving the higher order singlet currents. %Notice, before going on, that all divergences (and logarithms) of $S_S (l_+)$ and $S_{P1} (l_+)$ in (\ref{singular}) are sensible to the bound state dynamics, whereas those for $S_{P2} (l_+)$ are not.
Figure \ref{figmatch} shows the relevant diagrams which contribute to the IR
behavior we are eventually looking for. We need NNLO in $\als$, but only LO in
the $({\bf k}_{\perp}/M)^2$ and $k_+/M$ expansion. These diagrams are IR
finite, but they induce, in the second step, the IR behavior which matches the
UV of (\ref{singularSS})-(\ref{singularP2}). The second step amounts to
                                %integrating out the scale $M\sqrt{1-z}$ by 
calculating  the loops with collinear gluons and expanding smaller scales in the integrand. We have displayed in fig. \ref{figdiv} the two diagrams which provide the aforementioned IR divergences. For the UV divergences that do not depend on the bound state dynamics, we need the matching at LO in $\als$ (last diagram in Fig. \ref{figmatch}) but NLO in $k_+/M$ and $({\bf k}_{\perp}/M)^2$. 

%ei ei ei aquest ultim paragraf s'ha de repassar, sobretot el final

The above means that the scale dependence of the leading order contributions of the color-octet currents is of the same order as the NNLO contributions in $\als$ of the color-singlet current, a calculation which is not available. One might, alternatively, attempt to resum logs and use the NLO calculation \cite{Kramer:1999bf} as the boundary condition. This log resummation is non-trivial. One must take into account the correlation of scales inherent to the non-relativistic system \cite{Luke:1999kz}, which in the framework of pNRQCD has been implemented in \cite{Pineda:2001ra,Pineda:2002bv,Pineda:2001et}, and combine it with the resummation of Sudakov logs in the framework of SCET \cite{Bauer:2000ew,Bauer:2001rh,Fleming:2002rv,Fleming:2002sr} (see also \cite{Hautmann:2001yz}). Correlations within the various scales of SCET may start playing a role here as well  \cite{Manohar:2000mx}. In any case, it should be clear that by only resumming Sudakov logs, as it has been done so far \cite{Bauer:2001rh}, one does not resum all the logs arising in the color octet contributions of heavy quarkonium, at least in the weak coupling regime.

Keeping this in mind, we can proceed and write the renormalized expressions for the shape functions. These renormalized expressions, in an $MS$ scheme, read
\[
S_{S}^{MS}(k_+)={4\als (\mu_u)\over 3 \pi N_c} \left({ c_F\over 2m}\right)^2
\Bigg\{2 \psi_{10}( {\bf 0})\!\!\left(m\sqrt{\frac{\gamma}{\pi}}\frac{\alpha_sN_c}{2}\right)\Bigg(\!\!\int_0^{\infty}  \!\!\!\left(\widetilde{I}_{S}({k_+\over 2}
    +x)-\frac{1}{z'}-\left(-1+2\lambda\ln
      2\right)\cdot\right.
\]
\[
\left.\left.\cdot\frac{1}{z'^2}\right)dx-2\frac{\gamma}{\sqrt{m}}\sqrt{\frac{k_+}{2}+\frac{\gamma^2}{m}}%+\frac{\gamma^2}{m}\left(-1+2\lambda\log 2\right)\log\left(\frac{\mu_{pc}^2}{m\left(\frac{k_+}{2}+\frac{\gamma^2}{m}\right)}\right)
\right)-\left(m\sqrt{\frac{\gamma}{\pi}}\frac{\alpha_sN_c}{2}\right)^2\left(\int_0^{\infty}\!\!\!\left(\widetilde{I}_{S}^2({k_+\over 2}
        +x)-\frac{1}{z'^2}\right)dx%+\frac{\gamma^2}{m}\log\left(\frac{\mu_{pc}^2}{m\left(\frac{k_+}{2}+\frac{\gamma^2}{m}\right)}\right)
\right)\Bigg\}+
\]
\begin{equation}
+{c_F^2\als (\mu_u )\gamma^3 C_f^2 \als^2 (\mu_p)\over 3\pi^2 N_c m}(1-\lambda)\left( -2+\lambda (2\ln2 + 1)\right)\left(\ln\left(\frac{\mu_{pc}^2}{m\left(\frac{k_+}{2}+\frac{\gamma^2}{m}\right)}\right)\right)
\end{equation}
\[
S_{P1}^{MS}(k_+)={\als (\mu_u)\over 6 \pi N_c} 
\Bigg\{2 \psi_{10}( {\bf 0})\left(\sqrt{\frac{\gamma^3}{\pi}}\frac{8}{3}(2-\lambda)\right)\!\!\Bigg(\int_0^{\infty}  \!\!\!\left(\widetilde{I}_{P}({k_+\over 2}
    +x)\!-\!\frac{1}{2z'}\!-\!\left(-\frac{3}{4}+\lambda\ln
      2-\frac{\lambda}{4}\right)\cdot\right.
\]
\[
\left.\left.\cdot\frac{1}{z'^2}\right)dx-\frac{\gamma}{\sqrt{m}}\sqrt{\frac{k_+}{2}+\frac{\gamma^2}{m}}%+\frac{\gamma^2}{m}\left(-\frac{3}{4}+\lambda\log
\right)-\left(\sqrt{\frac{\gamma^3}{\pi}}\frac{8}{3}(2-\lambda)\right)^2\left(\int_0^{\infty}\!\!\!\left(\widetilde{I}_{P}^2({k_+\over 2}
        +x)-\frac{1}{4z'^2}\right)dx\right)\Bigg\}
  + 
\]
\[
+{\als (\mu_u )\gamma^3 m C_f^2 \als^2 (\mu_p)\over 9\pi^2 N_c }
\left(\!\!-\frac{31}{6}+\lambda (4\ln2+\frac{19}{6})-\lambda^2 (2\ln 2 + {1\over 6})\right)\ln\!\!\left(\frac{\mu_{pc}^2}{m\left(\frac{k_+}{2}+\frac{\gamma^2}{m}\right)}\right)
+
\]
\begin{equation}
+ {2\als (\mu_u )\gamma^5  \over 9\pi^2 N_c m}\left(-\ln\left(\frac{\mu_c^2}{k_+^2}\right)\right)%+\frac{\gamma^2}{4m}\log\left(\frac{\mu_{pc}^2}{m\left(\frac{k_+}{2}+\frac{\gamma^2}{m}\right)}\right)
\end{equation}
\[
S_{P2}^{MS}(k_+)={\als (\mu_u)\over 6 \pi N_c} 
\Bigg\{\psi^2_{10}( {\bf 0})k_+\left(-2+2\ln\left(\frac{\mu_c^2}{k_+^2}\right)\right)+ 
\]
\begin{equation}
+\int_0^{\infty}\!\!\!dx \frac{8k_+x}{\left(k_++2x\right)^2}\left(-2\psi_{10}( {\bf 0})I_P(\frac{k_+}{2}+x)+I_P^2(\frac{k_+}{2}+x)\right)\Bigg\}
\end{equation}
where
\begin{eqnarray}
\widetilde{I}_S(\frac{k_+}{2}+x) & := &
\left(m\sqrt{\frac{\gamma}{\pi}}\frac{\alpha_sN_c}{2}\right)^{-1}{I}_S(\frac{k_+}{2}+x)\nonumber\\
\widetilde{I}_P(\frac{k_+}{2}+x) & := & \left(\sqrt{\frac{\gamma^3}{\pi}}\frac{8}{3}(2-\lambda)\right)^{-1}{I}_P(\frac{k_+}{2}+x)
\end{eqnarray}
In ref. \cite{GarciaiTormo:2004kb} an additional subtraction related to linear divergencies was made. This subtraction was necessary in order to merge smoothly with the results in the central region. We will also need this subtraction when discussing the merging at LO in the following sections. We use
\begin{displaymath}
\int_0^{\infty}\!\!\!\!\!\!dx\,
\frac{1}{z'}
\longrightarrow
-2\frac{\gamma}{\sqrt{m}}\left[\sqrt{\frac{k_+}{2}+\frac{\gamma^2}{m}}-\sqrt%
{\frac{k_+}{2}}\right]
\end{displaymath}
which differ from the $MS$ scheme by the subtraction of the second term in the square brackets. In that other scheme (\emph{sub}) the expressions for the shape functions read
\bea
S_{S}^{sub}(k_+)&=& S_{S}^{MS}(k_+)+{4\als (\mu_u)\over 3 \pi N_c} \left({ c_F\over 2m}\right)^2
2 \psi_{10}( {\bf 0})\left(m\sqrt{\frac{\gamma}{\pi}}\frac{\alpha_sN_c}{2}\right)2\frac{\gamma}{\sqrt{m}}\sqrt{\frac{k_+}{2}}\\
S_{P1}^{sub}(k_+)&=& S_{P1}^{MS}(k_+)+{\als (\mu_u)\over 6 \pi N_c} 
2 \psi_{10}( {\bf 0})\left(\sqrt{\frac{\gamma^3}{\pi}}\frac{8}{3}(2-\lambda)\right)\frac{\gamma}{\sqrt{m}}\sqrt{\frac{k_+}{2}}\\
S_{P2}^{sub}(k_+)&=&S_{P2}^{MS}(k_+)
\eea

The validity of the formulas for the shape functions is limited by the perturbative treatment of the US gluons. The typical momentum of these gluons in light cone coordinates turns out to be:
\begin{equation}
(k_+, k_{\perp}, k_-)=\left(M(1-z),\sqrt{2M(1-z)\left(\frac{M(1-z)}{2}-E_1\right)},M(1-z)-2E_1\right)
\end{equation}
Note that the typical $k_{\perp}$ is not fixed by the bound state dynamics only but by a combination of the latter and the end-point kinematics. Hence, the calculation is reliable provided that $k_{\perp} \gtrsim 1 GeV$, which for the $\Upsilon(1S)$ system means $z<0.92$.

\subsection{Comparison with experiment}
We apply here the results in this section to the $\Upsilon (1S)$ system. There is good evidence that the  $\Upsilon (1S)$ state can be understood as a weak coupling (positronium like) bound state \cite{Titard:1993nn,Titard:1994id,Titard:1994ry,Pineda:1997hz,Pineda:1998ja,Pineda:2001zq,Brambilla:2001qk,Recksiegel:2002za,Kniehl:2002br,Penin:2002zv,Kniehl:2003ap}. Hence, ignoring $\mathcal{O}\left(\Lambda_{\rm QCD}\right)$ in the shape functions, as we have done, should be a reasonable approximation.

We will denote the contribution in the upper end-point region by $\frac{d\Gamma^e}{dz}$. It is given by
\begin{equation}\label{endp}
\frac{d\Gamma^e}{dz}=\frac{d\Gamma^{e}_{CS}}{dz}+\frac{d\Gamma^{e}_{CO}}{dz}
\end{equation}
where $CS$ and $CO$ stand for color singlet and color octet contributions respectively. The color singlet contribution is the expression with the Sudakov resummed coefficient (\ref{singres}). The color octet contribution is given by
\begin{equation}
\frac{d\Gamma_{CO}^{e}}{dz}=\alpha_s\left(\mu_u\right)\alpha_s\left(\mu_h\right)\left(\frac{16M\alpha}{81m^4%
}\right)\int_z^{\frac{M}{2m}}\!\!\! C(x-z) S_{S+P}(x)dx
\end{equation}
where $\mu_u$ is the US scale, that arises from the couplings of the US gluons (see below for the expression we use). $C(x-z)$ contains the Sudakov resummations explained in \ref{subsecresoc}\footnote{These matching coefficients, provided in reference \cite{Bauer:2001rh}, become imaginary for extremely small values of $z-1$, a region where our results do not hold anyway. We have just cut-off this region in the convolutions.}. The (tree level) matching coefficients (up to a global factor) and the various shape functions are encoded in $S_{S+P}(x)$,
\[
S_{S+P}(z):=z\left(-\left(\frac{4\alpha_s\left(\mu_u\right)}{3\pi
N_c}\left(\frac{c_F}{2m}\right)^2\right)^{-1}\!\!\!\!\!S_S(M(1-z))-\right.
\]
\begin{equation}\label{sp}
\left.-\left(\frac{\alpha_s\left(\mu_u\right)}{6\pi
N_c}\right)^{-1}\left(3S_{P1}(M(1-z))+S_{P2}(M(1-z))\right)\right)
\end{equation}
The shape functions $S_S$, $S_{P1}$ and $S_{P2}$ may become $S_S^{MS}$, $S_{P1}^{MS}$ and $S_{P2}^{MS}$ or  $S_S^{sub}$, $S_{P1}^{sub}$ and $S_{P2}^{sub}$ depending on the subtraction scheme employed. We will use the following values of the masses for the plots: $m=4.81$ GeV and $M=9.46$ GeV. The hard scale $\mu_h$ is set to $\mu_h=M$. The soft scale $\mu_s = m C_f\alpha_s$ is to be used for the $\alpha_s$ participating in the bound state dynamics, we have $\alpha_s(\mu_s)=0.28$. The ultrasoft scale $\mu_u$ is set to $\mu_u=\sqrt{2M(1-z)\left(\frac{M}{2}(1-z)-E_1\right)}$, as discussed in the previous subsection. We have used the \verb|Mathematica| package \verb|RunDec| \cite{Chetyrkin:2000yt} to obtain the (one loop) values of $\alpha_s$ at the different scales. %ei ei ei mirar que no falti definir el valor d'alguna quantitat

\begin{figure}
\centering
\includegraphics[width=14.5cm]{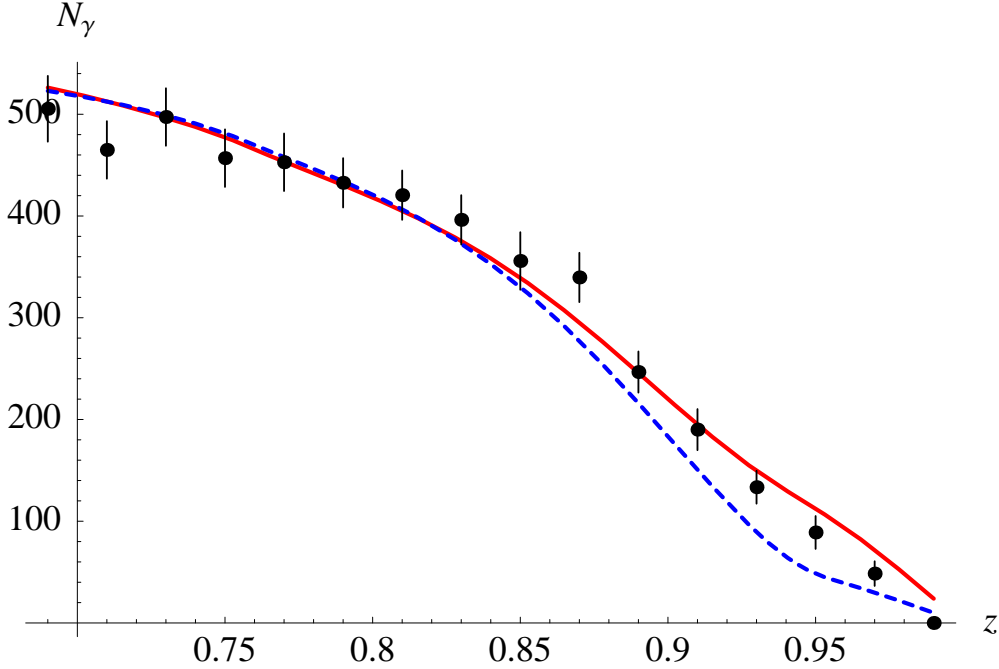}
\caption[End-point contribution of the spectrum]{End-point contribution of the spectrum, 
$d\Gamma^e/dz$, with
  the shape functions renormalized in an $MS$ scheme (blue dashed line) and
  in the $sub$ scheme 
(red solid line). The points are the CLEO data
  \cite{Nemati:1996xy}}\label{figepsubst}
\end{figure}

In figure \ref{figepsubst} we plot the end-point contribution (\ref{endp})
with the shape functions renormalized in an $MS$ scheme (blue dashed line) and
in the $sub$ scheme (red solid line), together with the experimental data
\cite{Nemati:1996xy} (we have convoluted the theoretical curves with the
experimental efficiency, the overall normalization of each curve is taken as a
free parameter). We see that a very good description of data is achieved and
that both schemes are equally good for the description of the shape of the
experimental data in the end-point region. This nice agreement with data is an encouraging result. But still remains to be
seen if it is possible to combine these results, for the end-point region, with the ones for
the central region (where the NRQCD formalism is expected to work). This will
be the subject of section \ref{secmerg}.

\subsection{Calculation of $\Upsilon (1S)$ NRQCD color octet matrix elements}
\begin{figure}
\centering
\includegraphics{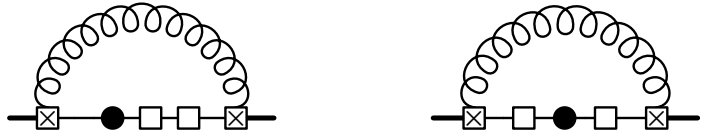}
\caption[Diagrams which require a ${\cal P}_1 (^3 S_1)$ operator for renormalization]{\label{figz2}Diagrams which require a ${\cal P}_1 (^3 S_1)$ operator for renormalization. The solid circle stands for either the
$\mathcal{O}_8 (^1 S_0)$ or $\mathcal{O}_8 (^3 P_J)$ operator, the crossed box for either the chromomagnetic (\ding{60} ) or chromoelectric
(\ding{182} ) interaction
in fig. \ref{figdos}, the empty box for the octet Coulomb potential, and the thin solid lines for free $Q\bar Q$ propagators.}
\end{figure}

\begin{figure}
\centering
\includegraphics{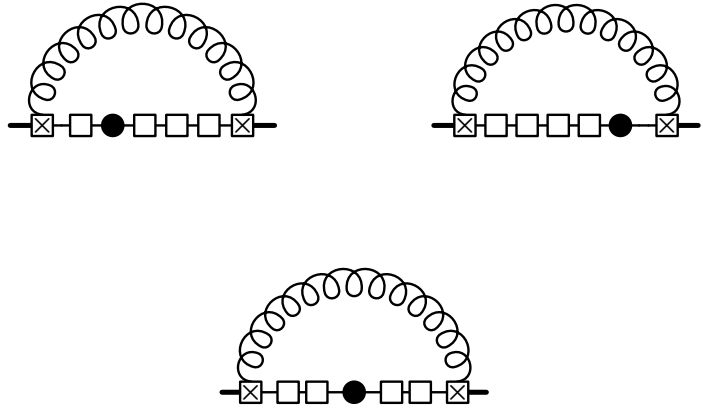}
\caption[Diagrams which require a $\mathcal{O}_1 (^3 S_1)$ operator for renormalization]{\label{figz4}Diagrams which require a $\mathcal{O}_1 (^3 S_1)$ operator for renormalization. Symbols are as in fig. \ref{figz2}. }
\end{figure}

The calculation of the shape functions can be easily taken over to provide a calculation of $\left<\Upsilon(1S)\right\vert $ $\left. \mathcal{O}_8(\phantom{}^1 S_0)\vert\Upsilon (1S) \right> $ and $\left<\Upsilon (1S)\vert \mathcal{O}_8(\phantom{}^3 P_J)\vert\Upsilon (1S)\right>$ assuming that $m\als^2\gg \Lambda_{QCD}$ is a reasonable approximation for this system. Indeed, we only have to drop the delta function (which requires a further integration over $k_{+}$) and arrange for the suitable factors in (\ref{ImTS}) and (\ref{ImTP}). We obtain
\begin{displaymath}
\left< \Upsilon (1S) \vert \mathcal{O}_8 (^1 S_0) \vert \Upsilon (1S) \right> =-2T_F^2 (N_c^2-1)
\int_0^{\infty}dk_+S_{S}(k_+)
\end{displaymath}
\begin{equation}
\left< \Upsilon (1S) \vert \mathcal{O}_8 (^3 P_J) \vert \Upsilon (1S) \right> =-{4(2J+1)T_F^2 (N_c^2-1)\over 3}
\int_0^{\infty}dk_+S_{P1}(k_+)
\end{equation}
where we have used
\be
\int_0^{\infty}dk_+S_{P2}(k_+)={2\over 3} \int_0^{\infty}dk_+S_{P1}(k_+)
\ee
The expressions above contain UV divergences which may be regulated by calculating the ultrasoft loop in $D$ dimensions. These divergences can be traced back to the diagrams in fig. \ref{figz2} and fig. \ref{figz4}. Indeed, if we expand $I_{S}$ and $I_{P}$ for $z^{\prime}$ large, we obtain
\begin{displaymath}
I_S\sim m\sqrt{\frac{\gamma}{\pi}}\frac{\als N_c}{2}\left\{\frac{1}{{z^{\prime}}}+\frac{1}{{z^{\prime}}^2}(-1+2\lambda\ln2)+\frac{1}{{z^{\prime}}^3}\left(1-2\lambda+\frac{\lambda^2\pi^2}{6}\right)+\right.
\end{displaymath}
\begin{equation}\label{IStaylor}
\left.+\frac{1}{{z^{\prime}}^4}\left(-1+\lambda(2\ln2+1)+\lambda^2(-4\ln2)+\frac{3}{2}\zeta(3)\lambda^3\right) +{\cal O}(\frac{1}{{z^{\prime}}^5})\right\}
\end{equation}
\begin{displaymath}
I_P\sim \sqrt{\frac{\gamma^3}{\pi}}
{8\over 3}(2-\lambda)
\left\{\frac{1}{2{z^{\prime}}}+\left(-\frac{3}{4}+\lambda\left(-\frac{1}{4}+\ln2\right)\right)\frac{1}{{z^{\prime}}^2}+\right.
\end{displaymath}
\begin{displaymath}
\left.+\left(1-\lambda+\frac{1}{12}(-6+\pi^2)\lambda^2\right)\frac{1}{{z^{\prime}}^3}+\frac{1}{4}\left(-5+\lambda+\lambda^2(2-8\ln2)+8\lambda\ln2+\right.\right.
\end{displaymath}
\begin{equation}\label{IPtaylor}
\left.\left.+\lambda^3\left(-4\ln2+3\zeta(3)\right)\right)\frac{1}{{z^{\prime}}^4} +{\cal O}(\frac{1}{{z^{\prime}}^5})\right\}
\end{equation}
It is easy to see that only powers of $1/z^{\prime}$ up to order four may give rise to divergences. Moreover, each power of $1/z^{\prime}$ corresponds to one Coulomb exchange. Taking into account the result of the following integral,
\be
\int_0^\infty dk_+\int_0^\infty dx(2k_+x)^{-\varepsilon}\frac{1}{{z'}^{\alpha}}=2^{1-2\varepsilon}\left(\frac{\gamma^2}{m}\right)^{2-2\varepsilon}\frac{\Gamma^2(1-\varepsilon)}{\Gamma\left(\frac{\alpha}{2}\right)}\Gamma\left(\frac{\alpha}{2}+2\varepsilon-2\right)
\ee
we see that only the $1/{z^{\prime}}^2$ and $1/{z^{\prime}}^4$ terms produce divergences. The former correspond to diagrams in fig. \ref{figz2} and the latter to fig. \ref{figz4}, which can be renormalized by the operators ${\cal P}_1 (^3 S_1)$ and $\mathcal{O}_1(^3S_1)$ respectively. It is again important to notice that these divergences are a combined effect of the ultrasoft loop and quantum mechanics perturbation theory ({\it potential} loops \cite{Beneke:1997zp}) and hence it may not be clear at first sight if they must be understood as ultrasoft (producing $\log \mu_u$ in the notation of refs. \cite{Pineda:2001ra,Pineda:2002bv,Pineda:2001et}) or potential (producing $\log \mu_p$ in the notation of refs. \cite{Pineda:2001ra,Pineda:2002bv,Pineda:2001et}). In any case, the logarithms they produce depend on the regularization and renormalization scheme used for both ultrasoft and potential loops. Remember that the scheme we use here is not the standard one in pNRQCD \cite{Pineda:1997ie,Pineda:1998kn,Pineda:2001ra,Pineda:2002bv,Kniehl:2002br,Penin:2002zv}. In the standard scheme the ultrasoft divergences (anomalous dimensions) are identified by dimensionally regulating both ultrasoft and potential loops and subsequently taking $D\rightarrow 4$ in the ultrasoft loop divergences only. If we did this in the present calculation we would obtain no ultrasoft divergence. Hence, in the standard scheme there would be contributions to the potential anomalous dimensions only. The singular pieces in our scheme are displayed below
\begin{displaymath}
\left.\left< \Upsilon (1S) \vert \mathcal{O}_8 (^1 S_0) \vert \Upsilon (1S) \right>\right\vert_{\varepsilon\rightarrow 0} \simeq -{1\over \varepsilon}\left({2\gamma^2\over \mu m}\right)^{-2\varepsilon}{1\over 24} c_{F}^2 N_c \als (\mu_u)\left(C_f\als (\mu_s) \right)^4 {\gamma^3\over \pi^2}
\Bigg( 2+
\end{displaymath}
\begin{displaymath}
\left.
+\lambda \left[ -7-4\log 2\right]
+\lambda^2\left[ 4+8\log 2 + 4\log^2 2 + {\pi^2\over 3}\right]
 + \lambda^3\left[ -4\log^2 2-{\pi^2\over 3}-{3\over 2}\zeta (3)\right]\right)
\end{displaymath}
\begin{displaymath}
\left.\left< \Upsilon (1S) \vert \mathcal{O}_8 (^3 P_J) \vert \Upsilon (1S) \right>\right\vert_{\varepsilon\rightarrow 0} \simeq -(2J+1){1\over \varepsilon}\left({2\gamma^2\over \mu m}\right)^{-2\varepsilon}{4\over 27} C_f\als (\mu_u)(C_f\als (\mu_s) )^2 {\gamma^5\over \pi^2} \times
\end{displaymath}
\begin{displaymath}
\times (2-\lambda )\left( -4 +
+\lambda\left[ {47\over 12}+5\log 2\right]+\lambda^2\left[ {5\over 6}-{2\pi^2\over 9}-{8\over 3}\log 2-{8\over 3}\log^2 2 \right]+\right.
\end{displaymath}
\begin{equation}
\left.+\lambda^3\left[ -{7\over 12}+{\pi^2\over 9}-{5\over 3}\log 2 + {4\over 3}\log^2 2 +{3\over 4}\zeta (3)\right] \right)
\end{equation}
With these expressions we obtain the following estimates for the value of the matrix elements
\begin{eqnarray}
\left.\left< \Upsilon (1S) \vert \mathcal{O}_8 (^1 S_0) \vert \Upsilon (1S) \right>\right|_{\mu=M} & \sim & 0.004\,GeV^3\label{estomeS}\\
\left.\left< \Upsilon (1S) \vert \mathcal{O}_8 (^3 P_0) \vert \Upsilon (1S) \right>\right|_{\mu=M} & \sim & 0.08\,GeV^5\label{estomeP}
\end{eqnarray}
remember that the above numbers are obtained in an $MS$ scheme from dimensionally regularized US loops only. The value we assign to the $S$-wave matrix element is compatible with the recent 
(quenched) lattice determination (hybrid algorithm) \cite{Bodwin:2005gg}.

\section{Merging the various contributions to the spectrum}\label{secmerg}
Now, with the, for a long time elusive, end-point region of the spectrum well described, it is the time to put together all the contributions to the spectrum and see if a good description of data is achieved. As was already explained, the contributions to the spectrum can be split into direct ($^{dir}$) and fragmentation ($^{frag}$)
\begin{equation}
\frac{d\Gamma}{dz}=\frac{d\Gamma^{dir}}{dz}+\frac{d\Gamma^{frag}}{dz}
\end{equation}
The fragmentation contributions are those in which the photon is emitted from
the decay products of the heavy quark (final state light quarks), these
contributions where first taken into account in \cite{Catani:1994iz} and
further studied in \cite{Maltoni:1998nh}; while the direct contributions are
those in which the photon is emitted directly from the heavy quark
itself. Although this direct-fragmentation splitting is correct at the order
we are working it should be refined at higher orders. We discuss each of these
contributions in turn, in the two following subsections.

\subsection[Direct contributions: merging the central and upper end-point regions]{Direct contributions}
The approximations required to calculate the QCD formula (\ref{QCDexpr}) are different in the lower end-point region ($z\rightarrow 0$), in the central region ($z\sim 0.5$) and in the upper end-point region ($z\rightarrow 1$). 

In the lower end-point region the emitted low energy photon can only produce transitions within the non-relativistic bound state without destroying it. Hence the direct low energy photon emission takes place in two steps: (i) the photon is emitted (dominantly by dipole electric and magnetic transitions) and (ii) the remaining (off-shell) bound state is annihilated into light hadrons. This lower end-point contribution goes to zero, for $z\to 0$, as $z^ 3$, while the leading order NRQCD result goes to zero as $z$ (see \cite{Manohar:2003xv,Voloshin:2003hh} for a recent analysis of this lower end-point region in QED). As was already mentioned, at some point the direct photon emission is overtaken by the fragmentation contributions \cite{Catani:1994iz,Maltoni:1998nh}. In practice this happens about $z\sim 0.4$, namely much before than the $z^3$ behavior of the low energy direct photon emission can be observed, and hence we shall neglect the latter in the following. 

For $z$ away from the lower and upper end-points ($0$ and $1$ respectively), no further scale is introduced beyond those inherent of the non-relativistic system. The integration of the scale $m$ in the time ordered product of currents in (\ref{QCDexpr}) leads to local NRQCD operators with matching coefficients which depend on $m$ and $z$. We will summarize here the known results for the central region (we denote the direct contributions in the central region by $\Gamma_c$). At leading order one obtains
\begin{equation}
\label{LOrate}
\frac1{\Gamma_0} \frac{d\Gamma^c
_{\rm LO}}{dz} =  
\frac{2-z}{z} + \frac{z(1-z)}{(2-z)^2} + 2\frac{1-z}{z^2}\ln(1-z) - 2\frac{(1-z)^2}{(2-z)^3} \ln(1-z),
\end{equation}
where\footnote{\emph{Note added:} The $\left(\frac{2m}{M}\right)^2$ factor was missing in formula (4) of \cite{GarciaiTormo:2005ch} (and in a previous version of the thesis). This was just a typo, not affecting any of the subsequent results. We thank A.Vairo for help in identifying it.}
\begin{equation}
\Gamma_0 = \frac{32}{27}\alpha\alpha_s^2e_Q^2
\frac{\langle  V_Q (nS)\vert {\cal O}_1(^3S_1)\vert V_Q (nS)\rangle}{m^2}\left(\frac{2m}{M}\right)^2,
\label{gamma0}
\end{equation}
The $\alpha_s$ correction to this rate was calculated numerically in ref.~\cite{Kramer:1999bf}. The expression corresponding to (\ref{gamma0}) in pNRQCD is obtained at lowest order in any of the possible regimes by just making the substitution 
\begin{eqnarray}
\label{singletWF}
\langle  V_Q (nS) \vert {\cal O}_1(^3S_1) \vert  V_Q (nS) \rangle &=&
2 N_c |\psi_{n0}({\bf 0})|^2,
\end{eqnarray}
where $\psi_{n0}({\bf 0})$ is the wave function at the origin. The final result coincides with the one of the early QCD calculations \cite{Brodsky:1977du,Koller:1978qg}. We will take the Coulomb form $\psi_{10}({\bf 0})=\gamma^3/\pi$ for the LO analysis of $\Upsilon (1S)$. 

The NLO contribution in the original NRQCD counting \cite{Bodwin:1994jh} is $v^2$ suppressed with respect to (\ref{LOrate}). It reads

\begin{equation}
\label{RelCo}
\frac{d\Gamma^c_{\rm NLO}}{dz}=C_{\mathbf{1}}'\left(\phantom{}^3S_1\right)\frac{\langle  V_Q (nS)\vert {\cal
P}_1(^3S_1)\vert V_Q (nS)\rangle}{m^4}
\end{equation}

In the original NRQCD counting or in the weak coupling regime of pNRQCD the new matrix element above can be written in terms of the original one \cite{Gremm:1997dq}\footnote{In the strong coupling regime of pNRQCD an additional contribution appears \cite{Brambilla:2002nu}}

\begin{equation}
\frac{\langle  V_Q (nS)\vert {\cal P}_1(^3S_1)\vert V_Q (nS)\rangle}{m^4}=\left(\frac{M-2m}{m}\right)\frac{\langle  V_Q
(nS)\vert {\cal O}_1(^3S_1)\vert V_Q (nS)\rangle}{m^2}\left(1+\mathcal{O}\left(v^2\right)\right)
\end{equation}

The matching coefficient can be extracted from an early calculation \cite{Keung:1982jb} (see also \cite{Yusuf:1996av}). It reads

\begin{equation}
C_{\mathbf{1}}'\left(\phantom{}^3S_1\right)=-\frac{16}{27}\alpha\alpha_s^2e_Q^2\left(F_B(z)+\frac{1}{2} F_W(z)\right)
\end{equation}
where ($\xi=1-z$)

\[
F_B(z)\!=\!\frac{2\!-\!16\xi+10\xi^2-48\xi^3 -10\xi^4+64\xi^5-2\xi^6 +(1-3\xi+14\xi^2-106\xi^3+17\xi^4 -51\xi^5)\ln
\xi}{2\,(1-\xi)^3 
(1+\xi)^4}
\]
\[
F_W(z)=\frac{-26+14\xi-210\xi^2+134\xi^3+274\xi^4-150\xi^5-38\xi^6+2\xi^7}{3(1-\xi)^3 (1+\xi)^5}-
\]
\begin{equation}
-\frac{(27+50\xi+257\xi^2-292\xi^3+205\xi^4-78\xi^5-41\xi^6)\ln \xi}{3(1-\xi)^3 (1+\xi)^5}
\end{equation}

The contributions of color octet operators start at order $v^4$. Furthermore, away of the upper end-point region, the lowest order color octet contribution identically vanishes \cite{Maltoni:1998nh}. Hence there is no $1/\als$ enhancement in the central region and we can safely neglect these contributions here.

If we use the counting $\als (\mu_h)\sim v^2$, $\als\left(\mu_s\right)\sim v$ (remember that $\mu_h\sim m$ and $\mu_s\sim mv$ are the hard and the soft scales respectively) for the $\Upsilon (1S)$, the complete result up to NLO (including $v^2$ suppressed contributions) can be written as
\be
\frac{d\Gamma^c}{dz}=\frac{d\Gamma^{c}_{LO}}{dz}+\frac{d\Gamma^{c}_{NLO}}{dz}+\frac{d\Gamma^{c}_{LO,\als}}{dz}
\label{central}
\ee
The first term consist of the expression (\ref{LOrate}) with the Coulomb wave function at the origin (\ref{singletWF}) including corrections up to $\mathcal{O}\left[\left(\als\left(\mu_s\right)\right)^2\right]$ \cite{Melnikov:1998ug,Penin:1998kx}, the second term is given in (\ref{RelCo}), and the third term consists of the radiative $\mathcal{O}\left(\als (\mu_h)\right)$ corrections to (\ref{LOrate}) which have been calculated numerically in \cite{Kramer:1999bf}. Let us mention at this point that the $\mathcal{O}\left[\left(\als\left(\mu_s\right)\right)^2\right]$ corrections to the wave function at the origin turn out to be as large as the leading order term. This will be important for the final comparison with data at the end of the section. Note that the standard NRQCD counting we use does not coincide with the usual counting of pNRQCD in weak coupling calculations, where $\als (\mu_h) \sim \als (\mu_s) \sim \als (mv^2)$. The latter is necessary in order to get factorization scale independent results beyond NNLO for the spectrum and beyond NLO for creation and annihilation currents. However, for the $\Upsilon (1S)$ system (and the remaining heavy quarkonium states) the ultrasoft scale $mv^2$ is rather low, which suggests that perturbation theory should better be avoided at this scale \cite{Pineda:2001zq}. This leads us to standard NRQCD counting. The factorization scale dependences that this counting induces can in principle be avoided using renormalization group techniques \cite{Luke:1999kz,Pineda:2001ra,Pineda:2001et,Pineda:2002bv,Hoang:2002yy}. In practice, however, only partial NNLL results exists for the creation and annihilation currents \cite{Hoang:2003ns,Penin:2004ay} (see \cite{Pineda:2003be} for the complete NLL results), which would fix the scale dependence of the  wave function at the origin at ${\cal O} (\als^2 (mv))$. We will not use them and will just set the factorization scale to $m$. 

The upper end-point region of the spectrum has been discussed in great detail in the previous section. As we have seen there, different approximations, with respect to the ones for the central region, are needed here. It is not, by any way, obvious how the results for the central and for the upper end-point regions must be combined in order to get a reliable description of the whole spectrum. When the results of the central region are used in the upper end-point region, one misses certain Sudakov and Coulomb resummations which are necessary because the softer scales $M\sqrt{1-z}$ and $M(1-z)$ become relevant. Conversely, when results for the upper end-point region are used in the central region, one misses non-trivial functions of $z$, which are approximated by their end-point ($z\sim 1$) behavior. We will explain, in the remaining of this subsection, how to merge these two contributions.

\subsubsection{Merging the central and upper end-point regions}
One way to proceed with the merging is the following. If we assume that the expressions for the end-point contain the ones of the central region up to a certain order in $(1-z)$, we could just subtract from the expressions in the central region the behavior when $z\rightarrow 1$ at the desired order and add the expressions in the end-point region. Indeed, when $z\rightarrow 1$ this procedure would improve on the central region expressions up to a given order in $(1-z)$, and when $z$ belongs to the central region, they would reduce to the central region expressions up to higher orders in $\als$. This method was used in ref. \cite{Fleming:2002sr} and in ref. \cite{GarciaiTormo:2004kb}. In ref. \cite{Fleming:2002sr} only color singlet contributions were considered and the end-point expressions trivially contained the central region expressions in the limit $z\rightarrow 1$. In ref. \cite{GarciaiTormo:2004kb} color octet contributions were included, which contain terms proportional to $(1-z)$. Hence, the following formula was used
\begin{equation}
\frac{1}{\Gamma_0}\frac{d\Gamma^{dir}}{dz}=\frac{1}{\Gamma_0}\frac{d\Gamma_{LO}^{c}}{dz}+\left(\frac{1}{\Gamma_0%
}\frac{d\Gamma_{CS}^{e}}{dz}-z\right)+\left(\frac{1}{\Gamma_0}\frac{d\Gamma_{CO}^{e}}{dz}-z\left(4+2\log \left(1-z\right)
\right) (1-z)\right)
\label{mergingLO}
\end{equation}
Even though a remarkable description of data was achieved with this formula (upon using a suitable subtraction scheme, the {\it sub} scheme described in the previous subsection), this method suffers from the following shortcoming. The hypothesis that the expressions for the end-point contain the ones for the central region up to a given order in $(1-z)$ is in general not fulfilled. As we will see below, typically, they only contain part of the expressions for the central region. This is due to the fact that some $\als (\mu_h)$ in the central region may soften as $\als (M(1-z))$, others as $\als (M\sqrt{1-z})$ and others may stay at $\als (\mu_h)$ when approaching the end-point region. In a LO approximation at the end-point region, only the terms with the $\als$ at low scales would be kept and the rest neglected, producing the above mentioned mismatch. We shall not pursue this procedure any further.

Let us look for an alternative. Recall first that the expressions we have obtained for the upper end-point region are non-trivial functions of $M(1-z)$, $M\sqrt{1-z}$, $m\als (mv)$ and $m\als^2 (mv)$, which involve $\als$ at all these scales. They take into account both Sudakov and Coulomb resummations. When $z$ approaches the central region, we can expand them in $\als (M\sqrt{1-z})$, $\als (M(1-z))$ and the ratio $m\als (mv)/M\sqrt{1-z}$. They should reduce to the form of the expressions for the central region, since we are just undoing the Sudakov and (part of) the Coulomb resummations. Indeed, we obtain
\bea%\label{expand} 
\frac{d\Gamma^{e}_{CS}}{dz} &\longrightarrow  \displaystyle{\left.\frac{d\Gamma^{e}_{CS}}{dz}\right\vert_c}= & \Gamma_0z\left(1+\frac{\als}{6\pi}\left(C_A\left(2\pi^2-17\right)+2n_f\right)\log (1-z) + \mathcal{O}(\als^2) \right)\label{expands}\\
\frac{d\Gamma^{e}_{CO}}{dz} &\longrightarrow  \displaystyle{\left.\frac{d\Gamma^{e}_{CO}}{dz}\right\vert_c} = & -z\als^2\left(\frac{16M\alpha}{81m^4}\right)2\left|\psi_{10}\left(\bf{0}\right)\right|^2
\bigg( m\als\sqrt{1-z} A
+\nonumber\\
 & & \left.+
%2\left|\psi\left(\bf{0}\right)\right|^2
M(1-z)\left(-1+\log\left(\frac{\mu_c^2}{M^2(1-z)^2}\right)\right)+\right.\nonumber\\
 & & \left.+
M\frac{\als}{2\pi}\left(-2C_A\left(\frac{1}{2}(1-z)\log^2(1-z)\left[\log\left(\frac{\mu_c^2}{M^2(1-z)^2}\right)-1\right]+\right.\right.\right.\nonumber\\
 & & \left.+\int_z^1\!\!\!dx\frac{\log(x-z)}{x-z}f(x,z)\right)-\nonumber\\
 & & \left.\left.-\left(\frac{23}{6}C_A-\frac{n_f}{3}\right)\left((1-z)\log(1-z)\left[\log\left(\frac{\mu_c^2}{M^2(1-z)^2}\right)-1\right]+\right.\right.\right.\nonumber\\
 & & \left.\left.+\int_z^1\!\!\!dx\frac{1}{x-z}f(x,z)\right)\right)-\nonumber\\
 & & \left.-
\frac{\gamma^2}{m}2\left(\log\left(\frac{\mu_c^2}{M^2(1-z)^2}\right)+1\right)+ \mathcal{O}\left( m\als^2, \als \frac{\gamma^2}{m}, \frac{\gamma^4}{m^3}\right) \right)\label{expando}
\eea
where
\begin{equation}
f(x,z)=\left(1-x\right)\log\left(\frac{\mu_c^2}{M^2(1-x)^2}\right)-(1-z)\log \left(\frac{\mu_c^2}{M^2(1-z)^2}\right)+x-z
\end{equation}
$A=-N_c-136C_f(2-\lambda)/9$ (in an $MS$ scheme; it becomes $A=-64C_f(2-\lambda)/9$ in the 
$sub$ scheme). In the next paragraph we explain how to obtain these expressions for the shape functions in the central region.

First consider the $S$-wave octet shape function
\begin{equation}\label{Swave}
I_{S}({k_+\over 2} +x):=\int d^3 {\bf x} \psi_{10}( {\bf x})\left( 1-{\frac{k_+}{2}+x \over h_o -E_1 +{k_+ \over 2}+x}\right)_{{\bf x},{\bf 0}}
\end{equation}
$h_o={\bf p}^2/m+V_o$, $V_o=\als/(2N_c\vert {\bf r}\vert )$.
When $z$ approaches the central region, $k_+\sim M(1-z) \gg -E_1$ and the larger three momentum scale is $M\sqrt{1-z}\gg\gamma$, the typical three momentum in the bound state. Therefore we can treat the Coulomb potential in (\ref{Swave}) as a perturbation when it is dominated by this scale. It is convenient to proceed in two steps. First we write $h_o=h_s + (V_o-V_s)$, where $h_s={\bf p}^2/m+V_s$, $V_s=-\als C_f/\vert {\bf r}\vert$,  and expand $V_o-V_s$. This allows to set $h_s -E_1$ to zero in the left-most propagator and makes explicit the cancellation between the first term in the series and the first term in (\ref{Swave}). It also makes explicit that the leading term will be proportional to $\als (M\sqrt{1-z})$. Second, we expand $V_s$ in $h_s={\bf p}^2/m+V_s$. In addition, since $M\sqrt{1-z} \gg\gamma$, the wave function can be expanded about the origin. Only the first term in both expansion is relevant in order to get (\ref{expando}). Consider next the $P$-wave shape functions
\begin{equation}\label{Pwave}
I_{P}({k_+\over 2} +x):=-\frac{1}{3}\int d^3 {\bf x} {\bf x}^i \psi_{10}( {\bf x})\left(\left(1- {\frac{k_+}{2}+x \over h_o -E_1 +{k_+ \over 2}+x} \right)\bfnabla^i \right)_{{\bf x},{\bf 0}}
\end{equation}
In order to proceed analogously to the $S$-wave case, we have first to move the ${\bf x}^i$ away from the wave function
\begin{displaymath}
I_{P}({k_+\over 2} +x)=\psi_{10}({\bf 0})+\frac{\frac{k_+}{2}+x}{3}\int d^3 {\bf x}\psi_{10}( {\bf x})\left\{\frac{1}{h_o -E_1 +{k_+ \over 2}+x}{\bf x} \bfnabla+\right.
\end{displaymath}
\begin{equation}\label{Pwave2}
\left. +{\frac{1}{h_o -E_1 +{k_+ \over 2}+x}}\left(-\frac{2\bfnabla^i}{m}\right){\frac{1}{h_o -E_1 +{k_+ \over 2}+x}}\bfnabla^i \right\}
\end{equation}
For the left-most propagators we can now proceed as before, namely expanding $V_o-V_s$. Note that the leading contribution in this expansion of the second term above exactly cancels against the first term. Of the remaining contributions of the second term only the next-to-leading one ($\mathcal{O}(\als)$) is relevant to obtain (\ref{expando}). Consider next the leading order contribution in this expansion of the last term. It reads 
\begin{displaymath}
-\frac{2}{3m}\!\int d^3 {\bf x}\psi_{10}( {\bf x})\left\{\bfnabla^i \frac{1}{h_o -E_1 +{k_+ \over 2}+x}\bfnabla^i\right\}=
-\frac{2}{3m}\int d^3 {\bf x}\psi_{10}( {\bf x})\left\{\!\left( \frac{1}{h_o -E_1 +{k_+ \over 2}+x}\bfnabla^i-\right.\right.
\end{displaymath}
\begin{equation}\label{Pwave3}
\left.\left.-\frac{1}{h_o -E_1 +{k_+ \over 2}+x}\bfnabla^iV_o\frac{1}{h_o -E_1 +{k_+ \over 2}+x}\right)\bfnabla^i \right\}
\end{equation}
Now we proceed as before with the left-most propagators, namely expanding $V_o-V_s$. The leading order contribution of the first term above produces the relativistic correction $\mathcal{O}(v^2)$ of (\ref{expando}). The next-to-leading contribution of this term and the leading order one of the second term are $\mathcal{O}(\als)$ and also relevant to (\ref{expando}). The next-to-leading order contribution of the last term in (\ref{Pwave2}) in the $V_o-V_s$ expansion of the left-most propagator is also $\mathcal{O}(\als )$ and relevant to (\ref{expando}).

Returning now to equations (\ref{expands})-(\ref{expando}), we see that the color singlet contribution reproduces the full LO expression for the central region in the limit $z\rightarrow 1$. The color octet shape functions $S_{P1}$  and $S_{P2}$ give contributions to the relativistic corrections (\ref{RelCo}), and $S_{P2}$ to terms proportional to $(1-z)$ in the limit $z\rightarrow 1$ of (\ref{LOrate}) as well. We have checked that, in the $z\rightarrow 1$ limit, both the $(1-z)\ln (1-z)$ of (\ref{LOrate}) and the $\ln (1-z)$ of the relativistic correction (\ref{RelCo}) are correctly reproduced if $\mu_c \sim M\sqrt{1-z}$, as it should. All the color octet shape functions contribute to the $\mathcal{O}(\als (\mu_h))$ correction in the first line of (\ref{expando}). There are additional $\mathcal{O}(\als (\mu_h))$ contributions coming from the expansion of the (Sudakov) resummed matching coefficients of the color singlet contribution and of the $S_{P2}$ color octet shape function. The $\als\log(1-z)$ in (\ref{expands}) reproduces the logarithm in ${d\Gamma^{c}}_{LO,\als}/{dz}$.

We propose the following formula
\be
\frac{1}{\Gamma_0}\frac{d\Gamma^{dir}}{dz}=\frac{1}{\Gamma_0}\frac{d\Gamma^{c}}{dz}+\left(\frac{1}{\Gamma_0
}\frac{d\Gamma_{CS}^{e}}{dz}-\left.{\frac{1}{\Gamma_0
}\frac{d\Gamma_{CS}^{e}}{dz}}\right\vert_c\right)+\left(\frac{1}{\Gamma_0}\frac{d\Gamma_{CO}^{e}}{dz}-\left.{\frac{1}{\Gamma_0
}\frac{d\Gamma_{CO}^{e}}{dz}}\right\vert_c\right)
\label{mergingNLO}
\ee
This formula reduces to the NRQCD expression in the central region. When we approach the upper end-point region the second terms in each of the parentheses are expected to cancel corresponding terms in the $z\to1$ limit of the expression for the central region up to higher order terms (in the end-point region counting). Thus, we are left with the resummed expressions for the end-point (up to higher order terms). 

There are of course other possibilities for the merging. For instance, one may choose a $z_1$ below which one trusts the calculation for the central region and a $z_2$ above which one trusts the end-point region calculation, and use some sort of interpolation between $z_1$ and $z_2$ (see for instance \cite{Lin:2004eu}). This would have the advantage of keeping the right approximation below $z_1$ and beyond $z_2$ unpolluted, at the expense of introducing further theoretical ambiguities due to the choice of $z_1$ and $z_2$, and, more important, due to the choice of the interpolation between $z_1$ and $z_2$. We believe that our formula (\ref{mergingNLO}) is superior because it does not introduce the above mentioned theoretical ambiguities. The price to be paid is that the expressions from the central region have an influence in the end-point region and vice-versa. This influence can always be chosen to be parametrically subleading but large numerical factors may make it noticeable in some cases, as we shall see below.

\subsubsection{Merging at LO}
\begin{figure}
\centering
\includegraphics[width=12.5cm]{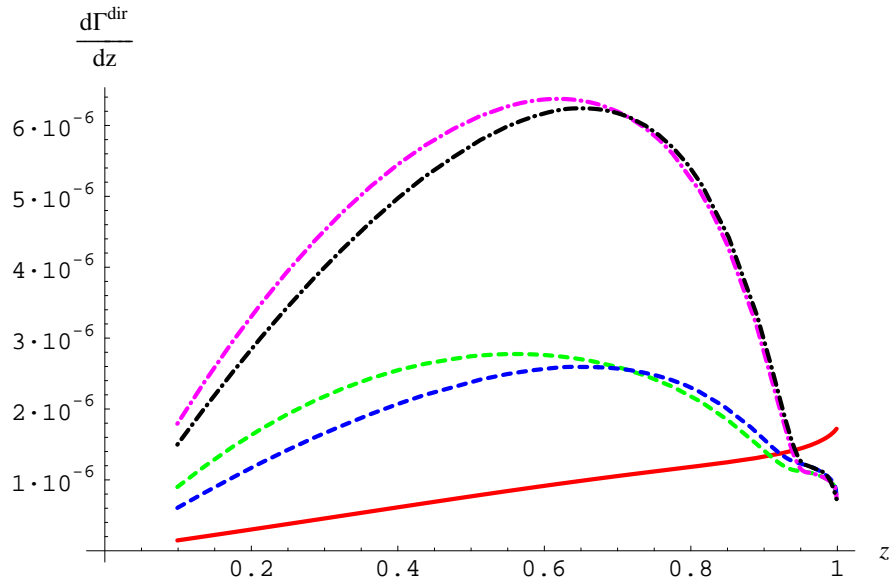}
\caption[Merging at LO]{Merging at LO. The solid red line is the NRQCD expression (\ref{LOrate}). The dot-dashed curves are obtained using an $MS$ scheme: the pink (light) curve is the end-point contribution (\ref{endp}) and the black (dark) curve is the LO merging. The dashed curves are obtained using the $sub$ scheme: the green (light) curve is the end-point contribution (\ref{endp}) and the blue (dark) curve is the LO merging.}\label{figcompLO}
\end{figure}

\begin{figure}
\centering
\includegraphics[width=12.5cm]{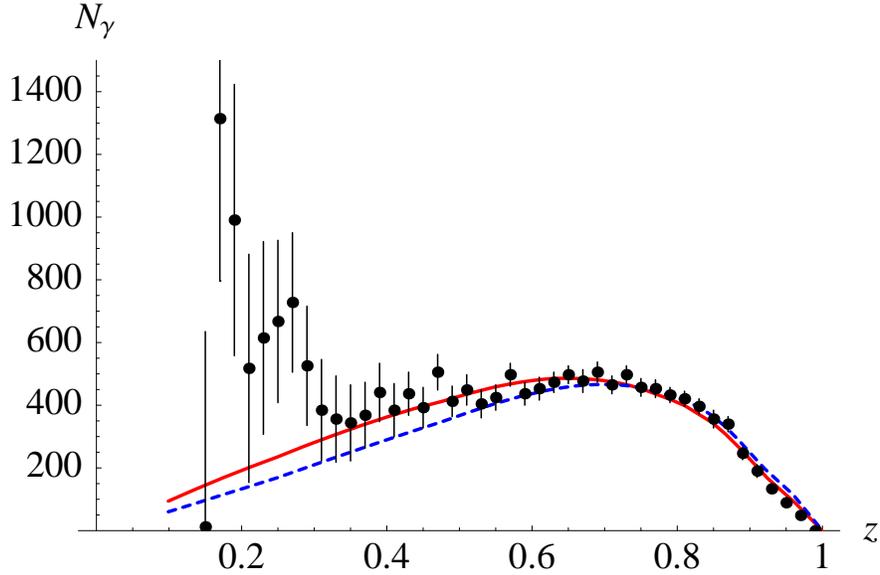}
\caption[Direct contribution to the spectrum: LO merging]{Direct contribution to the spectrum. The solid red line corresponds to the LO merging and the blue dashed line corresponds to equation (\ref{mergingLO}). The points are the CLEO data \cite{Nemati:1996xy}.}\label{figdibLO}
\end{figure}

If we wish to use only the LO expressions for the central region, we should take (\ref{expands}) and (\ref{expando}) at LO, namely
\be
\left. { \frac{1}{\Gamma_0
}\frac{d\Gamma_{CS}^{e}}{dz}}\right\vert_c = z \quad ,\quad \left. {\frac{1}{\Gamma_0
}\frac{d\Gamma_{CO}^{e}}{dz}}\right\vert_c= z\left(2-4\log \left(\frac{\mu_c}{M(1-z)}\right)
\right) (1-z) 
\ee
and substitute them in (\ref{mergingNLO}). Unexpectedly, the results obtained with this formula in the central region deviate considerably from those obtained with formula (\ref{LOrate}) (see fig. \ref{figcompLO}). This can be traced back to the fact that the $\als\sqrt{1-z} $ corrections in (\ref{expando}) are enhanced by large numerical factors, which indicates that the merging should better be done including $\als (\mu_h)$ corrections in the central region, as we discuss next. Alternatively, we may change our subtraction scheme in order to (partially) get rid of these contributions. With the new subtraction scheme ($sub$), described in the preceding section, the situation improves, although it does not become fully satisfactory (see fig. \ref{figcompLO}). This is due to the fact that some $\als\sqrt{1-z} $ terms remain, which do not seem to be associated to the freedom of choosing a particular subtraction scheme. In spite of this the description of data turns out to be extremely good. In figure \ref{figdibLO} we plot, using the $sub$ scheme, the merging at LO (solid red line) and also, for comparison, equation (\ref{mergingLO}) (blue dashed line). We have convoluted the theoretical curves with the experimental efficiency and the overall normalization is taken as a free parameter.

\subsubsection{Merging at NLO}
\begin{figure}
\centering
\includegraphics[width=12.5cm]{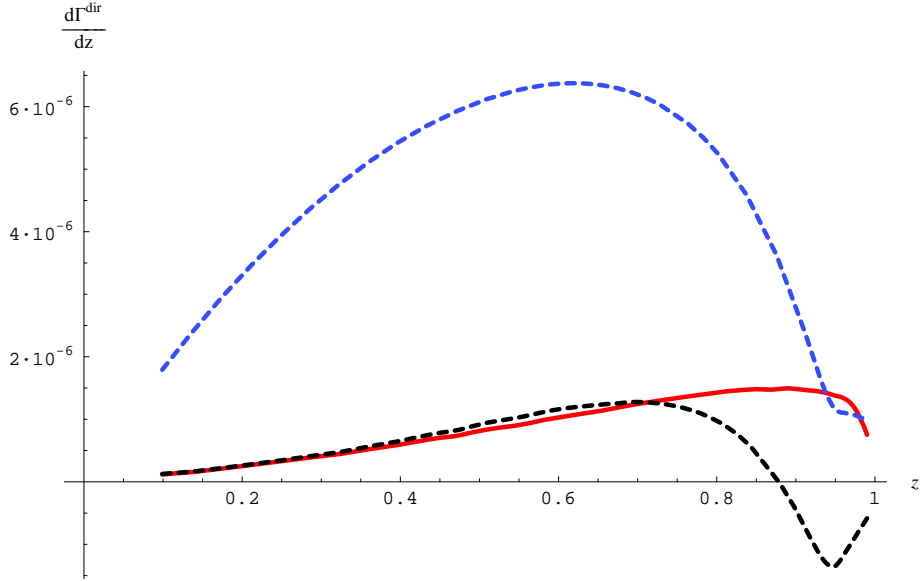}
\caption[Merging at NLO (wave function at the origin at LO)]{Merging at NLO (using an $MS$ scheme and the wave function at the origin at LO). The solid red line is the NRQCD result (\ref{central}), the blue (light) dashed curve is the end-point contribution (\ref{endp}) and the black (dark) dashed curve is the NLO merging.}\label{figcompNLOFO}
\end{figure}

\begin{figure}
\centering
\includegraphics[width=12.5cm]{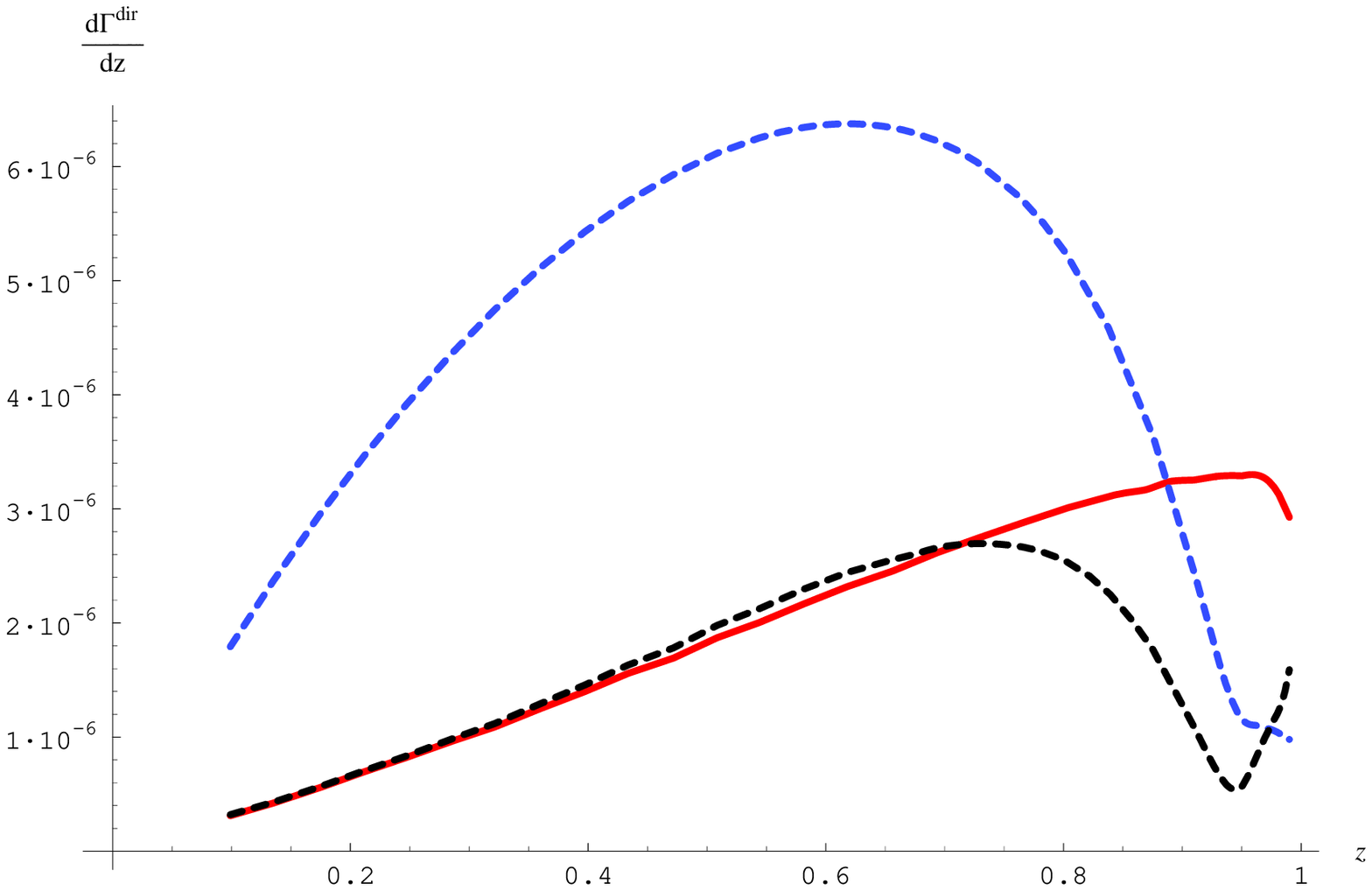}
\caption[Merging at NLO (wave function at the origin with $\als^2 (\mu_s)$ corrections)]{Merging at NLO (using an $MS$ scheme and the wave function at the origin with the $\als^2 (\mu_s)$ corrections included). The solid red line is the NRQCD result (\ref{central}), the blue (light) dashed curve is the end-point contribution (\ref{endp}) and the black (dark) dashed curve is the NLO merging.}\label{figcompNLOFO2}
\end{figure}

If we wish to use the NLO expressions for the central region (\ref{central}), we should take all the terms displayed in (\ref{expands})- (\ref{expando}) and substitute them in (\ref{mergingNLO}). Unlike in the LO case, for values of $z$ in the central region the curve obtained from (\ref{mergingNLO}) now approaches smoothly the expressions for the central region (\ref{central}) as it should. This is so no matter if we include the $\als^2 (\mu_s)$ corrections to the wave function at the origin in $d\Gamma^c_{LO}/dz$, as we in principle should, or not (see figs. \ref{figcompNLOFO} and \ref{figcompNLOFO2}). However, since the above corrections are very large, the behavior of the curve for $z\rightarrow 1$, strongly depends on whether we include them or not (see again figs. \ref{figcompNLOFO} and \ref{figcompNLOFO2}). We believe that the two possibilities are legitimate. If one interpretes the large $\als^2 (\mu_s)$ corrections as a sign that the asymptotic series starts exploding, one should better stay at LO (or including $\als (\mu_s)$ corrections). However, if one believes that the large $\als^2 (\mu_s)$ corrections are an accident and that the $\als^3 (\mu_s)$ ones (see \cite{Beneke:2005hg,Penin:2005eu} for partial results) will again be small, one should use these $\als^2 (\mu_s)$ corrections. We consider below the two cases.

If we stay at LO (or including $\als (\mu_s)$ corrections) for the wave function at the origin, the curve we obtain for $z\rightarrow 1$ differs considerably from the expressions for the end-point region (\ref{endp}) (see fig. \ref{figcompNLOFO}). This can be traced back to the $\als\sqrt{1-z}$ term in (\ref{expando}) again. This term is parametrically suppressed in the end-point region, but, since it is multiplied by a large numerical factor, its contribution turns out to be overwhelming. This term might (largely) cancel out against higher order contributions in the end-point region, in particular against certain parts of the NLO expressions for the color singlet contributions, which are unknown at the moment. 

\begin{figure}
\centering
\includegraphics[width=12.5cm]{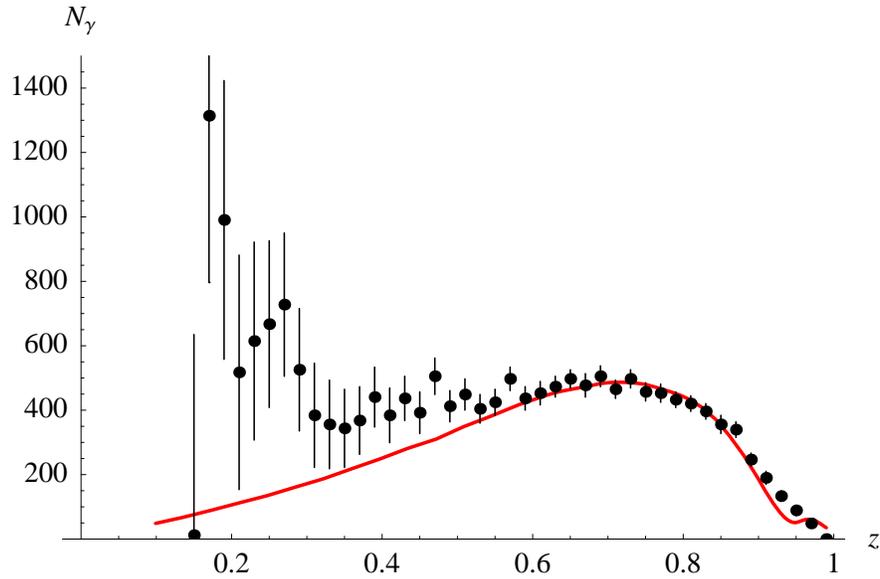}
\caption[Direct contribution to the spectrum: NLO merging]{Direct contribution to the spectrum using the NLO merging (in an $MS$ scheme and the wave function at the origin with the $\als^2 (\mu_s)$ corrections included). The points are the CLEO data \cite{Nemati:1996xy}.}\label{figdibNLO}
\end{figure}

If we use the wave function at the origin with the $\als^2 (\mu_s)$
corrections included, the curves we obtain for $z\rightarrow 1$ become much
closer to the expressions for the end-point region (\ref{endp}) (see fig
\ref{figcompNLOFO2}). Hence, a good description of data is obtained with no
need of additional subtractions\footnote{One might be worried about the big
  difference that the corrections to the wave function at the origin introduce
in the result. In that sense let us mention that when we analyze the
electromagnetic decay width of $\Upsilon(1S)$ ($\Gamma(\Upsilon(1S)\to e^+\;
e^-)$, the formulas needed to compute the width can be found, for instance, in
\cite{Vairo:2003gh}), with the same power counting
we have employed here, the result we obtain is $5.24\cdot10^{-7}$GeV if we do
not include the $\als^2 (\mu_s)$ corrections and $1.17\cdot10^{-6}$GeV if we do include
them. This is to be compared with the experimental result
$1.32\cdot10^{-6}$GeV \cite{Eidelman:2004wy}.}, as shown in figure \ref{figdibNLO} (as usual experimental efficiency has been taken into account and the overall normalization is a free parameter). This are now good news. Because this final curve incorporates all the terms that are supposed to be there according to the power counting.

\subsection{Fragmentation contributions}\label{subsecfragcon}
The fragmentation contributions can be written as
\begin{equation}
\frac{d\Gamma^{frag}}{dz}=\sum_{a = q,\bar q, g} \int_z^1\frac{dx}{x}C_a(x)D_{a\gamma}\left(\frac{z}{x},M\right),
\end{equation}
where $C_a$ represents the partonic kernels and $D_{a\gamma}$ represents the fragmentation functions. The partonic kernels can again be expanded in powers of $v$ \cite{Maltoni:1998nh}
\begin{equation}
C_a=\sum_{\mathcal{Q}}C_a[\mathcal{Q}]
\end{equation}
The leading order term in $v$ is the color singlet rate to produce three gluons
\[
C_g\left[{\cal O}_1(^3S_1)\right]=\frac{40}{81}\alpha_s^3
\left(\frac{2-z}{z} + \frac{z(1-z)}{(2-z)^2} + 2\frac{1-z}{z^2}\ln(1-z) - 2\frac{(1-z)^2}{(2-z)^3}
\ln(1-z)\right)\cdot
\]
\begin{equation}\label{fragsing} 
\cdot\frac{\langle  V_Q (nS)\vert {\cal O}_1(^3S_1)\vert V_Q (nS)\rangle}{m^2}
\end{equation}
The color octet contributions start at order $v^4$ but have a $\frac{1}{\alpha_s}$ enhancement with respect to
(\ref{fragsing})
\[
C_g\left[{\cal O}_8(^1S_0)\right]=\frac{5\pi\alpha_s^2}{3}
\delta(1-z)\frac{\langle V_Q (nS)\vert{\cal O}_8(^1S_0)\vert V_Q (nS)\rangle}{m^2}
\]
\[
C_g\left[{\cal O}_8(^3P_J)\right]=\frac{35\pi\alpha_s^2}{3}
\delta(1-z)\frac{\langle V_Q (nS)\vert{\cal O}_8(^3P_0)\vert V_Q (nS)\rangle}{m^4}
\]
\begin{equation}
C_q\left[{\cal O}_8(^3S_1)\right]=\frac{\pi\alpha_s^2}{3}
\delta(1-z)\frac{\langle V_Q (nS)\vert{\cal O}_8(^3S_1)\vert V_Q (nS)\rangle}{m^2}\label{octetf}
\end{equation}

Then the color singlet fragmentation contribution is of order $\alpha_s^3D_{g\to\gamma}$ and the color octet fragmentation are of order $v^4\alpha_s^2D_{g\to\gamma}$ ($\phantom{}^1S_0$ and $\phantom{}^3P_J$ contributions) or $v^4\alpha_s^2D_{q\to\gamma}$ ($\phantom{}^3S_1$ contribution). We can use, as before, the counting $v^2\sim\alpha_s$ to compare the relative importance of the different contributions together with the existing models for the fragmentation functions \cite{Aurenche:1992yc}. The latter tell us that $D_{q\to\gamma}$ is much larger than $D_{g\to\gamma}$. This causes the $\mathcal{O}(v^4\alpha_s^2D_{q\to\gamma})$ $\phantom{}^3S_1$ octet contribution to dominate in front of the singlet $\mathcal{O}(\alpha_s^3D_{g\to\gamma})$ and the octet $\mathcal{O}(v^4\alpha_s^2D_{g\to\gamma})$ contributions. In fact, $\alpha_sD_{q\to\gamma}$ is still larger than $D_{g\to\gamma}$, so we will include in our plots the $\alpha_s$ corrections to the color octet contributions (\ref{octetf}) proportional to $D_{q\to\gamma}$, which have been calculated in \cite{Maltoni:1998nh}. In addition, the coefficients for the octet $\phantom{}^3P_J$ contributions have large numerical factors, causing these terms to be more important than the color singlet contributions. Let us finally notice that the $\alpha_s$ corrections to the singlet rate  will produce terms of $\mathcal{O}(\alpha_s^4D_{q\to\gamma})$, which from the considerations above are expected to be as important as the octet $\phantom{}^3S_1$ contribution. These $\alpha_s$ corrections to the singlet rate are unknown, which results in a large theoretical uncertainty in the fragmentation contributions. 

For the quark fragmentation function we will use the LEP measurement \cite{Buskulic:1995au}  
\begin{equation}
D_{q\gamma}(z,\mu) = \frac{e_q^2\alpha(\mu)}{2\pi}
\left[P_{q\gamma}(z) \ln\left(\frac{\mu^2}{\mu_0^2(1-z)^2}\right) + C\right]
\end{equation}
where
\begin{equation}
C = -1-\ln(\frac{M_Z^2}{2\mu_0^2})\quad ;\quad P_{q\gamma}(z) = \frac{1+ (1-z)^2}{z}\quad
;\quad\mu_0=0.14^{+0.43}_{-0.12} {\rm\ GeV}
\end{equation} 
and for the gluon fragmentation function the model \cite{Owens:1986mp}. These are the same choices as in \cite{Fleming:2002sr}. However, for the $\mathcal{O}_8 (^1 S_0)$ and $\mathcal{O}_8 (^3 P_0)$ matrix elements we will use our estimates (\ref{estomeS})-(\ref{estomeP}). Notice that we do not assume that a suitable combination of these matrix elements is small, as it was done in \cite{Fleming:2002sr}. The $\mathcal{O}_8 (^3 S_1)$ matrix element can be extracted from a lattice determination of the reference \cite{Bodwin:2005gg}. Using the wave function at the origin with the $\als^2 (\mu_s )$ corrections included, we obtain (we use the numbers of the hybrid algorithm), 
\be
\left.\left< \Upsilon (1S) \vert \mathcal{O}_8 (^3 S_1) \vert \Upsilon (1S)
  \right>\right|_{\mu=M}  \sim 0.00026\,GeV^3 
\label{lattice}
\ee
which differs from the estimate using NRQCD $v$ scaling by more than two orders of magnitude:
\begin{equation}
\left.\left< \Upsilon (1S) \vert \mathcal{O}_8 (^3 S_1) \vert \Upsilon (1S) \right>\right|_{\mu=M}\sim
v^4\left.\left< \Upsilon (1S) \vert \mathcal{O}_1 (^3 S_1) \vert \Upsilon (1S) \right>\right|_{\mu=M}\sim 0.02\,GeV^3
\label{vscaling}
\end{equation}
(we have taken $v^2\sim0.08$), which was used in ref. \cite{Fleming:2002sr}. The description of data turns out to be better with the estimate (\ref{vscaling}). However, this is not very significant, since, as mentioned before, unknown NLO contributions are expected to be sizable.   

In the $z\rightarrow 0$ region soft radiation becomes dominant and the fragmentation contributions completely dominate the spectrum in contrast with the direct contributions \cite{Catani:1994iz}. Note that, since the fragmentation contributions have an associated bremsstrahlung spectrum, they can not be safely integrated down to $z=0$; that is $\int_0^1dz\frac{d\Gamma^{frag}}{dz}$ is not an infrared safe observable. In any case we are not interested in regularizing such divergence because the resolution of the detector works as a physical cut-off.

\subsection{The complete photon spectrum}
We can now compare the theoretical expressions with data in the full range of $z$. First note that formula (\ref{mergingNLO}) requires $d\Gamma^{e}/dz$ for all values of $z$. The color octet shape functions, however, were calculated in the end-point region under the assumption that $M\sqrt{1-z}\sim \gamma$, and the scale of the $\als$ was set accordingly. When $z$ approaches the central region $M\sqrt{1-z}\gg \gamma$, and hence some $\als$ will depend on the scale $M\sqrt{1-z}$ and others on $\gamma$ (we leave aside the global $\als (\mu_u)$). In order to decide the scale we set for each $\als$ let us have a closer look at the formula (\ref{expando}). We see that all terms have a common factor $\gamma^3$. This indicates that one should extract $\gamma^3$ factors in the shape functions, the $\als$ of which should stay at the scale $\mu_s$. This is achieved by extracting $\gamma^{3/2}$ in $I_S$ and $I_P$. If we set the remaining $\als$ to the scale $\mu_p=\sqrt{m(M(1-z)/2-E_1)}$, we will reproduce (\ref{expando}) when approaching to the central region, except for the relativistic correction, the $\als$ of which will be at the scale $\mu_p$ instead of at the right scale $\mu_s$. We correct for this by making the following substitution
\be
S_{P1}\longrightarrow S_{P1}+{\als (\mu_u)\over 6\pi N_c}{\gamma^3\over\pi}\left(\log {k_+^2\over \mu_c^2}-1\right)\left( {4\gamma^2\over 3m}- {m C_f^2\als^2 (\mu_p)\over 3} \right)
\ee
Notice that the replacements above are irrelevant as far as the end-point region is concerned, but important for the shape functions to actually (numerically) approach the expressions (\ref{expando}) in the central region, as they should.

\begin{figure}
\centering
\includegraphics[width=14.5cm]{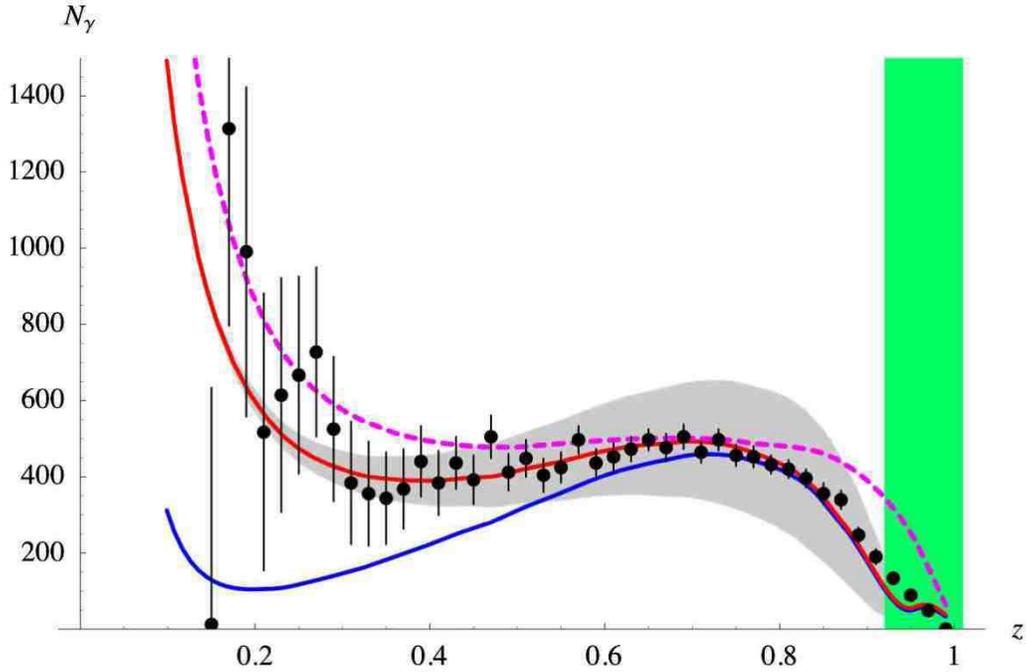}
\caption[Photon spectrum]{Photon spectrum. The points are the CLEO data \cite{Nemati:1996xy}. The solid lines are the NLO merging plus the fragmentation contributions: the red (light) line and the blue (dark) line are obtained by using (\ref{vscaling}) and (\ref{lattice}) for $\left< \Upsilon (1S) \vert \mathcal{O}_8 (^3 S_1) \vert \Upsilon (1S) \right>$  respectively. The grey shaded region is obtained by varying $\mu_{c}$ by $\sqrt{2^{\pm 1}}\mu_{c}$. The green shaded region on the right shows the zone where the calculation of the shape functions is not reliable (see subsection \ref{subseccalshpfct}). The pink dashed line is the result in \cite{Fleming:2002sr}, where only color singlet contributions were included in the direct contributions.}\label{figtotal}
\end{figure}

\begin{figure}
\centering
\includegraphics[width=14.5cm]{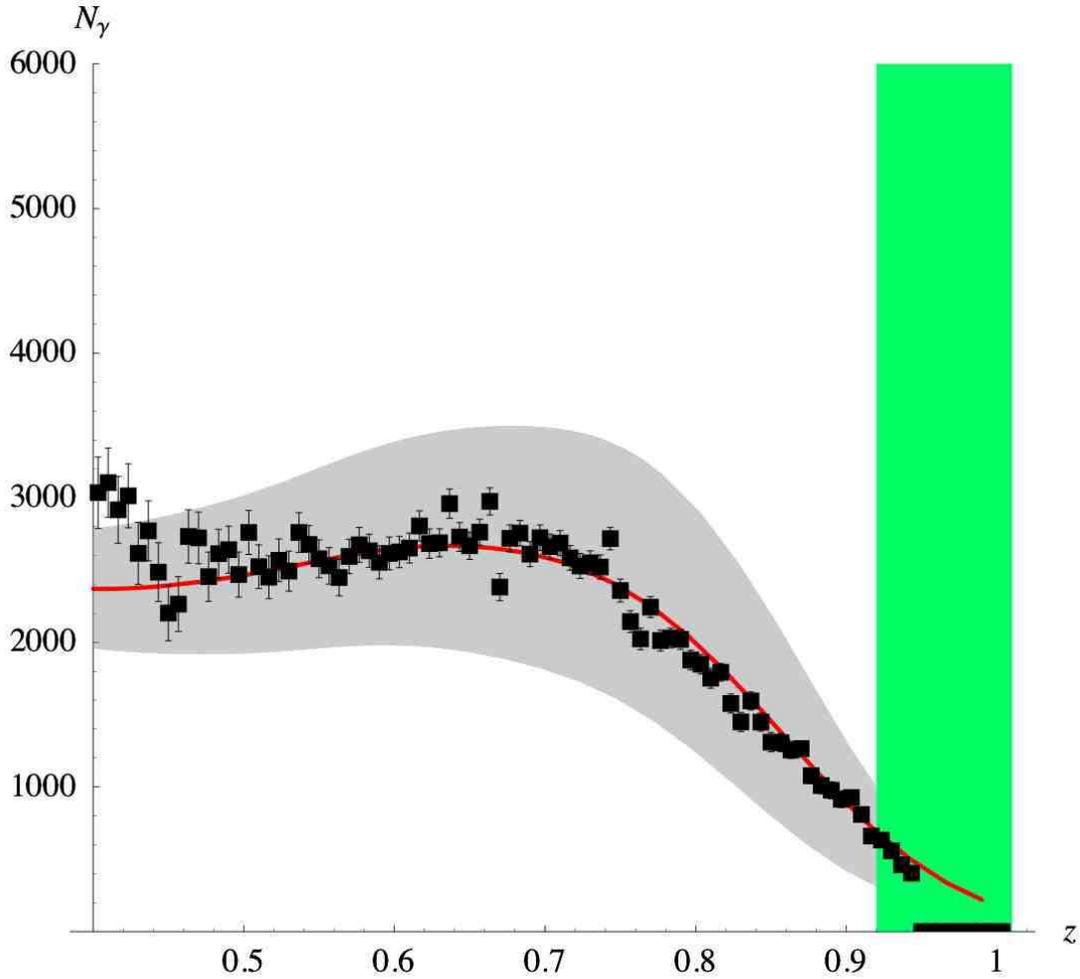}
\caption[Photon spectrum (most recent data)]{Photon spectrum. The points are the new CLEO data \cite{Besson:2005jv}. The red solid line is the NLO merging plus the fragmentation contributions, using (\ref{vscaling}) for $\left< \Upsilon (1S) \vert \mathcal{O}_8 (^3 S_1) \vert \Upsilon (1S) \right>$. The grey shaded region is obtained by varying $\mu_{c}$ by $\sqrt{2^{\pm 1}}\mu_{c}$. The green shaded region on the right shows the zone where the calculation of the shape functions is not reliable (see subsection \ref{subseccalshpfct}).}\label{figtotalnou}
\end{figure}

The comparison with the experiment is shown in figures \ref{figtotal} and
\ref{figtotalnou}. These plots are obtained by using the merging formula
(\ref{mergingNLO}) at NLO with the $\als^2 (\mu_s)$  corrections to the wave
function at the origin included for the direct contributions plus the
fragmentation contributions in subsection \ref{subsecfragcon} including the
first $\als$ corrections in $C_q$ and using the estimate (\ref{vscaling}) for
the $\left< \Upsilon (1S) \vert\mathcal{O}_8 (^3 S_1) \vert \Upsilon (1S)
\right>$ matrix element. The error band is obtained by replacing $\mu_{c}$ by
$\sqrt{2^{\pm 1}}\mu_{c}$. Errors associated to the large $\als^2 (\mu_s)$
corrections to the wave function at the origin, to possible large NLO color
singlet contributions in the end-point region and to the fragmentation
contributions are difficult to estimate and not displayed (see the
corresponding sections in the text for discussions). The remaining error
sources are negligible. In figure \ref{figtotal}, as usual, experimental
efficiency has been taken into account and the overall normalization is a free
parameter. Figure \ref{figtotalnou} compares our results with the new (and very precise) data from CLEO \cite{Besson:2005jv}. This plot takes into account the experimental efficiency and also the resolution of the experiment (the overall normalization is a free parameter).

We can see from figures \ref{figtotal} and \ref{figtotalnou} that, when we put together the available
theoretical results, an excellent description of data is achieved for the
whole part of the spectrum where experimental errors are reasonable
small (recall that the error bars showed in the plots only take into account the statistical errors, and not the systematic ones \cite{Nemati:1996xy,Besson:2005jv}). Clearly then, our results indicate that the introduction of a finite gluon mass \cite{Field:2001iu} is unnecessary. One should keep in mind, however, that in order to have the theoretical errors under control higher order calculations are necessary both in the direct (end-point) and fragmentation contributions.

Let us mention that the inclusion of color octet contributions in the end-point region, together with the merging with the central region expression explained here, may be useful for production processes like inclusive $J/\psi$ production in $e^+e^-$ machines \cite{Fleming:2003gt,Lin:2004eu,Hagiwara:2004pf}.

\section{Identifying the nature of heavy quarkonium}
As we have just seen in the previous section, the photon spectrum in the
radiative decay of the $\Upsilon (1S)$ can be well explained
theoretically. This fact, together with the recent appearance of measurements of the
photon spectra for the $\Upsilon (2S)$ and $\Upsilon (3S)$ states \cite{Besson:2005jv}, motivates
us to try to use these radiative decays to uncover the properties of the
decaying heavy
quarkonia.

As it has already been explained, the
interplay of $\lQ$  with the scales $mv$ and $mv^2$ dictates the degrees of
freedom of pNRQCD. Two regimes have been identified: the weak coupling
regime, $\lQ \lesssim mv^2$, and the strong coupling regime, $mv^2 \ll \lQ
\lesssim mv$. Due to the fact that none of the scales involved in these hierarchies are directly accessible experimentally, given a heavy quarkonium state, it is not obvious to which regime it must be assigned. Only the $\Upsilon (1S)$ 
appears to belong to the weak coupling regime, since weak coupling calculations in $\als (mv)$ converge reasonably well. 
The fact that the spectrum of excitations is not Coulombic suggests that the
higher excitations are not in the weak coupling regime, which can be
understood from the fact that $\mathcal{O}(\lQ)$ effects in this regime are
proportional to a high power of the principal quantum number
\cite{Voloshin:1978hc,Leutwyler:1980tn}. Nevertheless, there have been claims in the
literature, using renormalon-based approaches, that also $\Upsilon (2S)$ and
even $\Upsilon (3S)$ can also be understood within the weak coupling regime
\cite{Brambilla:2001fw,Brambilla:2001qk,Recksiegel:2003fm}. We will see that the photon spectra in semi-inclusive radiative decays of heavy quarkonia to light hadrons provide important information which may eventually settle this question. 

We start by writing the radiative decay rate for a state with generic principal
quantum number $n$. Again we split the decay rate into direct and
fragmentation contributions
\begin{equation}
\frac{d\Gamma_n}{dz}=\frac{d\Gamma^{dir}_n}{dz}+\frac{d\Gamma^{frag}_n}{dz}
\end{equation}
here $z=2E_\gamma /M_n$ ($M_n$ is the mass of the heavy quarkonium state). We
shall now restrict our discussion to $z$ in the central region, in which no further scale is introduced beyond those
inherent of the non-relativistic system. We write the spectrum in the
following compact form
\begin{equation}\label{dir}
\frac{d\Gamma^{dir}_n}{dz}=\sum_{\mathcal{Q}}C[\mathcal{Q}](z)\frac{\langle \mathcal{Q}\rangle_n}{m^{\delta_{\mathcal{Q}}}}
\end{equation}
\begin{equation}\label{frag}
\frac{d\Gamma_n^{frag}}{dz} = \sum_{a = q,\bar q, g} \int_z^1\frac{dx}{x}\sum_{\mathcal{Q}}C_a[\mathcal{Q}](x)D_{a\gamma}\left(\frac{z}{x},m\right)\
 :=\sum_{\mathcal{Q}}f_\mathcal{Q}(z)\frac{\langle \mathcal{Q}\rangle_n}{m^{\delta_{\mathcal{Q}}}}
\end{equation}
where $\mathcal{Q}$ is a local NRQCD operator, $\delta_{\mathcal{Q}}$ is an
integer which follows from the dimension of $\mathcal{Q}$ and $\langle
\mathcal{Q}\rangle_n:=\langle  V_Q (nS)\vert \mathcal{Q}\vert V_Q
(nS)\rangle$. It is important for what follows that the $f_\mathcal{Q}(z)$ are
universal and do not depend on the specific bound state $n$. Due to the
behavior of the fragmentation functions above, the fragmentation contributions
are expected to dominate the spectrum in the lower $z$ region and to be
negligible in the upper $z$ one. In the central region, in which we will focus on,
they can always be treated as a perturbation, as we will show below.

Let us first consider the weak coupling regime, for which the original NRQCD
velocity counting holds \cite{Bodwin:1994jh} (this is the situation described
in the previous sections of this chapter, we recall here some of the arguments
for an easier reading). The direct contributions are given at leading order by
the $\mathcal{O}_1\left(\phantom{}^3S_1\right)$ operator; the next-to-leading
order (NLO) ($v^2$ suppressed) term is given by the $\mathcal{P}_1\left(\phantom{}^3S_1\right)$ operator. The contributions of color octet operators start at order $v^4$
and are not $\als^{-1}(m)$ enhanced in the central region.
The fragmentation contributions are more difficult to organize since the importance of each term is not only fixed by the velocity counting alone but also involves the size of the fragmentation functions. It will be enough for us to restrict ourselves to the LO operators both in the singlet and octet sectors. The LO color singlet operator is $\mathcal{O}_1\left(\phantom{}^3S_1\right)$ as well.
The leading color octet contributions are $v^4$  suppressed but do have a $\als^{-1}(m)\sim 1/v^2$ enhancement with respect to the singlet ones here. They involve $\mathcal{O}_8\left(\phantom{}^3S_1\right)$, $\mathcal{O}_8\left(\phantom{}^1S_0\right)$ and $\mathcal{O}_8\left(\phantom{}^3P_0\right)$.
Then in the central region, the NRQCD expression (at the order described above) reads
\[
\frac{d\Gamma_n}{dz} =\left(C_1\left[\phantom{}^3S_1\right](z)+f_{\mathcal{O}_1\left(\phantom{}^3S_1\right)}(z)\right)\frac{\langle \mathcal{O}_1(^3S_1)\rangle_n}{m^2}+C_1'\left[\phantom{}^3S_1\right](z)\frac{\langle\mathcal
  {P}_1(^3S_1)\rangle_n}{m^4}+f_{\mathcal{O}_8\left(\phantom{}^3S_1\right)}(z)\cdot
\]
\begin{equation}\label{width}
\cdot\frac{\langle\mathcal{ O}_8(^3S_1)\rangle_n}{m^2}++f_{\mathcal{O}_8\left(\phantom{}^1S_0\right)}(z)\frac{\langle\mathcal{O}_8(^1S_0)\rangle_n}{m^2}+f_{\mathcal{O}_8\left(\phantom{}^3P_J\right)}(z)\frac{\langle\mathcal{O}_8(^3P_0)\rangle_n}{m^4}
\end{equation}
If we are in the strong coupling regime and use the so called conservative counting, 
the color octet matrix elements are suppressed by $v^2$ rather than by $v^4$. Hence we should include the color octet operators in the direct contributions as well. In practise, this only amounts to the addition of $C_8$'s to the $f_{\mathcal{O}_8}$'s.
Furthermore, $f_{\mathcal{O}_1\left(\phantom{}^3S_1\right)}(z)$, $f_{\mathcal{O}_8\left(\phantom{}^1S_0\right)}(z)$ and $f_{\mathcal{O}_8\left(\phantom{}^3P_J\right)}(z)$ are proportional to $D_{g\gamma}\left(x,m\right)$, which is small (in the central region) according to the widely accepted model \cite{Owens:1986mp}.  $f_{\mathcal{O}_8\left(\phantom{}^3S_1\right)}(z)$ is proportional to $D_{q\gamma}\left(x,m\right)$, which has been measured at LEP \cite{Buskulic:1995au}. It turns out that numerically $f_{\mathcal{O}_8\left(\phantom{}^3S_1\right)}(z)\sim C_8[\phantom{}^3S_1](z)$ in the central region. Therefore, all the LO fragmentation contributions can be treated as a perturbation. Consequently, the ratio of decay widths of two states with different principal quantum numbers is given at NLO by
\[
\frac{\displaystyle\frac{d\Gamma_n}{dz}}{\displaystyle\frac{d\Gamma_r}{dz}}=\frac{\langle\mathcal{O}_1(^3S_1)\rangle_n}{\langle\mathcal{O}_1(^3S_1)\rangle_r}\left(\!\!1\!+\!\frac{C_1'\left[\phantom{}^3S_1\right](z)}{C_1\left[\phantom{}^3S_1\right](z)}\frac{\mathcal{R}_{\mathcal{P}_1(\phantom{}^3S_1)}^{nr}}{m^2}+\frac{f_{\mathcal{O}_8\left(\phantom{}^3S_1\right)}(z)}{C_1\left[\phantom{}^3S_1\right](z)}\!\mathcal{R}_{\mathcal{O}_8(\phantom{}^3S_1)}^{nr}\!\!+\!\!\frac{f_{\mathcal{O}_8\left(\phantom{}^1S_0\right)}(z)}{C_1\left[\phantom{}^3S_1\right](z)}\!\mathcal{R}_{\mathcal{O}_8(\phantom{}^1S_0)}^{nr}+\right.
\]
\begin{equation}\label{nrqcd}
\left.+\frac{f_{\mathcal{O}_8\left(\phantom{}^3P_J\right)}(z)}{C_1\left[\phantom{}^3S_1\right](z)}\frac{\mathcal{R}_{\mathcal{O}_8(\phantom{}^3P_0)}^{nr}}{m^2}\right)
\end{equation}
where
\begin{equation}
\mathcal{R}_{\mathcal{Q}}^{nr}=\left(\frac{\langle\mathcal{Q}\rangle_n}{\langle\mathcal{O}_1(^3S_1)\rangle_n}-\frac{\langle\mathcal{Q}\rangle_r}{\langle\mathcal{O}_1(^3S_1)\rangle_r}\right)
\label{r}
\end{equation}

Note that the $\als (m)$ corrections to the matching coefficients give rise to
negligible next-to-next-to-leading order (NNLO) contributions in the ratios above.
No further simplifications can be achieved at NLO without explicit assumptions on the counting. If the two states $n$ and $r$ are in the weak coupling regime, then
$\mathcal{R}_{\mathcal{P}_1(\phantom{}^3S_1)}^{nr}=m(E_{n}-E_{r})$ \cite{Gremm:1997dq}. 
In addition,
the ratio of matrix elements in front of the rhs of (\ref{nrqcd}) can be expressed in terms of the measured leptonic decay widths
\begin{equation}\label{qucO1}
 \frac{\langle\mathcal{O}_1(^3S_1)\rangle_n}{\langle\mathcal{O}_1(^3S_1)\rangle_r}=\frac{\Gamma\left(V_Q(nS)\to e^+e^-\right)}{\Gamma\left(V_Q(rS)\to e^+e^-\right)}\left[\!1\!-\!\frac{\mathrm{Im}g_{ee}\left(\phantom{}^3S_1\right)}{\mathrm{Im}f_{ee}\left(\phantom{}^3S_1\right)}\frac{E_{n}-E_{r}}{m}\right]
\end{equation} 
$\mathrm{Im}g_{ee}$ and $\mathrm{Im}f_{ee}$ are short distance matching coefficient which may be found in \cite{Bodwin:1994jh}.
Eq.(\ref{qucO1}) and the expression for $\mathcal{R}_{\mathcal{P}_1(\phantom{}^3S_1)}^{nr}$ also hold  if both $n$ and $r$ are in the strong coupling coupling regime \cite{Brambilla:2002nu,Brambilla:2001xy,Brambilla:2003mu}, but none of them does if one of the states is in the weak coupling regime and the other in the strong coupling regime. In the last case the NRQCD expression depends on five unknown parameters, which depend on $n$ and $r$.
If both $n$ and $r$ are in the strong coupling regime further simplifications
occur. The matrix elements of the color octet NRQCD operators are proportional
to the wave function at the origin times universal (bound state independent)
non-perturbative parameters \cite{Brambilla:2002nu,Brambilla:2001xy,Brambilla:2003mu} (see appendix \ref{appME}). Since $\langle\mathcal{O}_1(^3S_1)\rangle_n$ is also proportional to the wave function at the origin, the latter cancels in the ratios involved in (\ref{r}).
Hence, 
$\mathcal{R}_{\mathcal{Q}}^{nr}=0$ for the octet operators appearing in (\ref{nrqcd}). Then, the pNRQCD expression for the ratio of decay widths reads
\begin{equation}\label{scsc}
\frac{\displaystyle\frac{d\Gamma_n}{dz}}{\displaystyle\frac{d\Gamma_r}{dz}} =\frac{\Gamma\left(V_Q(nS)\to e^+e^-\right)}{\Gamma\left(V_Q(rS)\to e^+e^-\right)}\left[\!1\!-\!\frac{\mathrm{Im}g_{ee}\left(\phantom{}^3S_1\right)}{\mathrm{Im}f_{ee}\left(\phantom{}^3S_1\right)}\frac{E_{n}-E_{r}}{m}\right]\left(1+\frac{C_1'\left[\phantom{}^3S_1\right](z)}{C_1\left[\phantom{}^3S_1\right](z)}\frac{1}{m}\left(E_{n}-E_{r}\right)\right)
\end{equation}
Therefore, in the strong coupling regime 
we can predict
, using pNRQCD, 
the ratio of photon spectra at NLO (in the $v^2$, $(\lQ /m )^2$
\cite{Brambilla:2002nu,Brambilla:2001xy} and $\als (\sqrt{m\lQ})\times \sqrt{ \lQ /m} $
\cite{Brambilla:2003mu} expansion). On the other hand, if one of the states $n$ is in the weak coupling regime, $\mathcal{R}_{\mathcal{Q}}^{nr}$ will have a non-trivial dependence on the principal quantum number $n$  and hence it is not expected to vanish.
Therefore, expression (\ref{scsc}) provides invaluable help for identifying the 
nature of heavy quarkonium states. If the two states are in the strong coupling regime, the ratio must follow the formula (\ref{scsc}); on the other hand, if (at least) one of the states is in the weak coupling regime the ratio is expected to deviate from (\ref{scsc}), 
and should follow the general formula (\ref{nrqcd}). We illustrate the expected deviations in the plots (dashed curves) by assigning to the unknown $\cal R$s in (\ref{nrqcd}) the value $v^4$ ($v^2 \sim 0.1$), according to the original NRQCD velocity scaling.

%(since eq.(\ref{nrqcd}) depends on several unknown parameters, we cannot discard the possibility that all these parameters conspire among themselves to 
%give a numerical result similar to eq. (\ref{scsc}); but this is extremely unlikely, note that the different terms in (\ref{nrqcd}) have different $z$ dependences).

\begin{figure}
\centering
\includegraphics[width=12.5cm]{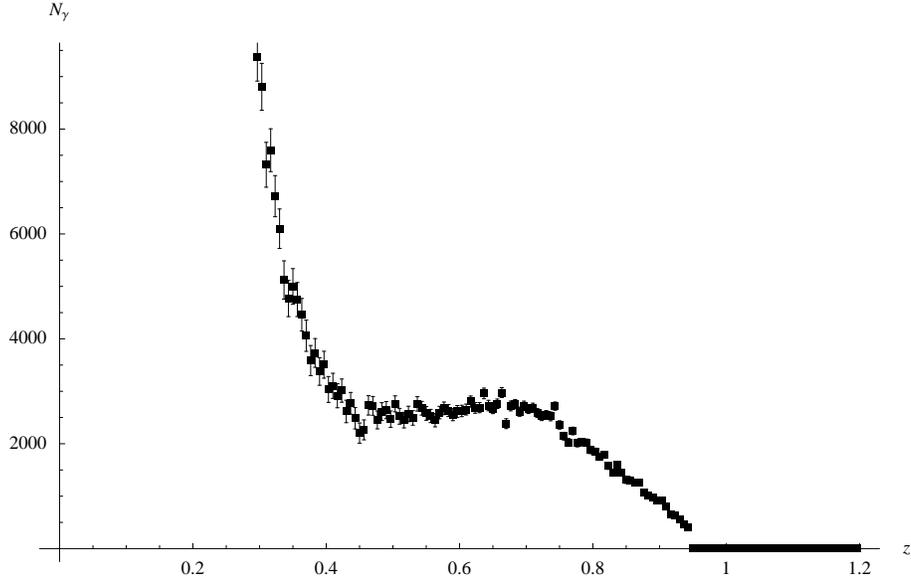}
\caption[CLEO data for $\Upsilon (1S)$ photon spectrum]{Background-subtracted
  CLEO data for the $\Upsilon (1S)$ photon spectrum \cite{Besson:2005jv}.}
\label{fignou1s}
\end{figure}

\begin{figure}
\centering
\includegraphics[width=12.5cm]{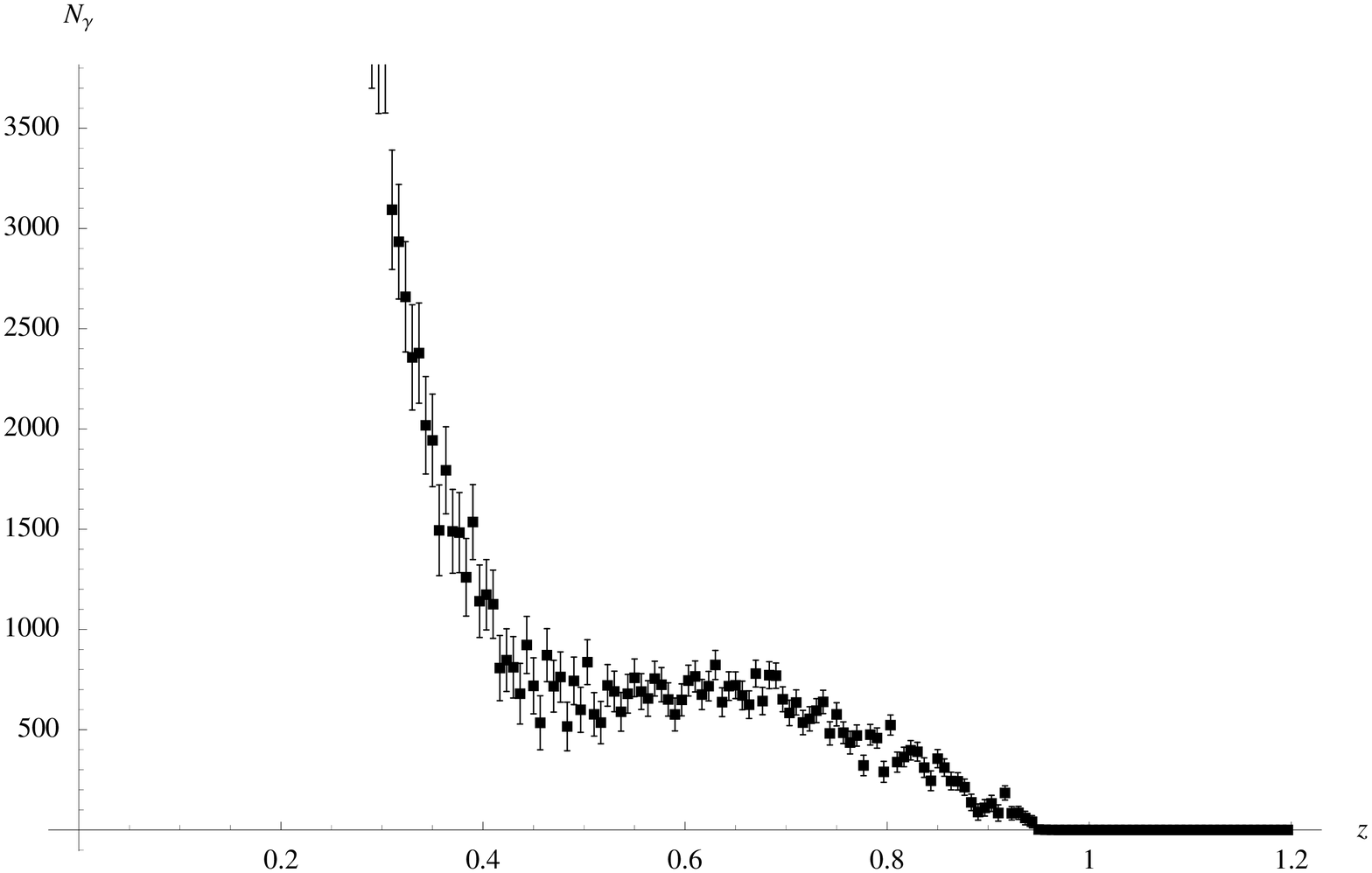}
\caption[CLEO data for $\Upsilon (2S)$ photon spectrum]{Background-subtracted
  CLEO data for the $\Upsilon (2S)$ photon spectrum \cite{Besson:2005jv}.}
\label{fignou2s}
\end{figure}

\begin{figure}
\centering
\includegraphics[width=12.5cm]{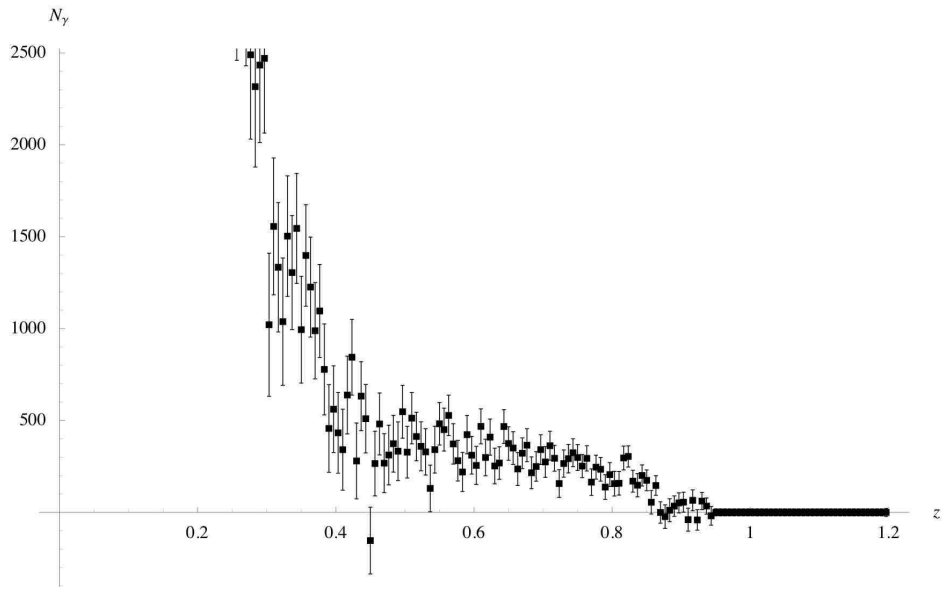}
\caption[CLEO data for $\Upsilon (3S)$ photon spectrum]{Background-subtracted
  CLEO data for the $\Upsilon (3S)$ photon spectrum \cite{Besson:2005jv}.}
\label{fignou3s}
\end{figure}

\begin{figure}
\centering
\includegraphics[width=12.5cm]{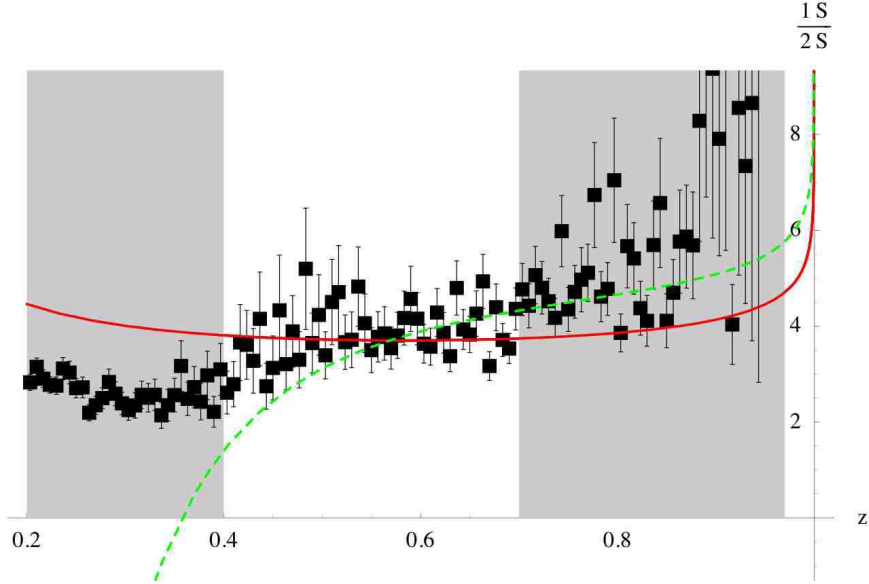}
\caption[Ratio of the $\Upsilon (1S)$ and $\Upsilon (2S)$ photon spectra]{Ratio of the $\Upsilon (1S)$ and $\Upsilon (2S)$ photon spectra. The points are obtained from the CLEO data \cite{Besson:2005jv}. The solid line is eq.(\ref{scsc}) (overall normalization fitted), the dashed line is the estimate of (\ref{nrqcd}) (see text). Agreement between the solid curve and the points in the central (unshaded) region would indicate that the two states are in the strong coupling regime.}
\label{fig1s2s}
\end{figure}

\begin{figure}
\centering
\includegraphics[width=12.5cm]{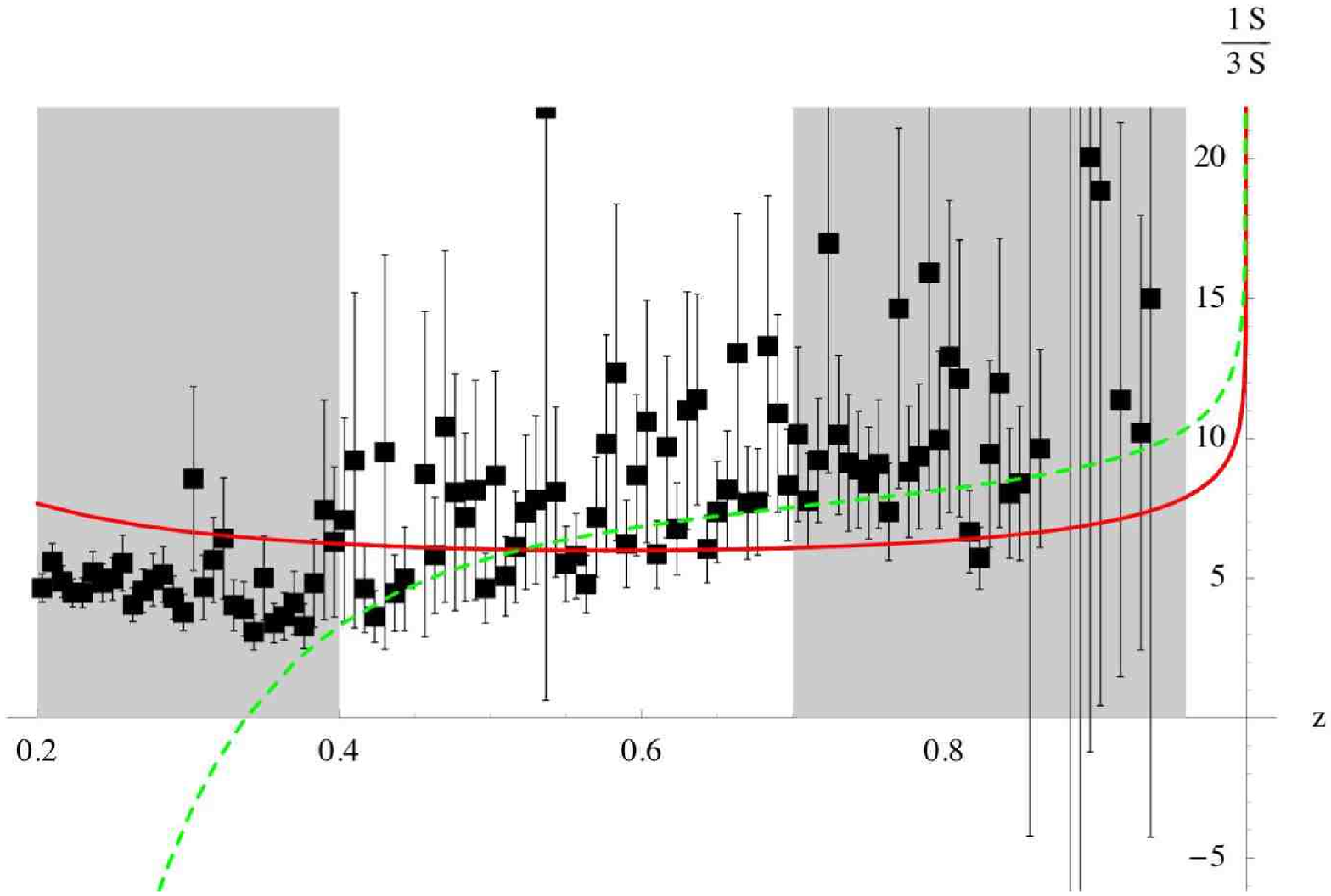}
\caption[Ratio of the $\Upsilon (1S)$ and $\Upsilon (3S)$ photon spectra]{Same as fig. \ref{fig1s2s} for 
$\Upsilon (1S)$ and $\Upsilon (3S)$ 
.} 
\label{fig1s3s}
\end{figure}

\begin{figure}
\centering
\includegraphics[width=12.5cm]{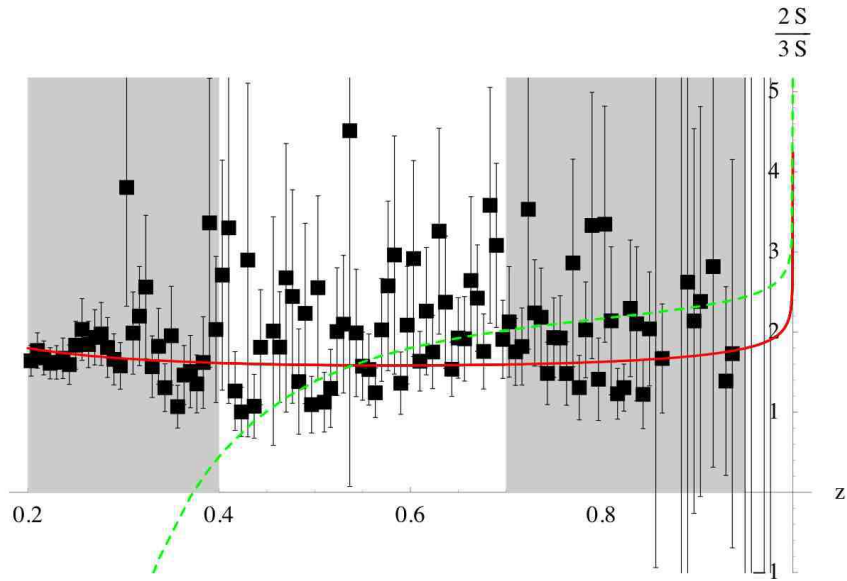}
\caption[Ratio of the $\Upsilon (2S)$ and $\Upsilon (3S)$ photon spectra]{Same as fig. \ref{fig1s2s} for 
$\Upsilon (2S)$ and $\Upsilon (3S)$ 
.} 
\label{fig2s3s}
\end{figure}

We will use the recent data from CLEO \cite{Besson:2005jv} (which includes a
very precise measurement of the $\Upsilon (1S)$ photon spectrum, as well as
measurements of the $\Upsilon (2S)$ and $\Upsilon (3S)$ photon spectra, see
figures \ref{fignou1s}, \ref{fignou2s} and \ref{fignou3s}) to
check our predictions. 
In order to do the comparison we use the following
procedure. First we efficiency correct the data (using the efficiencies
modeled by CLEO). Then we perform the ratios $1S/2S$, $1S/3S$ and $2S/3S$ (we
add the errors of the different spectra in quadrature). Now we want to discern
which of these ratios follow eq.(\ref{scsc}) and which ones deviate from it;
to do that we fit eq.(\ref{scsc}) to each of the ratios leaving only the
overall normalization as a free parameter (the experimental normalization is unknown). The fits are done in the central region, that is $z\in[0.4,0.7]$, where eq.(\ref{scsc}) holds. A good (bad) $\chi^2$ obtained from the fit will indicate that the ratio 
does (not) follow the shape dictated by eq.(\ref{scsc}).
In figures \ref{fig1s2s}, \ref{fig1s3s} and \ref{fig2s3s} we plot the ratios
$1S/2S$, $1S/3S$ and $2S/3S$ (respectively) together with eq.(\ref{scsc}) and
the estimate of (\ref{nrqcd}) mentioned above (overall normalizations fitted
for all curves, the number of d.o.f. is then $45$). The figures show the
spectra for $z\in[0.2,1]$ for an easier visualization but remember that we are
focusing in the central $z$ region, denoted by the unshaded region in the
plots. The theoretical errors due to higher orders in $\als (m)$ and in the expansions below (\ref{scsc}) are negligible with respect to the experimental ones. For the $1S/2S$ ratio we obtain a $\chi^2/\mathrm{d.o.f.}\vert_{1S/2S}\sim 1.2$, which corresponds to an $18\%$ CL.
The errors for the $\Upsilon (3S)$ photon spectrum are considerably larger
than those of the other two states (see
figures \ref{fignou1s}, \ref{fignou2s} and \ref{fignou3s}), this causes the ratios involving the $3S$ state to be less conclusive than the other one. In any case we obtain 
$\chi^2/\mathrm{d.o.f.}\vert_{1S/3S}\sim 0.9$, 
which corresponds to a $68\%$ CL,
and  $\chi^2/\mathrm{d.o.f.}\vert_{2S/3S}\sim 0.75$, which corresponds to an $89\%$ CL. Hence, the data disfavors $\Upsilon (1S)$ in the strong coupling regime but is consistent with $\Upsilon (2S)$ and $\Upsilon (3S)$ in it. %For completeness we also quote the numbers corresponding to the estimate of eq. (\ref{nrqcd}) (dashed lines):  $\chi^2/\mathrm{d.o.f.}\vert_{1S/2S}\sim 1.93$ ($.02\%$ CL), $\chi^2/\mathrm{d.o.f.}\vert_{1S/3S}\sim .56$ ($99\%$ CL) and $\chi^2/\mathrm{d.o.f.}\vert_{2S/3S}\sim .66$ ($96\%$ CL). Recall that these curves are only estimates to illustrate what the differences from eq. (\ref{scsc}) to eq. (\ref{nrqcd}) may be and do not intend to best fit data.

In summary, using pNRQCD we have worked out a model-independent formula which involves the photon spectra of two heavy quarkonium states and holds at NLO in the strong coupling regime.
When this formula is applied to the Upsilon system, current data indicate that
the $\Upsilon (2S)$ and the $\Upsilon (3S)$  are consistent as states in the
strong coupling regime\footnote{$\Upsilon (2S)$ also seems difficult to
  accomodate in a weak coupling picture in the analysis of the radiative
  transition $\Upsilon (2S)\to\eta_b\gamma$ \cite{Brambilla:2005zw}.} whereas the $\Upsilon (1S)$ in this regime is
disfavor. A decrease of the current experimental errors for $\Upsilon (2S)$
and, specially, for the $\Upsilon (3S)$ is necessary to confirm this
indication. This is important, since it would validate the use of the formulas in \cite{Brambilla:2002nu,Brambilla:2001xy,Brambilla:2003mu}, and others which may be derived in the future under the same assumptions, not only for the $\Upsilon (2S)$ and $\Upsilon (3S)$ but also for the $\chi_b (2P)$s, since their masses lie in between, as well as for their pseudoscalar partners.

%\section{A partir d'aqui venen els papers}

%\input{shpfct}
%\input{merg}

%%% Local Variables: 
%%% mode: latex
%%% TeX-master: t
%%% End: 

\chapter{Conclusions/Overview}\label{chapconcl}
In this thesis we have employed Effective Field Theory techniques to study the
heavy quark sector of the Standard Model. We have focused in three different
subjects. First, we have studied the singlet static QCD potential, employing
potential Non-Relativistic QCD. With the help of that effective theory we have
been able to determine the sub-leading infrared dependence of that static
potential. Among other possible applications, this calculation will enter in
the third order analysis of $t-\bar{t}$ production near threshold. An analysis
which will be needed for a future $e^+-e^-$ linear collider. After that we
have studied an anomalous dimension in Soft-Collinear Effective Theory. That
effective theory has very important applications in the field of
$B$-physics. A field which is of crucial importance for the indirect searches
of new physics effects (through the study of $CP$ violation and the CKM
matrix). And finally we have studied the semi-inclusive radiative decays of
heavy quarkonium to light particles, employing a combined use of potential
Non-Relativistic QCD and  Soft-Collinear Effective Theory. Viewed in
retrospect, that process can be seen as a nice example on how a process is
well described theoretically once one includes all the relevant degrees of
freedom (in the effective theory) and uses a well defined power counting. When we have the radiative decay understood, it can be used to determine properties of the decaying heavy quarkonia, as we have also shown in the thesis.

%%% Local Variables: 
%%% mode: latex
%%% TeX-master: t
%%% End: 

\appendix
\chapter{Definitions}
In this appendix we collect the definitions of some factors appearing
throughout the thesis.

\phantom{}

\noindent$\gamma_E$ is the Euler constant $\gamma_E=0.577216...$ . $\zeta(s)$
is the Riemann zeta function, with $\zeta(3)=1.2021...$ . The Euler beta
function is given by
\begin{equation}
B(\tau,\omega)=\frac{\Gamma(\tau)\Gamma(\omega)}{\Gamma(\tau+\omega)}
\end{equation}

\section{Color factors}
The color factors for an $SU(N_c)$ group are given by
\begin{equation}
C_f=\frac{N_c^2-1}{2N_c}\qquad C_A=N_c\qquad T_F=\frac{1}{2}
\end{equation}
\section{QCD beta function}
The strong coupling $\als=g^2/(4\pi)$ constant runs according to
\begin{equation}
\mu\frac{d\als}{d\mu}=-2\als\left\{\beta_0\frac{\als}{4\pi}+\beta_1\left(\frac{\als}{4\pi}\right)^2+\cdots\right\}
\end{equation}
The first coefficients of the QCD beta function are given by
\begin{equation}
\beta_0=\frac{11}{3}C_A-\frac{4}{3}T_Fn_f\qquad\beta_1=\frac{34}{3}C_A^2-\frac{20}{3}C_AT_Fn_f-4C_fT_Fn_f
\end{equation}
where, here and throughout the thesis, $n_f$ is the number of light flavors.

%%% Local Variables: 
%%% mode: latex
%%% TeX-master: t
%%% End: 

\chapter{Feynman rules}\label{appFR}
\section{pNRQCD}
\begin{figure}
\centering
\includegraphics{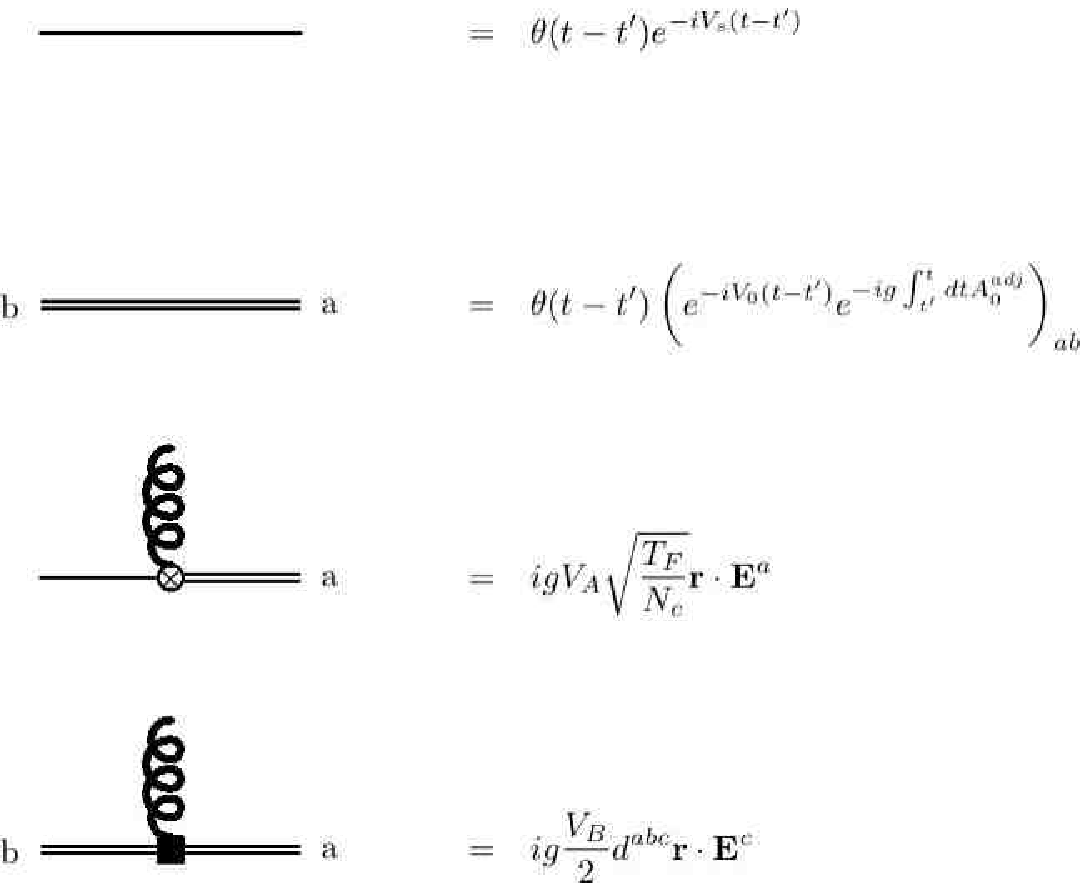}
\caption[Propagators and vertices in pNRQCD. Position space]{Propagators and vertices in pNRQCD in position space. We have displayed the rules at leading order in $1/m$ and order $r$ in the multipole expansion. If one wants to perform a perturbative calculation these rules must be expanded in $g$.}\label{figposs}
\end{figure}

\begin{figure}
\centering
\includegraphics{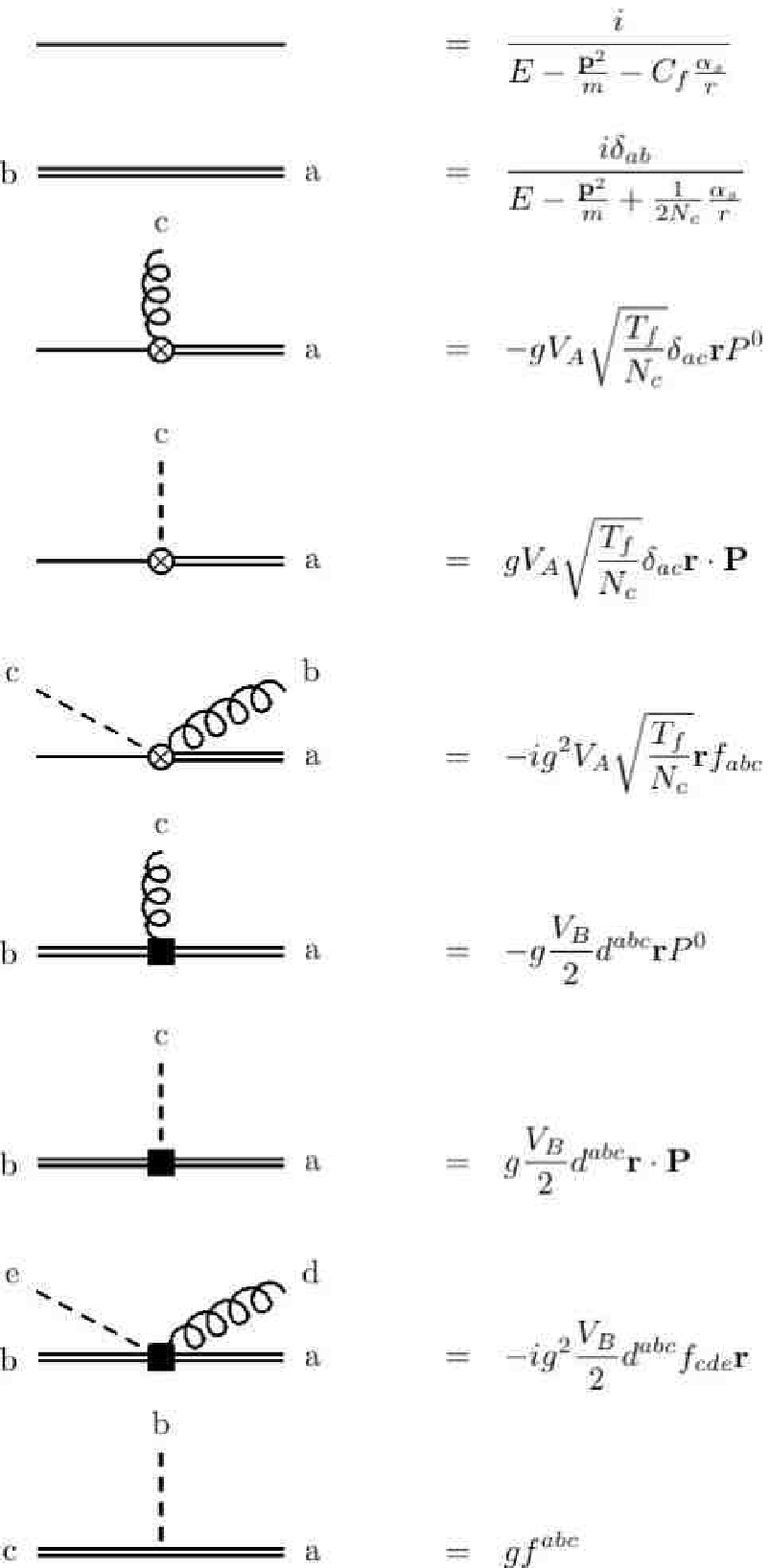}
\caption[Propagators and vertices in pNRQCD. Momentum space]{Propagators and vertices in pNRQCD in ultrasoft momentum space. $P^{\mu}$ is the gluon incoming momentum. Dashed lines represent longitudinal gluons and springy lines transverse ones.}\label{figmoms}
\end{figure}

The pNRQCD Lagrangian (\ref{pNRSO}) gives the position space rules for the vertices and propagators displayed in figure \ref{figposs}. Feynman rules in ultrasoft momentum space are also useful. These are displayed in figure \ref{figmoms} (additionally an insertion of a correction to the potential $\delta V$ in a singlet or octet propagator will give rise to a $-i\delta V$ factor).

\section{SCET}
\begin{figure}
\centering
\includegraphics{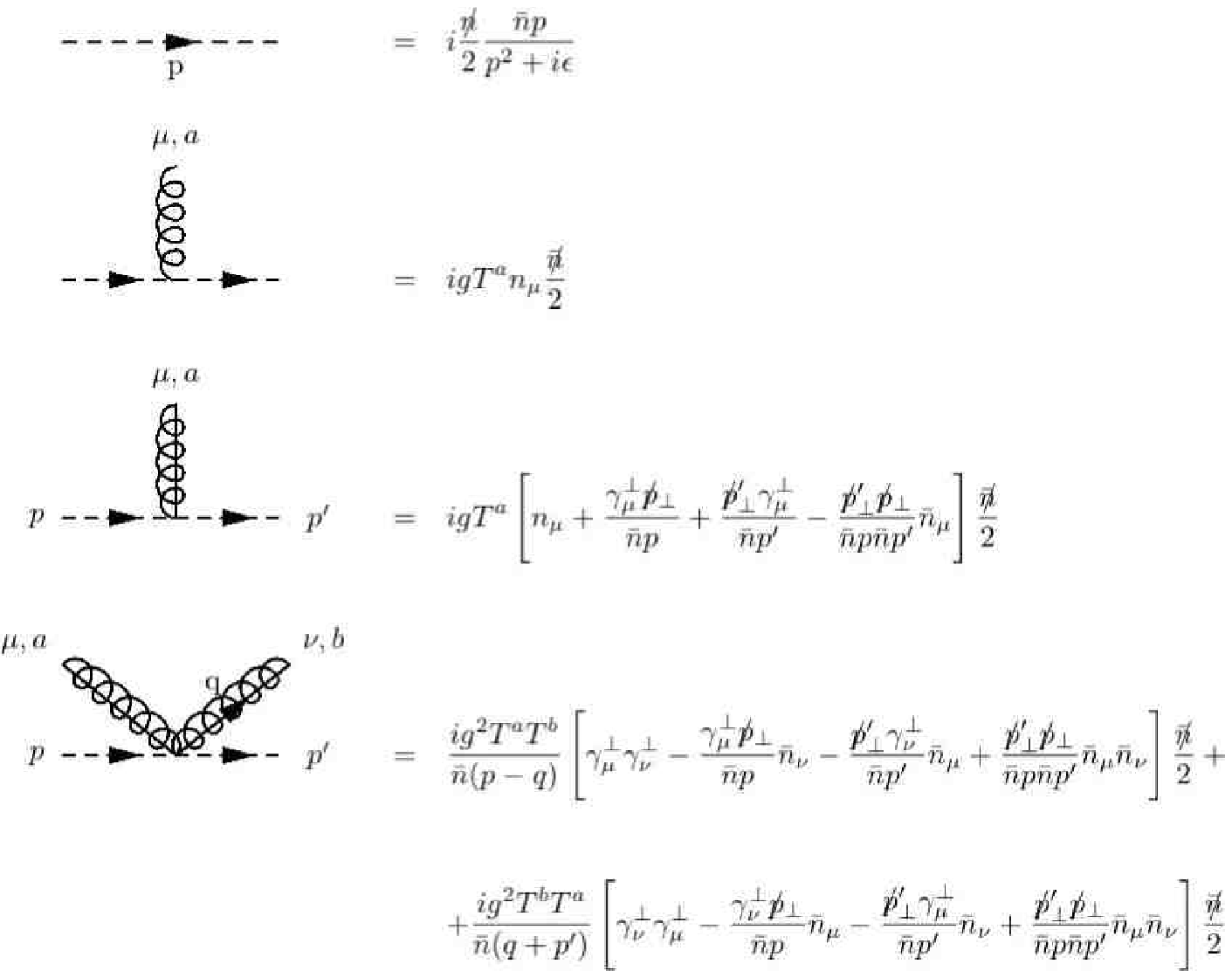}
\caption[$\mathcal{O}(\lambda^0)$ SCET Feynman rules]{Propagators and vertices arising from the SCET Lagrangian (\ref{LSCET}). We have just displayed the interactions with one and two collinear gluons, although interactions with an arbitrary number of them are allowed. Dashed lines are collinear quarks, springy lines are ultrasoft gluons and springy lines with a line inside are collinear gluons.}\label{figfrSCET}
\end{figure}

The Feynman rules that arise from the Lagrangian (\ref{LSCET}) are represented in figure \ref{figfrSCET}.

%\section{HQET}

%%% Local Variables: 
%%% mode: latex
%%% TeX-master: t
%%% End: 

\chapter{NRQCD matrix elements in the strong coupling regime}\label{appME}
First we list the four-fermion NRQCD operators that appear in the subsequent formulas.
\begin{eqnarray}
O_1({}^3S_1) &=& \psi^\dagger \mbox{\boldmath $\sigma$} \chi \cdot
\chi^\dagger \mbox{\boldmath $\sigma$} \psi \\
O_8({}^1S_0) &=& \psi^\dagger T^a \chi \, \chi^\dagger T^a \psi \\
O_8({}^3S_1) &=& \psi^\dagger \mbox{\boldmath $\sigma$} T^a \chi \cdot
\chi^\dagger \mbox{\boldmath $\sigma$} T^a \psi \\
{\cal P}_1({}^1S_0) &=& {1\over 2}
\left[\psi^\dagger \chi \, \chi^\dagger (-\mbox{$\frac{i}{2}$} \stackrel{\leftrightarrow}{\bf
D})^2 \psi \;+\; {\rm H.c.}\right] \\
{\cal P}_1({}^3S_1) &=& {1\over 2}\left[\psi^\dagger \mbox{\boldmath $\sigma$}
\chi
        \cdot \chi^\dagger \mbox{\boldmath $\sigma$} (-\mbox{$\frac{i}{2}$}
\stackrel{\leftrightarrow}{\bf D})^2 \psi \;+\; {\rm H.c.}\right] \\
O_8({}^1P_1) &=& \psi^\dagger (-\mbox{$\frac{i}{2}$} \stackrel{\leftrightarrow}{\bf D}) T^a\chi
        \cdot \chi^\dagger (-\mbox{$\frac{i}{2}$} \stackrel{\leftrightarrow}{\bf D}) T^a\psi \\
O_8({}^3P_{0}) &=&  {1 \over 3} \;
\psi^\dagger (-\mbox{$\frac{i}{2}$} \stackrel{\leftrightarrow}{\bf D} \cdot \mbox{\boldmath $\sigma$}) T^a\chi
        \, \chi^\dagger (-\mbox{$\frac{i}{2}$} \stackrel{\leftrightarrow}{\bf D} \cdot \mbox{\boldmath
$\sigma$}) T^a\psi \\
O_8({}^3P_{1}) &=&  {1 \over 2} \;
\psi^\dagger (-\mbox{$\frac{i}{2}$} \stackrel{\leftrightarrow}{\bf D} \times \mbox{\boldmath
$\sigma$}) T^a\chi
        \cdot \chi^\dagger (-\mbox{$\frac{i}{2}$} \stackrel{\leftrightarrow}{\bf D} \times
\mbox{\boldmath $\sigma$}) T^a\psi \\
O_8({}^3P_{2}) &=& \psi^\dagger (-\mbox{$\frac{i}{2}$} \stackrel{\leftrightarrow}{{\bf D}}{}^{(i}
\bfsigma^{j)}) T^a\chi
        \, \chi^\dagger (-\mbox{$\frac{i}{2}$} \stackrel{\leftrightarrow}{{\bf D}}{}^{(i} \bfsigma^{j)}) T^a\psi \\
O_{\rm EM}(^3S_1) &=& 
\psi^\dagger {\bfsigma} \chi |{\rm vac}\rangle \langle {\rm vac}|
\chi^\dagger {\bfsigma} \psi \\
{\cal P}_{\rm EM}(^1S_0) &=& {1\over 2} \left[ \psi^\dagger \chi 
|{\rm vac}\rangle \langle {\rm vac}|
\chi^\dagger \left( -{i\over 2} {\bf D}^2 \right)
\psi + {\rm H.c.} \right] \\
{\cal P}_{\rm EM}(^3S_1) &=& {1\over 2} \left[ \psi^\dagger {\bfsigma} \chi 
|{\rm vac}\rangle \langle {\rm vac}|
\chi^\dagger {\bfsigma} \left( -{i\over 2} {\bf D}^2 \right)
\psi + {\rm H.c.} \right]
\end{eqnarray}

In the strong coupling regime ($\Lambda_{\mathrm{QCD}}\gg E$) the following factorized formulas can be derived for the NRQCD matrix elements. The analytic contributions in $1/m$ for some $S$-wave states (we just display here expressions involving $S$-wave states, since are the only ones really used in the thesis, see \cite{Brambilla:2002nu} for a complete list), up to corrections of ${\cal O}(p^3/m^3 \times (\lQ^2/m^2, E/m))$, are given by 

\begin{eqnarray}
\label{O13S1}
&&\hspace{-8mm}
\langle V_Q(nS)|O_1(^3S_1)|V_Q(nS)\rangle^{1/m}=
C_A {|R^V_{n0}({0})|^2 \over 2\pi}
\left(1-{E_{n0}^{(0)} \over m}{2{\cal E}_3 \over 9}
+{2{\cal E}^{(2,t)}_3 \over 3 m^2 }+{c_F^2{\cal B}_1 \over 3 m^2 }\right)
\\
%&&\hspace{-8mm}
%\langle P_Q(nS)|O_1(^1S_0)|P_Q(nS)\rangle^{1/m}=
%C_A {|R^P_{n0}({0})|^2 \over 2\pi}
%\left(1-{E_{n0}^{(0)} \over m}{2{\cal E}_3 \over 9}
%+{2{\cal E}^{(2,t)}_3 \over 3 m^2}+{c_F^2{\cal B}_1 \over m^2}\right),
%\\
&&\hspace{-8mm}
\langle V_Q(nS)|O_{\rm EM}(^3S_1)|V_Q(nS)\rangle^{1/m}=
C_A {|R^V_{n0}({0})|^2 \over 2\pi}
\left(1-{E_{n0}^{(0)} \over m}{2{\cal E}_3 \over 9}
+{2{\cal E}^{(2,{\rm EM})}_3 \over 3 m^2}+{c_F^2{\cal B}_1 \over 3 m^2}\right)
\\
\label{OEM1S0}
%&&\hspace{-8mm}
%\langle P_Q(nS)|O_{\rm EM}(^1S_0)|P_Q(nS)\rangle^{1/m}=
%C_A {|R^P_{n0}({0})|^2 \over 2\pi}
%\left(1-{E_{n0}^{(0)} \over m}{2{\cal E}_3 \over 9}
%+{2{\cal E}^{(2,{\rm EM})}_3 \over 3 m^2}+{c_F^2{\cal B}_1 \over m^2}\right),
%\\
%&&\hspace{-8mm}
%\langle \chi_Q(nJS) | O_1(^{2S+1}P_J ) | \chi_Q(nJS) \rangle^{1/m} = 
%\langle \chi_Q(nJS) | O_{\rm EM}(^{2S+1}P_J ) | \chi_Q(nJS) \rangle^{1/m}  
%\nonumber\\
%&&\qquad\qquad\qquad\qquad\qquad\qquad\quad
%={3 \over 2}{C_A \over \pi} |R^{(0)\,\prime}_{n1}({0})|^2,
%\label{chio1}
%\\
&&\hspace{-8mm}
\langle V_Q(nS)|{\cal P}_1(^3S_1)|V_Q(nS)\rangle^{1/m}=
\langle P_Q(nS)|{\cal P}_1(^1S_0)|P_Q(nS)\rangle^{1/m}=
\nonumber\\
&&\hspace{-8mm}
\langle V_Q(nS)|{\cal P}_{\rm EM}(^3S_1)|V_Q(nS)\rangle^{1/m}=
\langle P_Q(nS)|{\cal P}_{\rm EM}(^1S_0)|P_Q(nS)\rangle^{1/m}
\nonumber\\
&&\qquad\qquad\qquad\qquad\qquad\qquad\quad
=C_A {|R^{(0)}_{n0}({0})|^2 \over 2\pi}
\left(m E_{n0}^{(0)} -{\cal E}_1 \right)
\label{P13S1}
\\
&&\hspace{-8mm}
\langle V_Q(nS)|O_8(^3S_1)|V_Q(nS)\rangle^{1/m}=
\langle P_Q(nS)|O_8(^1S_0)|P_Q(nS)\rangle^{1/m}
\nonumber\\
&&\qquad\qquad\qquad\qquad\qquad\qquad\quad
=C_A {|R^{(0)}_{n0}({0})|^2 \over 2\pi}
\left(- {2 (C_A/2-C_f) {\cal E}^{(2)}_3 \over 3 m^2 }\right)
\\
&&\hspace{-8mm}
\langle V_Q(nS)|O_8(^1S_0)|V_Q(nS)\rangle^{1/m}=
{\langle P_Q(nS)|O_8(^3S_1)|P_Q(nS)\rangle^{1/m} \over 3}
\nonumber\\
&&\qquad\qquad\qquad\qquad\qquad\qquad\quad
=C_A {|R^{(0)}_{n0}({0})|^2 \over 2\pi}
\left(-{(C_A/2-C_f) c_F^2{\cal B}_1 \over 3 m^2 }\right)
\\
&&\hspace{-8mm}
{\langle V_Q(nS)|O_8(^3P_J)|V_Q(nS)\rangle^{1/m} \over 2J+1}=
{\langle P_Q(nS)|O_8(^1P_1)|P_Q(nS)\rangle^{1/m} \over 9}
\nonumber\\
&&\qquad\qquad\qquad\qquad\qquad\qquad\quad
=
\,C_A {|R^{(0)}_{n0}({0})|^2 \over 2\pi}
\left(-{(C_A/2-C_f) {\cal E}_1 \over 9 }\right)
\\
%&&\hspace{-8mm}
%\langle \chi_Q(nJS)\vert O_8(^1S_0)\vert \chi_Q(nJS) \rangle^{1/m}
%= {T_F\over 3}
%{\vert R^{(0)\,\prime}_{n1}({0})\vert^2 \over \pi m^2} {\cal E}_3,
%\label{matoct}
\end{eqnarray}
There are also non-analytic contributions in $1/m$. Up to corrections of order ${\cal O}(p^3/m^3 \times \lQ/m \times m\als/\sqrt{m\,\lQ})$ they are given by (see \cite{Brambilla:2003mu} for the complete list of known corrections)
\begin{eqnarray}
\langle V_Q(nS)|O_1(^3S_1)|V_Q(nS)\rangle^{1/\sqrt{m}}&=&
\langle V_Q(nS)|O_{\rm EM}(^3S_1)|V_Q(nS)\rangle^{1/\sqrt{m}}
\nonumber\\
&=& C_A {|R^V_{n0}({0})|^2 \over 2\pi}
\left(1+  {4 (2\,C_f+C_A)\over 3\Gamma(7/2)} \, \; {\als\,
 {\cal E}^E_{5/2}\over m^{1/2}}\right)
\label{O3S1nonan}
\end{eqnarray}
In all those expressions $R$ represents the radial part of the wave function, $E$ the binding energy and all the $\mathcal{E}$ and $\mathcal{B}$ are universal (bound state independent) non-perturbative parameters.

%%% Local Variables: 
%%% mode: latex
%%% TeX-master: t
%%% End: 

%%%%%%%%%%%%%%%%%%%%%%%%%%%%%%%%%%%%%%%%%%%%%%%%
%%%%%%%%%%%%%%%%%%%%%%%%%%%%%%%%%%%%%%%%%%%%%%%%

%%% Local Variables: 
%%% mode: latex
%%% TeX-master: t
%%% End: 

\end{document}